\documentclass[twocolumn]{aastex63}
\usepackage{longtable}
\usepackage{tabularx}
\usepackage{enumitem}   
\usepackage{tablefootnote}
\usepackage{xcolor}

\usepackage{amsmath}
\usepackage{slashbox}
\usepackage{subfigure}
\usepackage{hyperref}
\usepackage[mathscr]{euscript}
\usepackage{booktabs}
\usepackage{placeins}
\submitjournal{ApJ}

\shorttitle{CSP Hubble Constant}
\shortauthors{Uddin et al.}
\graphicspath{{./}{figures/}}
\begin{document}

\title{Carnegie Supernova Project-I $\&$ -II: Measurements of $H_0$ using Cepheid, TRGB, and SBF Distance Calibration to Type Ia Supernovae  \footnote{This paper includes data gathered with the 6.5 meter Magellan Telescopes located at Las Campanas Observatory, Chile.}}

\correspondingauthor{Syed~A.~Uddin}
\email{saushuvo@gmail.com}

\author[0000-0002-9413-4186]{Syed~A.~Uddin}
\affiliation{Observatories of the Carnegie Institution for Science, 813 Santa Barbara St, Pasadena, CA, 91101, USA}

\affiliation{George P. and Cynthia Woods Mitchell Institute for Fundamental Physics and Astronomy, Texas A\&M University,
Department of Physics and Astronomy,  College Station, TX 77843, USA}
\affiliation{US Naval Observatory, 3450 Massachusetts Ave NW, Washington, DC 20392}
\affiliation{Center for Space Studies, American Public University System, 111 W. Congress Street, Charles Town, WV 25414, USA }

\author[0000-0003-4625-6629]{Christopher~R.~Burns}
\affiliation{Observatories of the Carnegie Institution for Science, 813 Santa Barbara St, Pasadena, CA, 91101, USA}

\author[0000-0003-2734-0796]{M~M.~Phillips}
\affiliation{Carnegie Observatories, Las Campanas Observatory, Casilla 601, La Serena, Chile}

\author[0000-0002-8102-181X]{Nicholas~B.~Suntzeff}
\affiliation{George P. and Cynthia Woods Mitchell Institute for Fundamental Physics and Astronomy, Texas A\&M University,
Department of Physics and Astronomy,  College Station, TX 77843, USA}

\author[0000-0003-3431-9135]{Wendy~L.~Freedman}
\affiliation{Observatories of the Carnegie Institution for Science, 813 Santa Barbara St., Pasadena, CA 91101, USA}
\affiliation{Department of Astronomy and Astrophysics, University of Chicago, 5640 S. Ellis Ave, Chicago, IL 60637, USA}

\author[0000-0001-6272-5507]{Peter~J.~Brown}
\affiliation{George P. and Cynthia Woods Mitchell Institute for Fundamental Physics and Astronomy, Texas A\&M University,
Department of Physics and Astronomy,  College Station, TX 77843, USA}

\author[0000-0003-2535-3091]{Nidia~Morrell}
\affiliation{Carnegie Observatories, Las Campanas Observatory, Casilla 601, La Serena, Chile}

\author[0000-0001-7981-8320]{Mario~Hamuy}
\affiliation{Hagler Institute for Advanced Studies, Texas A\&M University, College Station, TX 77843, USA}
\affiliation{Fundación Chilena de Astronomía, El Vergel 2252, Santiago, Chile}

\author[0000-0002-6650-694X]{Kevin~Krisciunas}
\affiliation{George P. and Cynthia Woods Mitchell Institute for Fundamental Physics and Astronomy, Texas A\&M University,
Department of Physics and Astronomy,  College Station, TX 77843, USA}

\author[0000-0001-7092-9374]{Lifan~Wang}
\affiliation{George P. and Cynthia Woods Mitchell Institute for Fundamental Physics and Astronomy, Texas A\&M University,
Department of Physics and Astronomy,  College Station, TX 77843, USA}

\author[0000-0003-1039-2928]{Eric~Y.~Hsiao}
\affiliation{Department of Physics, Florida State University, 77 Chieftan Way, Tallahassee, FL  32306, USA}

\author[0000-0002-4163-4996]{Ariel~Goobar}
\affiliation{The Oskar Klein Centre, Department of Physics, Stockholm University, SE-106 91 Stockholm, Sweden}

\author[0000-0002-4436-4661]{Saul~Perlmutter}
\affiliation{Lawrence Berkeley National Laboratory, Department of Physics, 1 Cyclotron Road, Berkeley, CA 94720, USA}
\affiliation{Astronomy Department, University of California at Berkeley, Berkeley, CA 94720, USA}

\author[0000-0002-3900-1452]{Jing~Lu}
\affiliation{Department of Physics, Florida State University, 77 Chieftan Way, Tallahassee, FL  32306, USA}

\author[0000-0002-5571-1833]{Maximilian~Stritzinger}
\affiliation{Department of Physics and Astronomy, Aarhus University, Ny Munkegade 120, DK-8000 Aarhus C, Denmark}

\author[0000-0003-0227-3451]{Joseph~P.~Anderson}
\affiliation{European Southern Observatory, Alonso de C\'{o}rdova 3107, Casilla 19, Santiago, Chile}

\author[0000-0002-5221-7557]{Chris~Ashall}
\affiliation{Department of Physics, Virginia Tech, Blacksburg, VA 24061, USA}

\author[0000-0002-4338-6586]{Peter~Hoeflich}
\affiliation{Department of Physics, Florida State University, 77 Chieftan Way, Tallahassee, FL  32306, USA}

\author[0000-0003-4631-1149]{Benjamin~J.~Shappee}
\affiliation{Institute for Astronomy, University of Hawaii, 2680 Woodlawn Drive, Honolulu, HI 96822, USA}

\author[0000-0003-0554-7083]{S.~E.~Persson}
\affiliation{Observatories of the Carnegie Institution for Science, 813 Santa Barbara St., Pasadena, CA 91101, USA}

\author[0000-0001-6806-0673]{Anthony~L.~Piro}
\affiliation{Observatories of the Carnegie Institution for Science, 813 Santa Barbara St., Pasadena, CA 91101, USA}

\author[0000-0001-5393-1608]{E~Baron}
\affiliation{Planetary Science Institute, 1700 East Fort Lowell Road, Suite 106, Tucson, AZ 85719-2395, USA}
\affiliation{Hamburger Sternwarte, Gojenbergsweg 112, D-21029 Hamburg, Germany}
\affiliation{Homer L. Dodge Department of Physics and Astronomy, University of Oklahoma, 440 W. Brooks, Rm 100, Norman, OK 73019-2061, USA}

\author[0000-0001-6293-9062]{Carlos~Contreras}
\affiliation{Carnegie Observatories, Las Campanas Observatory, Casilla 601, La Serena, Chile}
\author[0000-0002-1296-6887]{Lluís~Galbany}
\affiliation{Institute of Space Sciences (ICE, CSIC), Campus UAB, Carrer de Can Magrans, s/n, E-08193 Barcelona, Spain}
\affiliation{Institut d’Estudis Espacials de Catalunya (IEEC), E-08034 Barcelona, Spain}

\author[0000-0001-8367-7591]{Sahana~Kumar}
\affiliation{Department of Physics, Florida State University, 77 Chieftan Way, Tallahassee, FL  32306, USA}
\author[0000-0002-9301-5302]{Melissa~Shahbandeh}
\affiliation{Department of Physics, Florida State University, 77 Chieftan Way, Tallahassee, FL  32306, USA}

\author[0000-0002-2806-5821]{Scott~Davis}
\affiliation{Department of Physics, University of California, 1 Shields Avenue, Davis, CA 95616-5270, USA}

\author[0000-0001-9051-1338]{Jorge~Anais}
\affiliation{Centro de Astronom\'ia (CITEVA), Universidad de Antofagasta, Av. Angamos 601, Antofagasta, Chile}

\author[0000-0001-9952-0652]{Luis~Busta}
\affiliation{Carnegie Observatories, Las Campanas Observatory, Casilla 601, La Serena, Chile}

\author[0000-0002-3829-9920]{Abdo~Campillay}
\affiliation{Carnegie Observatories, Las Campanas Observatory, Casilla 601, La Serena, Chile}
\affiliation{Departamento de F\'{i}sica, Universidad de La Serena, Cisternas 1200, La Serena, Chile}

\author{Sergio~Castell\'{o}n}
\affiliation{Carnegie Observatories, Las Campanas Observatory, Casilla 601, La Serena, Chile}

\author{Carlos~Corco}
\affiliation{Carnegie Observatories, Las Campanas Observatory, Casilla 601, La Serena, Chile}
\affiliation{SOAR Telescope, Casilla 603, La Serena, Chile}

\author[0000-0002-0805-1908]{Tiara~Diamond}
\affiliation{Department of Physics, Florida State University, 77 Chieftan Way, Tallahassee, FL  32306, USA}
\affiliation{Laboratory of Observational Cosmology, Code 665, NASA Goddard Space Flight Center, Greenbelt, MD 20771, USA}

\author[0000-0002-8526-3963]{Christa~Gall}
\affiliation{DARK, Niels Bohr Institute, University of Copenhagen, Jagtvej 128, 2200 Copenhagen, Denmark}

\author{Consuelo~Gonzalez}
\affiliation{Carnegie Observatories, Las Campanas Observatory, Casilla 601, La Serena, Chile}

\author[0000-0002-3415-322X]{Simon~Holmbo}
\affiliation{Department of Physics and Astronomy, Aarhus University, Ny Munkegade 120, DK-8000 Aarhus C, Denmark}

\author[0000-0001-5179-980X]{Miguel~Roth},
\affiliation{Carnegie Observatories, Las Campanas Observatory, Casilla 601, La Serena, Chile}
\affiliation{GMTO Corporation, Presidente Riesco 5335, Of. 501, Nueva Las Condes, Santiago}

\author[0000-0002-8303-776X]{Jacqueline~Ser\'{o}n}
\affiliation{Cerro Tololo Inter-American Observatory/NSF’s NOIRLab, Casilla 603, La
Serena, Chile}

\author[0000-0002-2387-6801]{Francesco~Taddia}
\affiliation{The Oskar Klein Centre, Department of Astronomy, Stockholm University, SE-106 91 Stockholm, Sweden}

\author{Sim\'{o}n~Torres}
\affiliation{SOAR Telescope, Casilla 603, La Serena, Chile}

\author[0000-0003-0424-8719]{Charles~Baltay}
\affiliation{Department of Physics, Yale University, 217 Prospect Street, New Haven, CT 06511, USA}

\author[0000-0001-5247-1486]{Gast\'{o}n Folatelli}
\affiliation{Facultad de Ciencias Astron\'{o}micas y Geof\'{i}sicas, Universidad Nacional de La Plata, Instituto de Astrof\'{i}sica de La Plata (IALP), 
CONICET, Paseo del Bosque S/N, B1900FWA La Plata, Argentina}

\author{Ellie~Hadjiyska}
\affiliation{Department of Physics, Yale University, 217 Prospect Street, New Haven, CT 06511, USA}

\author[0000-0002-5619-4938]{Mansi~Kasliwal}
\affiliation{Caltech, 1200 East California Boulevard, MC 249-17, Pasadena, CA 91125, USA}

\author[0000-0002-3389-0586]{Peter~E.~Nugent}
\affiliation{Lawrence Berkeley National Laboratory, Department of Physics, 1 Cyclotron Road, Berkeley, CA 94720, USA}
\affiliation{Astronomy Department, University of California at Berkeley, Berkeley, CA 94720, USA}

\author{David~Rabinowitz}
\affiliation{Department of Physics, Yale University, 217 Prospect Street, New Haven, CT 06511, USA}

\author[0000-0003-4501-8100]{Stuart~D.~Ryder}
\affiliation{School of Mathematical and Physical Sciences, Macquarie University, NSW 2109, Australia}
\affiliation{Astronomy, Astrophysics and Astrophotonics Research Centre, Macquarie University, Sydney, NSW 2109, Australia}

%\author[0000-0001-6589-1287]{Brian~P.~Schmidt}
%\affiliation{The Research School of Astronomy and Astrophysics, Australian National University, ACT 2601, Australia}

%\author{Emma~S.~Walker}
%\affiliation{Department of Physics, Yale University, 217 Prospect Street, New Haven, CT 06511, USA}



\begin{abstract}

We present an analysis of Type Ia Supernovae (SNe~Ia) from both the Carnegie Supernova Project~I (CSP-I) and II (CSP-II), and extend the Hubble diagram from the optical to the near-infrared wavelengths ($uBgVriYJH$). We calculate the Hubble constant, $H_0$, using various distance calibrators: Cepheids, Tip of the Red Giant Branch (TRGB), and Surface Brightness Fluctuations (SBF). Combining all methods of calibrations, we derive $\rm H_0=71.76 \pm 0.58 \ (stat) \pm 1.19 \ (sys) \ km \ s^{-1} \ Mpc^{-1}$ from $B$-band, and  $\rm H_0=73.22 \pm 0.68 \ (stat) \pm 1.28 \ (sys) \ km \ s^{-1} \ Mpc^{-1}$ from $H$-band. By assigning equal weight to the Cepheid, TRGB, and SBF calibrators, we derive the systematic errors required for consistency in the first rung of the distance ladder, resulting in a systematic error of $1.2\sim 1.3 \rm \ km \ s^{-1} \ Mpc^{-1}$ in $H_0$. As a result, relative to the statistics-only uncertainty, the tension between the late-time $H_0$ we derive by combining the various distance calibrators and the early-time $H_0$ from the Cosmic Microwave Background is reduced. The highest precision in SN~Ia luminosity is found in the $Y$ band ($0.12\pm0.01$ mag), as defined by the intrinsic scatter ($\sigma_{int}$). 
We revisit SN~Ia Hubble residual-host mass correlations and recover previous results that these correlations do not change significantly between the optical and the near-infrared wavelengths. 
%We recommend further studies on this correlation in the ultraviolet wavelengths to see if metallicity drives these correlations, since explanations based on dust law or progenitor age are not supported by our analysis. 
Finally, SNe~Ia that explode beyond 10 kpc from their host centers exhibit smaller dispersion in their luminosity, confirming our earlier findings. Reduced effect of dust in the outskirt of hosts may be responsible for this effect. 

%While the origin of this finding is not clearly known, we find that SNe~Ia exploding beyond 10 kpc are relatively redder, and explode in more massive hosts.

%\crb{Right now, this is very heavy on the $H_0$ and calibrators. We want
%to highlight the SNe~Ia, especially the host galaxy properties. I suggested a beginning of the abstract. Maybe build on that.}

\end{abstract}
%From multi-band analysis we find the TRGB calibration in $YJH$ gives a larger value of $H_0$ compared to those from the optical.
\section{Introduction}\label{sec:intro}

A solution to the field equations of the General Theory of Relativity (\citealt{einstein1915}) points to a Universe that must be expanding. Aleksandr Friedmann and George Lema\^{i}tre independently predicted this Universal expansion in the early twentieth century (\citealp{friedmann22, lemaitre27}).
%While George Lema\^{i}tre predicted an expansion in the form of $V/c = (R'/R)\ r$,
%Edwin Hubble observationally established a relation between the distances and the recessional velocities of extra-galactic objects \citep{hubble29}, demonstrating that the Universe was expanding as predicted. Such a relation is known as the Hubble-Lema\^{i}tre law, expressed as $V=H_0D$, where $V$ is the recessional velocity and $D$ is the distance\footnote{In Lema\^{i}tre's expression $c$ is the speed of light in vacuum, $r$ is the distance from earth, and $R'/R$ is a constant which he calculated to be $\rm 0.68\times 10^{-27} \ cm^{-1}.$}.The slope of the Hubble-Lema\^{i}tre law, $H_0$, corresponds to the current expansion rate of the Universe, and is known as the Hubble constant. This constant is one of the six fundamental parameters in the $\rm \Lambda CDM$ cosmological framework, where the expansion of the Universe is accelerating due to an unknown force called Dark Energy (\citealt{perlmutter99}, \citealt{riess98}).
Edwin Hubble used the Period-Luminosity relation of classical Cepheid variable stars (Leavitt Law; \citealt{leavitt12}) to measure distances to external galaxies (\citealt{hubble1926}), revolutionizing the understanding of the distance scale of the Universe. These new distance measurements and techniques, in turn, led to estimates of the local expansion of the Universe which is commonly parameterized as the Hubble constant, $H_0$. This constant is one of the six fundamental parameters in the $\rm \Lambda CDM$ cosmological framework, where the expansion of the Universe is accelerating due to an unknown force called dark energy (\citealt{perlmutter99}, \citealt{riess98}).

The first measurements of $H_0$ using distances tied to the 1926 work of Hubble and velocities from Slipher (\citealt{slipher1915,slipher1917}) were made by Lema{\^i}tre (\citealt{lemaitre27}) and Robertson (\citealt{robertson1928}), who found values of $\rm 625\ and\ 490\  km\ s^{-1} Mpc^{-1}$. Until the 1927 work  of Lema{\^i}tre, where an expansion in the form of $V/c = (R'/R)\ r$ was predicted\footnote{In Lema\^{i}tre's expression $V$ is the recession velocity, $c$ is the speed of light in vacuum, $r$ is the distance from earth, and $R'/R$ is a constant in which $R$ is the scale factor. Lema{\^i}tre calculated $R'/R$ to be $\rm 0.68\times 10^{-27} \ cm^{-1}.$}, there was some confusion whether there was a quadratic term in the distance-velocity relationship in  de Sitter's ``Solution B" (\citealt{desitter1916a, desitter1916b}), an empty and expanding universe with a cosmological constant. Both Lema{\^i}tre and Robertson showed that in their world models, the local expansion should be linear in theory, and they fit the distance-velocity relationship as such. Neither stressed that the data {\it required} a linear fit.

Hubble published his value for the expansion of the universe in 1929 of $\rm 500\ km \ s^{-1} Mpc^{-1}$ ({\citealt{hubble29}}). He did not cite the source of the distances and velocities used, but it is clear that most of the velocities came from Slipher. These values of $H_0$ are much higher than the modern values due to a number of factors including stellar crowding, the admixture of W Virginis (Pop II) Cepheids, and refinement in the photometric scale. As with the previous work of Lema{\^i}tre and Robinson, Hubble fit a linear law but did not show that the fit must be linear. What was new was that he plotted the data in his famous Figure 1 which showed that a linear relation was a good fit to the data. The issue of the order of the fit was laid to rest in the monumental paper of \cite{hubble_humason1931} when the new data on galaxy velocities and magnitudes from Mt. Wilson became available, which extended out to the Leo and Ursa Majoris clusters, and velocities up to 20,000   $\rm km\ s^{-1}$. We now refer to the linear relationship between distance and velocity as the ``Hubble-Lema{\^i}tre" law\footnote{This law is expressed as $V=H_0D$, where $V$ is the recession velocity and $D$ is the distance to external galaxies.}. Excellent reviews of the history of early measurements of the Hubble constant can be found in \cite{kragh_smith2003} and 
somewhat tongue-in-cheek in \cite{trimble2013}.

The value of $H_0$ published by Hubble gave the Universe an age of 2 billion years, while the radioactive dating showed that the Earth was 3 billion years old\footnote{\href{https://lweb.cfa.harvard.edu/~dfabricant/huchra/hubble/}{https://lweb.cfa.harvard.edu/~dfabricant/huchra/hubble/}}. This apparent contradiction was not resolved until the mid-1950s when \cite{humason56} published a reduced value of $\rm H_0 =180\pm36 \ km \ s^{-1} Mpc^{-1}$. While \cite{sandage58} reduced it further down to $\rm 75\pm25 \ km \ s^{-1} Mpc^{-1}$, after a series of papers, \cite{sandage82} reached a global value of $\rm H_0 =50\pm 7 \ km \ s^{-1} Mpc^{-1}$. A number of $H_0$ measurements ranging between 100 and 200 $\rm\ km \ s^{-1} Mpc^{-1}$ were also published during this time by Sydney van den Berg and Gerard deVaucouleurs (see \citealt{trimble96} for a historical perspective on this).

Besides Cepheids, a number of other methods have been developed to measure extra-galactic distances (and $H_0$). \textbf{For example,  \cite{sandage74} introduced a distance ladder concept for extra-galactic H~II regions.} \cite{tully77} proposed a method in which the line profile widths of global neutral hydrogen in spiral galaxies are correlated with their luminosity, and obtained $\rm\ H_0=80\pm 8 \ km \ s^{-1} Mpc^{-1}$. \cite{mould80} also used the Tully-Fisher relation in the near-infrared to obtain $\rm H_0=65\pm 4 \ km \ s^{-1} Mpc^{-1}$. \textbf{\cite{jacoby92} summarizes various methods on extra-galactic distance measurements.}   

In recent times, a number of other methods have emerged such as the Tip of the Red Giant Branch method (TRGB; \citealt{madore95}), Surface Brightness Fluctuations (SBF; \citealt{tonry88}, \citealt{ferrarese00}), time-delay cosmography (\citealt{refsdal64}, \citealt{shajib20}), and more recently, gravitational-wave standard sirens (\citealt{holz05}, \citealt{vitale18}). These extra-galactic distance measurements (except the time-delay cosmography\footnote{Time-delay cosmography is also a geometrical distance indicator.} and the gravitational-wave standard sirens) are anchored to geometrical distances, such as stellar parallaxes, detached eclipsing binaries (\citealt{deb19}), and masers (\citealt{reid19}). However, most of these methods are for nearby galaxies where peculiar velocities from local matter density fluctuations are of a similar order of magnitude as the cosmological redshift. Therefore, more distant cosmological probes are required to accurately measure the expansion rate.

%\crb{Time-delay cosmography doesn’t depend on geometrical distance indicators. It is *itself* a geometrical distance indicator or sorts.}
%The pioneering discovery of the luminosity-light-curve width relation by \citealp{phillips93} paved the way to standardize SN Ia  luminosity within $\sim 0.15$ mag.

The first use of SNe to measure the expansion rate of the Universe was published in \cite{zwicky61} who reported an upper limit of $H_0$ of $\rm 175 \ km \ s^{-1} Mpc^{-1}$ from the observations of 10 SNe\footnote{It was not clear from the paper that whether these 10 SNe were SNe~Ia or a mix of various types.} in the Virgo cluster. The first use of SN~Ia in the Hubble flow was presented in \cite{kowal68}, who derived an average absolute magnitude of $\rm -16.8+5log(H_0/100)$ from 33 SNe~Ia, with a dispersion of 0.6 mag. Later, \cite{branch73} made an extensive spectroscopic study of SNe~Ia, determined a peak luminosity of $-20.8^{+0.9}_{-0.7}$ mag, and estimated  $\rm H_0=40^{+25}_{-13} \ km \ s^{-1} Mpc^{-1}$. In the following year, \cite{kirshner74} used Type II SNe and the expanding photosphere method to obtain $\rm H_0=65\pm 15 \ km \ s^{-1} Mpc^{-1}$. While most values of $H_0$ using SNe~Ia were hovering between $\rm 50 - 60 \ km \ s^{-1} Mpc^{-1}$ until the early 1990s (e.g., \citealp{cadonau85, tammann90, branch93, sandage93}), \cite{fukugita91} presented values of $H_0$ that ranged between $\rm 75 - 100 \ km \ s^{-1} Mpc^{-1}$. As \cite{arnett85} provided theoretical reasoning for SNe~Ia to be good standard candles, improvements in SNe~Ia luminosity calibration would be the key to make precision measurements of $H_0$, and other cosmological parameters.

\cite{rust1974} and \citet{pskovskii1977, pskovskii1984} suggested there existed a relationship between the peak luminosity in the $B$-band and light curve width for SNe~Ia. \cite{phillips93} pioneered the widely used "luminosity-width" relationship for the light curves of SNe~Ia and showed that SN~Ia luminosity can be precisely standardized within $\pm 0.15$ mag using the luminosity-width relation with the later addition of a luminosity-color relation (\citealt{tripp98})\footnote{Recently, \cite{boone21} published a spectrophotometry-based standardization method trained on 173 nearby SNe~Ia that yielded an rms dispersion of $\pm0.084$ mag in a carefully blinded analysis.}. The relationship was further refined by the Cal\'an-Tololo Supernova Project (\citealt{hamuy1996a}). With improved standardization of SN~Ia luminosity, \cite{hamuy95} derived $\rm H_0\simeq 62-67 \ km \ s^{-1} Mpc^{-1}$ and \cite{riess95} derived $\rm H_0=67\pm 7 \ km \ s^{-1} Mpc^{-1}$. A few other studies, such as the one in \cite{hoeflich96}, used theoretical models to predict SNe~Ia luminosity, and derived $\rm H_0=67\pm 9 \ km \ s^{-1} Mpc^{-1}$.

SNe~Ia thus became the most reliable tool to measure luminosity distances at higher redshifts, corresponding to look-back times of billions of years, where the observed velocity becomes cosmologically dominated (see \citealp{leibundgut18, goobar11} for  general reviews on SN~Ia cosmology).
Thanks to their brightness and the use of rolling searches, numerous SNe~Ia can be discovered in a short time.  They can be used to measure the expansion rate far into the Universe, fairly accurately, when calibrated with Cepheids, TRGB, SBF, or any other appropriate distance calibrators (e.g., \citealp{burns18, riess21, freedman21, khetan21, garnavich22}).

The Hubble Space Telescope (HST) Key Project initiated a campaign to achieve a $10\%$ precision measurement of $H_0$ by observing Cepheid variables in the nearby universe ($\rm \lesssim 20 \ Mpc$; \citealt{freedman01}). When using these Cepheid variables to calibrate SNe~Ia, they obtained $\rm H_0 = 71\pm 2 \rm \  (stat)\pm 6 \ (sys) \ km \ s^{-1} \ Mpc^{-1}$. After the success of the HST Key Project, two major campaigns have been ongoing to determine $H_0$ using SNe~Ia : Supernovae, H0 for the Equation of State of dark energy (SH0Es; \citealt{riess16}) and the Carnegie-Chicago Hubble Program (CCHP; \citealt{freedman19}). 

While the SH0Es program has continued to rely primarily on Cepheid variables to calibrate the SN~Ia distance scale and determine $H_0$, the CCHP has been using TRGB as another independent distance anchor. Along with low halo reddening, another advantage of using TRGB $I$-band observations is that they are less sensitive to metallicity effects (\citealt{freedman19}). 
Moreover, it can be applied in the halos of galaxies, where crowding and blending effects are negligible.
The CCHP has also extended TRGB data to the near-infrared $JHK$ bands (\citealt{madore18}) to further improve SN Ia calibration. More recently, a third method of calibrating the SN Ia distance ladder has been put forward using SBF distances (\citealp{blakeslee21, khetan21}). While the SBF distance scale is itself calibrated with Cepheids, it is nonetheless an independent path to calibrating the SNe~Ia distances. 

%Recently, \cite{jensen21} updated a number of SBF distances.

%\crb{I like the intro you've put together, but this is a SN paper and SNe are barely mentioned. I think you need a bit in here about the SN Ia specific efforts made. Mark's work, Calan/Tololo, etc. Also the latest SH0Es and TRGB efforts using the Tripp method mostly in the optical and how NIR has advantages.}

Alternatively, analysis of the Cosmic Microwave Background (CMB) can provide $H_0$ directly from the angular scale of the sound horizon (\citealt{planck18}). This method requires extrapolating a cosmological model fit to the CMB over three orders of magnitude in redshift into a local value of $H_0$. Similarly, $H_0$ can be derived using the Baryon Acoustic Oscillation (BAO) standard ruler in a distance ladder tied to the CMB (\citealt{maccaulay19}). We summarize representative measurements of $H_0$ from various methods in Table~\ref{tab:h0}.

\begin{table*}
    \centering
   \caption{Recent $H_0$ measurements with statistical uncertainties. Systematic uncertainty in the Cepheid calibration has the least value of 1.04 $\rm km \ s^{-1}Mpc^{-1}$, while for TRGB and SBF-based calibrations, they range between 1.6 and 3.4 $\rm km \ s^{-1}Mpc^{-1}$.}
    \begin{tabular}{llcl}
     
    \hline
    \hline
        Method &Calibration &  $\rm H_0$ & Reference\\
        & & $(\rm km \ s^{-1}Mpc^{-1})$ &\\
        \hline
         SN~Ia &Cepheid &$73.04\pm 1.01$  & \cite{riess22}\\
         SN~Ia &TRGB  & $69.80\pm 0.60$ & \cite{freedman21}\\
         SN~Ia &Cepheid  & $72.70\pm 1.60$ & \cite{burns18}\\
         SN~Ia &SBF  &$70.50 \pm 2.40$ & \cite{khetan21} \\
         SN~Ia &SBF  &$74.60 \pm 0.90$ & \cite{garnavich22} \\
         
        SBF& Cepheid &$73.30\pm 0.07$ & \cite{blakeslee21}\\
        \hline
        SN~Ia &BAO+CMB  & $67.80\pm 1.20$ & \cite{maccaulay19} \\
         CMB &- &$67.40 \pm 0.50$ & \cite{planck18} \\
         %BAO  &- & $65.7\pm 2.4$ & \cite{maccaulay19} \\
         \hline
    \end{tabular}
    
    \label{tab:h0}
\end{table*}

It is clear from Table~\ref{tab:h0} that there are differences in $H_0$ obtained using various methods. In particular, $H_0$ inferred using methods tied to the CMB are significantly lower than those that are determined using local distance measurements. In addition, Cepheid-calibrated $H_0$ are also significantly higher than TRGB-calibrated $H_0$. \textbf{Recently, \cite{tully23} provided a historical review on $H_0$.}

A number of solutions have been proposed to resolve the tension in $H_0$. \cite{valentino21} presented a detailed review of possible solutions to the differences in $H_0$ that range from neutrino interactions to dynamical dark energy. \cite{bernal21} proposed a ``new cosmic triangle'' approach to break $H_0$ tension using the ages of older globular clusters. Alternatively, \citet{hamuy21} compared Cepheid and TRGB calibrations to various SN~Ia samples and concluded that the mismatch in $H_0$ using these two calibrators is a consequence of the systematic difference in the distance moduli. Previously, \cite{dhawan20} showed that the value of $H_0$ does not change with various cosmological model selections. %Recently, \cite{mortsell21} showed that relaxing the assumptions on the color calibration of Cepheid variables has a large effect on $H_0$. Re-calibrating these Cepheids, \cite{mortsell22} reports $\rm H_0=68.1\pm \ 2.6 km \ s^{-1}Mpc^{-1}$ by excluding the reddest Cepheids. 
\cite{addison21} found $H_0$ to be consistent with local calibrator methods using E-mode data of CMB. New datasets can bring new insights to $H_0$, even with existing calibration methods. 

The Carnegie Supernova Project (CSP; \citealt{hamuy06}) obtained high-quality light-curves of SNe~Ia in $uBgVriYJH$ bands from 2004 to 2015. Data from the CSP-I campaign, which ran from 2004 to 2009, played an important role in determining $H_0$ measurements for some of the above-mentioned studies (e.g., \citealt{freedman19}, \citealt{burns18}). CSP-I SN~Ia photometry was released in three papers: \cite{contreras10}, \cite{stritzinger11}, and  \cite{krisciunas17}; the latter included a re-calibration of the
entire CSP-I dataset. Spectroscopic data is presented in \cite{folatelli13}.
In the CSP-I campaign, most of the SNe~Ia observed were discovered in targeted galaxy searches, such as, Lick Observatory Supernova Search (LOSS; \citealt{li00}), and therefore are biased towards massive galaxies\footnote{See \cite{uddin17a} for a study on host galaxy bias in SN~Ia cosmology.}.   

%In order to address these shortcomings, 
 
A second follow-up campaign, the CSP-II, began in 2011 with the goal of extending the NIR SN Ia sample further out into the Hubble flow ($0.01 < z < 0.1$). The CSP-II ran for 4 years, observing SNe discovered by rolling searches, primarily the La Silla-Quest Low Redshift Survey (LSQ; \citealt{baltay13}). The project is summarized in \citet{phillips19} and \citet{hsiao19}.

%A total of 214 SNe~Ia were observed in the redshift range $0.004 < z < 0.137$ during the CSP-II campaign. Of them, there are 125 ``cosmology sample'' SNe~Ia in the smooth Hubble flow ($0.03\lesssim z \lesssim 1.0$), and 90 ``physics sample'' SNe~Ia in the nearby universe ($z\lesssim 0.04$).
A total of 214 SNe Ia were observed in the redshift range $0.004 < z < 0.137$ during the CSP-II campaign. Of these, there are 125 ``cosmology sample'' SNe Ia in the smooth Hubble flow ($0.03\lesssim z \lesssim 1.0$).  Of the same 214 SNe, NIR spectra were obtained for 90 ``physics sample'' SNe Ia in the nearby universe ($z\lesssim 0.04$). Both optical and near-infrared ($uBgVriYJH$) light-curves were obtained out to higher redshifts to reduce the error due to peculiar velocities. This also enables one to compare SN~Ia standard candles between optical and near-infrared wavelengths. 
In addition, 339 near-infrared spectra of 98 SNe were obtained in order to improve near-infrared K-corrections.

%\crb{Even tough the sample is laid out in Eric and Mark’s papers, it wouldn’t hurt to list the number of objects and the redshift range.}.
%CSP-II is described in detail by \cite{phillips19} and \cite{hsiao19}, and CSP-I is described in \cite{hamuy06}. Photometric data for CSP-I is published in three stages (\citealp{contreras10,stritzinger11,krisciunas17}), and spectroscopic data is presented in \cite{folatelli13}.

Combining the CSP-I and CSP-II samples has increased the total number of CSP SNe~Ia by almost a factor of three compared to CSP-I alone. We will denote this combined sample as the CSP SN~Ia sample for the rest of the paper. We will publish the CSP-II SN~Ia photometry data and optical spectra in separate papers (Suntzeff et al. in preparation; Morrell et al., submitted). 
The near-infrared spectra have been published by \citet{jing22}. 

%\crb{I think there needs to be a summary paragraph or two of the CSP. Particularly, how CSPI was follow-up to targeted surveys and CSPII is follow-up of un-targeted. And that we’ve pushed the NIR out to higher redshift to reduce the errors in peculiar velocities. And mention the goal of determining if the NIR is really that much better than optical. As it stands, the introduction talks more about the calibrators than the SNe~Ia. }

In this paper, we aim to determine $H_0$ by calibrating the full CSP SN~Ia sample using a combination of two primary distance calibrators: Cepheids and TRGB, and a secondary distance indicator, SBF. Combining all calibrators may reveal the systematic uncertainties that originate due to the variation of the SN~Ia distance scales among various calibrators, and may explain the $H_0$ tension from various methods. We also aim to determine $H_0$ in the $uBgVriYJH$ bands using various calibrators separately, and investigate intrinsic scatter of SNe~Ia at these wavelengths. Finally, we revisit correlations between SN~Ia Hubble residuals and host galaxy properties.

The outline of this paper is as follows: in \S\ref{sec:data} we describe sources of data; in \S\ref{sec:ana} we perform analysis to calibrate SNe~Ia and obtain values of $H_0$; in \S\ref{sec:dis} we discuss the results; and finally we present our conclusions in \S\ref{sec:con}. 

\section{Data} \label{sec:data}

\subsection{Light-curve Fit}\label{sec:lcfit}

SNe~Ia from the CSP-I and CSP-II samples are tabulated in \cite{krisciunas17} and \cite{phillips19}, respectively. We present the number of SN~Ia that we use in this work in each filter, along with the total number of calibrators, in Table~\ref{tab:sample}. Selection of CSP-I SNe~Ia is described by \cite{uddin20}. From the CSP-II sample, we exclude iPTF13dym, and iPTF13dyt because of the peculiar behavior of their near-infrared light-curves. We also exclude PS1-13eao due to its high extinction. We did not use 03fg-like SNe~Ia: ASASSN-15hy, SN~2007if, SN~2009dc, SN~2013ao, CSS140501-170414+174839 (\citealp{ashall21, lu21, hsiao20}), and also no Type Iax SNe~Ia were included.

%: CSP14acx, PTF14ans, SN2008ae, SN2008ha, SN2009J, SN2010ae, SN2012Z, SN2013gr, SN2015H.

%\crb{We should state why these are peculiar, is spectroscopically, photometrically, both? We should also state that we exclude a number of type-IaX and super-chan objects}\textbf{???}.

We fit SN~Ia light-curves using \texttt{SNooPy} (\citealt{burns11}), using the \texttt{max\_model} method, which provides peak magnitudes for each observed filter, the time of maximum, $(B-V)$ color\footnote{Throughout the paper our definition of $(B-V)$ color of SNe~Ia corresponds to $(B_{max}-V_{max})$ color.}, and the color-stretch parameter $s_{BV}$ \citep{burns14} as well as all the co-variances between the parameters.
\texttt{SNooPy} computes K-corrections and S-corrections (when needed) using the optical \citet{hsiao07} and the new near-infrared SN Ia spectral energy templates from \citet{jing23} along with the CSP filter functions (\citealp{krisciunas17, phillips19}).
From the peak magnitudes, we compute colors after applying Milky-Way reddening corrections using the dust maps of \citet{Schlafly2011}. \texttt{SNooPy} fitting is also done for SNe~Ia from the literature that have distance measurements from Cepheid, TRGB, and/or SBF methods. In all cases, we use the same version of \texttt{SNooPy} using the same spectral templates. We show distributions of SN~Ia color and color-stretch parameters in Figure~\ref{fig:lc}. We also present them as a function of redshift to reveal the selection effects in Figure~\ref{fig:z_st_bv}.
The most notable difference is the larger number of slow-declining SNe~Ia with bluer colors in the CSP-II sample compared to CSP-I. 

%Results of the \texttt{SNooPy} fits are given in table \ref{csptable}.
\begin{table}
%\hspace{-1cm}
\centering    
\caption{Number of SNe~Ia and calibrators available in each filter that are used in this work.}
\begin{tabular}{cccccc}
\hline
\hline
 &&&Number&&\\
\cmidrule{2-6}
Band& SNe~Ia\footnote{CSPI \& II} & Cepheids\footnote{\cite{riess21}} & TRGB\footnote{\cite{freedman19}} & SBF\footnote{\cite{khetan21}} & SBF\footnote{\cite{jensen21}}\\
%& (CSP-I \& II)& (\citealt{riess16}) & (\citealt{freedman19}) & (\citealt{khetan21}) & (\citealt{jensen21})\\
\hline
$u$ & 216& 17 & 11&12 & 11\\
$B$  & 322&25  & 18&24 &22\\
$g$ & 235&11  &5 &4 &5\\
$V$ &323 &25  & 18&24 &22\\
$r$ & 323& 24 &16 &19 & 19\\
$i$ &322 & 24 & 16&19 & 19\\
$Y$ & 275&11  & 5&2 &4\\
$J$ & 246&20  & 15&9 &8\\
$H$ & 213&  20& 15&9 &8\\
\hline
\end{tabular}
\label{tab:sample}
\end{table}

\subsection{Distance Calibrators}\label{calibrators}
Cepheid and TRGB distances to SN~Ia hosts are taken from \cite{riess22} and \cite{freedman19} respectively. Distances of two TRGB hosts are updated to \cite{hoyt21}. They are the hosts of SN2007on and SN2011iv (NGC 1401) and SN2013aa and SN2017cbv (NGC 5643). 

SBF distances to SNe~Ia hosts come from two sources: one is a collection compiled in \cite{khetan21} from various sources, and the other consists of systematic measurements using the WFC3/IR camera on the HST published in \cite{jensen21}. The \cite{jensen21} observations extend to greater distances than the \cite{khetan21} measurements --- the average of the \cite{jensen21} distance moduli is 1.5 mag larger than the average of the \cite{khetan21} values.
%Comparing these two sources, we find that the average of the distance moduli in \cite{jensen21} is $1.5$ mag larger than the average of the distance moduli in \cite{khetan21}.

While the multi-band light-curves of all 24 SBF-calibrated SNe~Ia from \cite{khetan21} are available publicly, the same is true for only three out of the 22 from \cite{jensen21}.
These are the hosts of SN~2016arc (NGC~1272), SN~2007gi (NGC~4036), and SN~2018aaz (NGC~3158). There are seven objects that are common between \citet{khetan21} and \citet{jensen21}. These are the hosts of SN~1970J (NGC~7619), SN~1995D (NGC~2962), SN~1997E (NGC~2258), SN~2003hv (NGC~1201), SN~2014bv (NGC~4386), SN~2015bp (NGC~5839), and SN~2000cx (NGC~0524). We compared the distance moduli of the common SBF hosts between \citet{khetan21} and \citet{jensen21} and found them to be comparable. 
%in \citet{jensen21} are relatively larger by $0.13$ magnitude.

%The mean difference in distance moduli of these seven SBF hosts between the two sources is  $0.13$ mag. 
%While the distances of these seven common hosts are different between the complication, SBF distance moduli published in \cite{jensen21} have better accuracy than the compilation in \cite{khetan21}.

A detailed discussion on various calibration methods is beyond the scope of this paper, and we encourage readers to consult the references in the footnotes of Table~\ref{tab:sample} for details about various calibration processes. The distribution of calibrator host distances is shown in Figure~\ref{fig:caldist}. SBF calibrators extend to larger distances, while TRGB hosts are found to be relatively nearer than the other two. As we have noted earlier, that SBF distances are calibrated using Cepheid distances to a set of nearby early-type galaxies in the Virgo cluster, so they have an additional rung in the distance ladder. %While this certainly can contribute to some co-variance due to any systematic errors in Cepheid variables as a population, the specific Cepheids in question are not part of the SH0Es sample, and therefore constitute an independent path to calibrating the SNe~Ia beyond the first rung.

\subsection{Host Galaxy Stellar Mass}\label{hostmass}

The host galaxy stellar masses of CSP SNe~Ia have been derived using procedures described by \cite{uddin20}. Briefly, we measure $uBgVriYJH$ photometry using host galaxy follow-up images taken after the SNe~Ia have faded, and build spectral energy distributions. We then use $\texttt{Z-PEG}$ (\citealt{leborgne02}) to determine the best-fit templates and derive stellar masses. Stellar masses of the hosts containing calibrating SNe~Ia are taken from \cite{burns18} and \cite{neill09}. Although host masses of galaxies from these two papers have offsets of 0.3 dex and 0.18 dex, respectively, with respect to those calculated by \cite{uddin20}, our results have not changed (see \S \ref{h0val}).
For consistency, the host masses of the calibrating SNe were adjusted for these offsets.
Distributions of the host galaxy stellar masses of distant and calibrating SNe~Ia are shown in Figure~\ref{fig:massdist}.

There are 19 SNe~Ia from CSP-II for which no hosts are identified. This could be either the hosts are very faint or there are no hosts at all. We randomly assign stellar masses between $\rm Log \ (M_*/M_{\odot})=7.1-7.9$ for the hosts of these apparently hostless SNe~Ia. We also calculate stellar masses of 10 SBF hosts from \cite{jensen21} for which we do not have multi-band photometry. We use $K$-band photometry from the Two Micron all Sky Survey, and use an empirical formula described in \cite{ma14}, to estimate their stellar mass. 

It is evident that host stellar mass in the CSP-I sample is biased towards higher mass, since most SNe~Ia followed were discovered in targeted surveys. Host mass in CSP-II sample is spread over a wider mass range, since most of the observed SNe~Ia were followed from untargeted surveys, such as, LSQ, Catalina Real Time Transit Survey (CRTS; \citealt{djorgovski11}), All-Sky Automated Survey for SuperNovae (ASAS-SN; \citealt{shappee14}), and a few other supernova surveys\footnote{\cite{phillips19} describes all surveys from which CSP-II sample was generated.}.  

We present the CSP sample used in this paper in Appendix \S \ref{alldata}. We also present Cepheid, TRGB, and SBF calibrators. Appendix \S \ref{alldata} lists peak magnitudes in the $B$-band. Similar data in all other bands are available online.\footnote{\href{https://github.com/syeduddin/h0csp}{https://github.com/syeduddin/h0csp}} 
%\crb{Not sure what you mean by ``different calibration'' in this context.}

\begin{figure*}[htbp]
    \centering
    \begin{tabular}{cc}
     \includegraphics[width=\columnwidth]{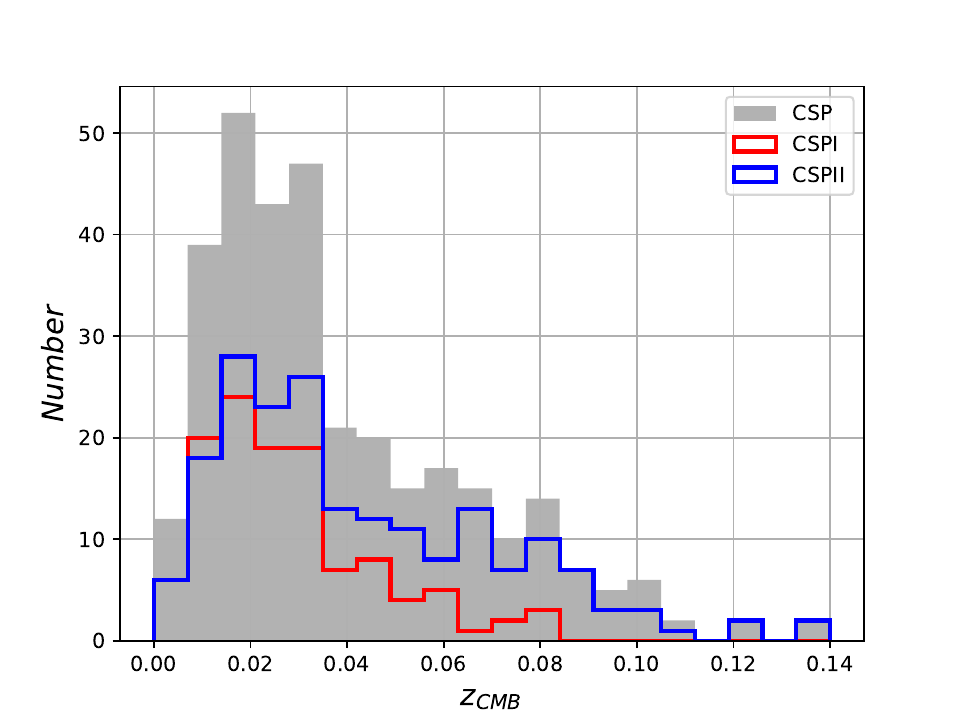} &
    
     \includegraphics[width=\columnwidth]{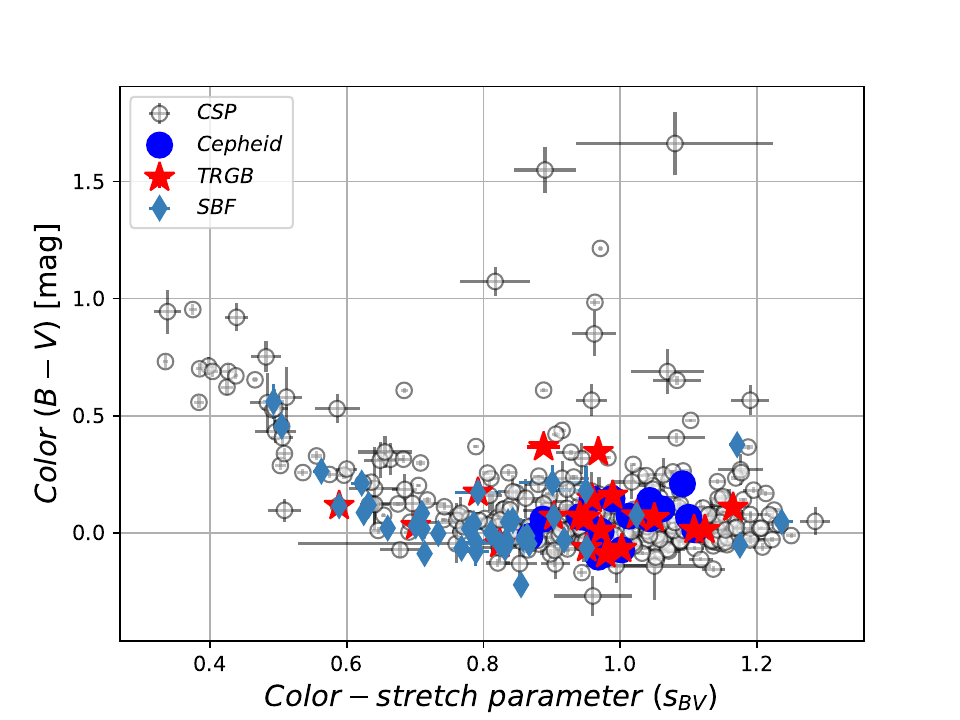}
     \end{tabular}
    \caption{\emph{Left:} Distribution of SN~Ia redshifts in CSP-I and in CSP-II. \emph{Right: }Distribution of SN~Ia color and color-stretch parameters used in this study.}
    \label{fig:lc}
\end{figure*}

 \begin{figure*}[htbp]
    \centering
    \includegraphics[width=\textwidth]{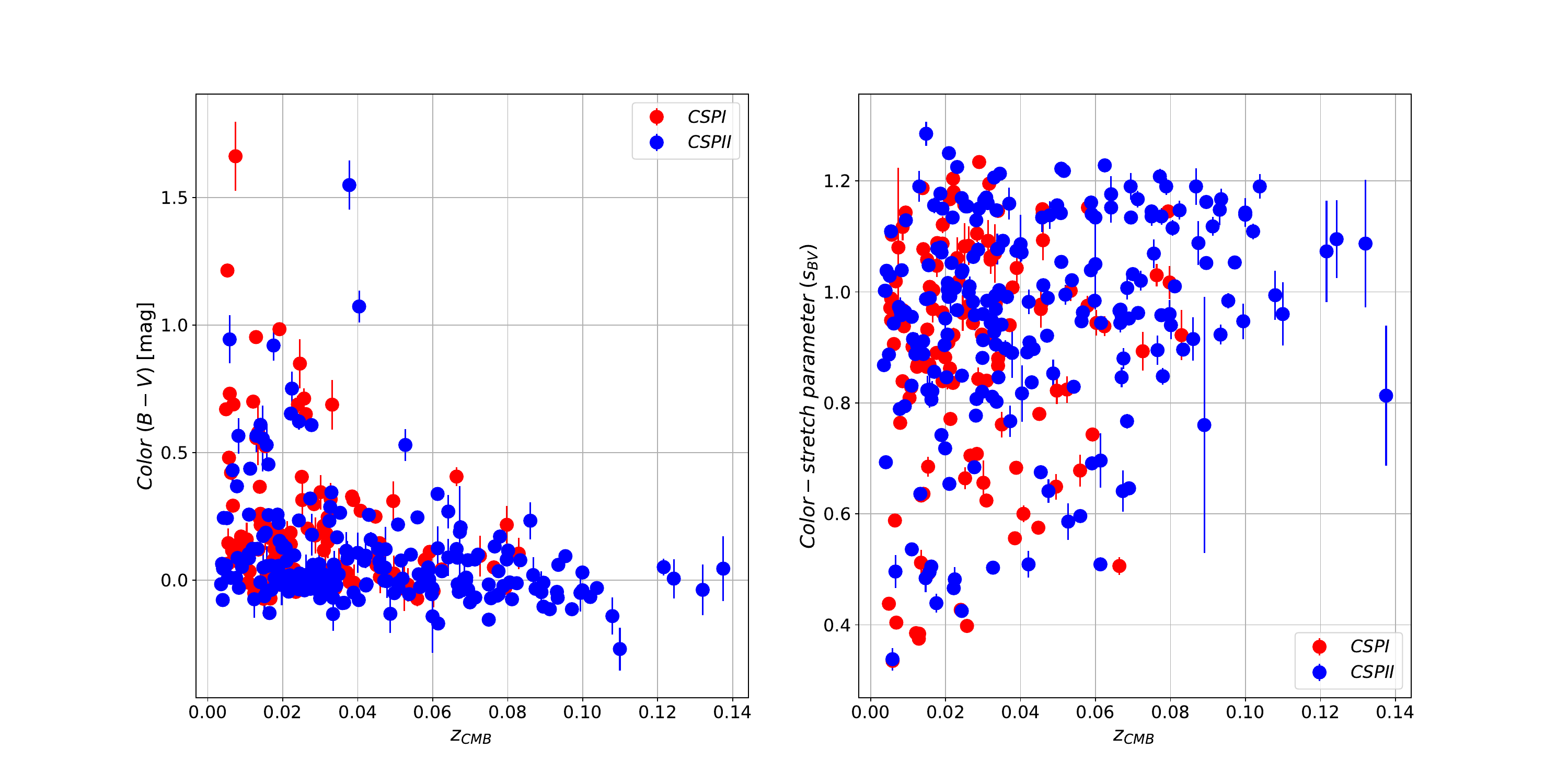}
    \caption{Distribution of SN~Ia color (left) and the color-stretch parameter (right) as a function of redshift. Due to selection effects, we find bluer and slower-declining SNe~Ia at higher redshift.}
    \label{fig:z_st_bv}
\end{figure*}

\begin{figure*}[htbp]
    \centering
    \begin{tabular}{cc}

    \includegraphics[width=\columnwidth]{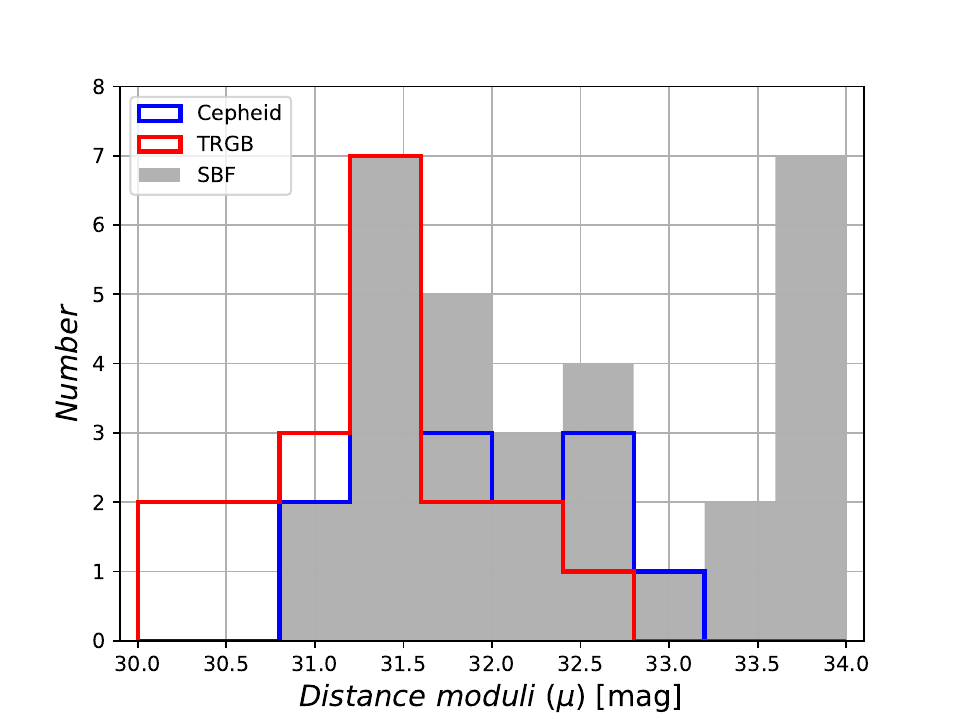}&
    \includegraphics[width=\columnwidth]{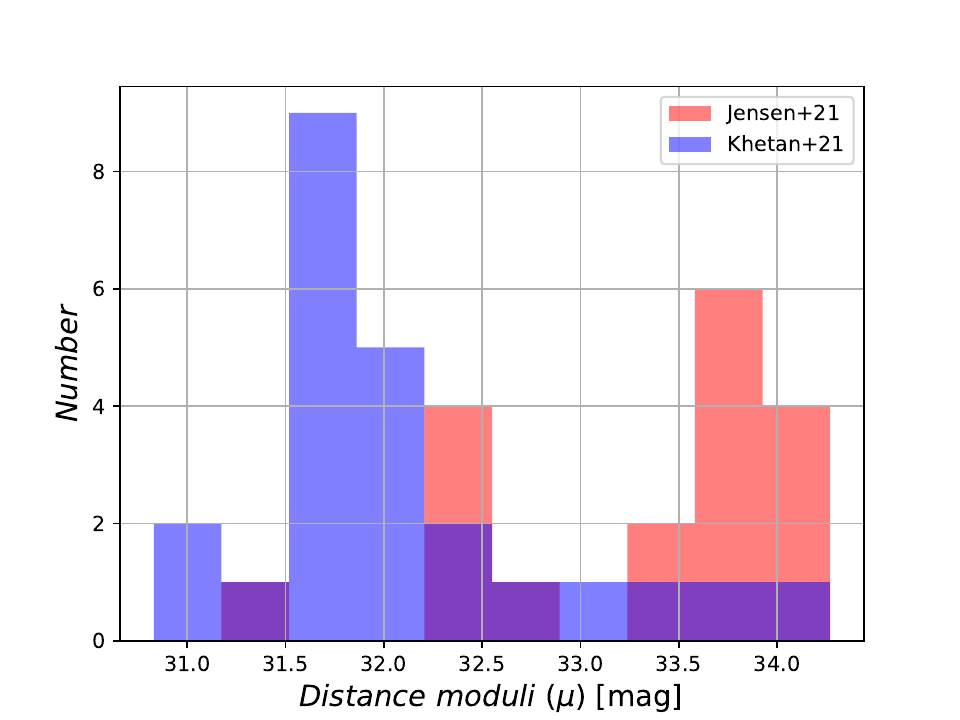}
    \end{tabular}
    \caption{\emph{Left:} Distribution of calibrator distance moduli. We note that SBF distances extend to higher values than TRGB and Cepheid distances. \emph{Right:} Comparison of two SBF calibration samples. Note that SBF hosts from \cite{jensen21} are peaked at higher distances.}
    \label{fig:caldist}
    
\end{figure*}

\begin{figure}[htbp]
    \centering
    \includegraphics[width=\columnwidth]{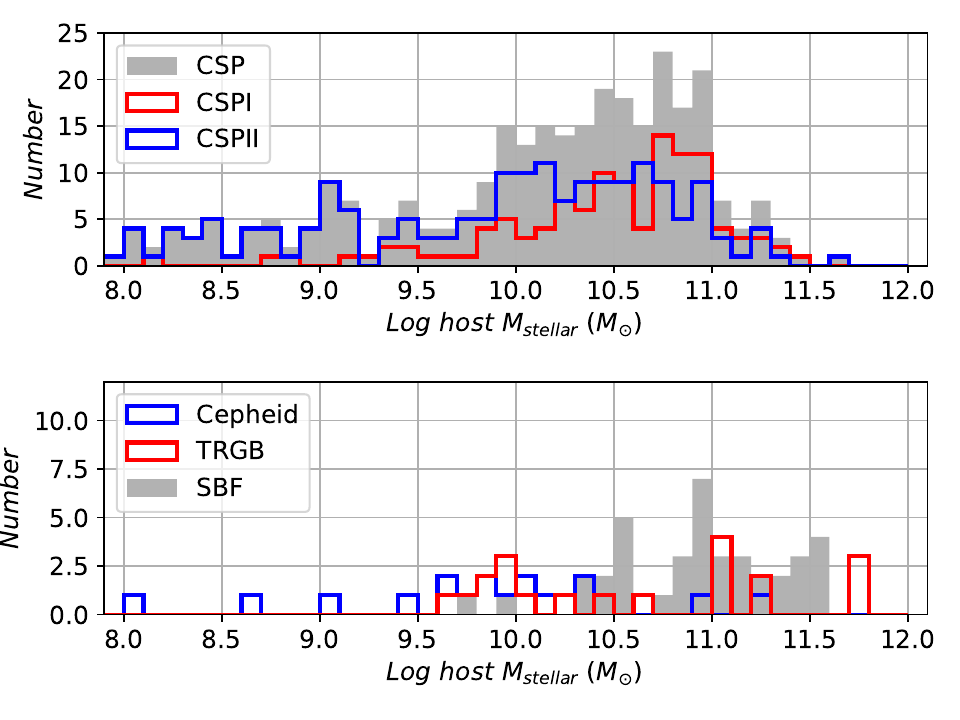}
    \caption{Distribution of host galaxy stellar mass for the CSP sample (top) and the calibrating SNe~Ia (bottom). See text for discussion.}
    \label{fig:massdist}
\end{figure}

\section{Analysis} \label{sec:ana}

\subsection{SN~Ia Luminosity Standardization}\label{standard}

The luminosity-decline rate relation \citep{phillips93, phillips99} paved the way to standardize SN~Ia luminosity to an accuracy of $\sim \rm 0.15 \ mag$. This relation shows that luminous SNe~Ia decline more slowly than less luminous events. 
However, this relation shows increasing scatter for faster-declining events.
This is primarily due to the fact that the light-curve width parameter, $\Delta m_{15}(B)$ (defined as the change in magnitude in the $B$-band between peak and 15 days after peak) was not able to accurately measure the rate of evolution of the sub-luminous SNe (\citealt{phillips12}). \cite{burns14} discuss issues related to $\Delta m_{15}(B)$ and introduced a color-stretch parameter $s_{BV}$, which better captures the diversity of the fastest-declining events by reducing scatter. In this analysis, we use $s_{BV}$ to measure the decline rate of our SNe~Ia.

Also, to deal with the effect of dust extinction and intrinsic color dependence, a correction is required. A simple approach is to apply a color correction as proposed by \cite{tripp98}, which we shall refer to as the Tripp method. Bluer SNe~Ia are found to be more luminous, though this can be a combination of both the effects of dust, and intrinsic color variations, since the measured color is known to correlate with $s_{BV}$ (see Figure \ref{fig:lc}, right panel).  One could also use such correlations between observed color and $s_{BV}$ to compute and apply a proper extinction correction, as was done by \citet{burns18}. However, this introduces large correlated errors between the magnitudes in each filter. One of the goals of this analysis is to compare the dispersion and host mass correction in different bands. We thus wish to avoid such correlated errors as much as possible.

The final standardization correction that we apply is due to an empirical correlation between SN~Ia luminosity and host mass (see \citealt{uddin20} and references therein). This is often applied as a mass step, where the sample is split in two at a particular mass threshold. This approach, while simple, has several problems. First, the choice of the mass threshold is somewhat arbitrary and due to the typically large uncertainties in stellar mass which are not normally distributed, assigning hosts to one side or the other of this threshold is itself an uncertain task and could introduce biases. Second, it is not obvious how to deal with the well-known bias in regression analysis introduced by large uncertainties in the independent variable \citep{Kelly2007}, whereas there exists well-developed software for handling these errors in linear regression. We therefore opt to apply a linear correction for the host stellar mass with slope $\alpha$.

We apply the above-mentioned corrections in the observed light-curves of SNe~Ia. The observed distance moduli ($\mu_{obs}$) of SNe~Ia are defined by:

\begin{equation}\label{muobs}
%\begin{split}
 %   \mu_{obs} = m_x -P^N(s_{BV}-1) -\beta(B-V) \\ -\alpha %\bigg(log_{10}\frac{M_*}{M_{\odot}} - M_0\bigg)
  %  \end{split}
\begin{split}
    \mu_{obs} = m_x -P^N(s_{BV}-1) -\beta(B-V) \\ -\alpha \big(\mathscr{M_*} - \mathscr{M}_0\big)
    \end{split}  
\end{equation}

Here, $m_x$ is the peak magnitude in band $x$. $P^N(s_{BV}-1)$ is a polynomial of order $N$. The coefficient of the zeroth term of this polynomial, $P0$, denotes the absolute magnitude, $M_x$ of a SN~Ia with zero $(B-V)$ color, color-stretch $s_{BV} = 1$, and host galaxy with stellar mass $\mathscr{M_*}=\mathscr{M}_0$. Here, $\rm \mathscr{M*} = log_{10}(M_*/M_{\odot})$ is the host stellar mass, and $\rm \mathscr{M}_0 = log_{10}(M_0/M_{\odot})$ is a mass zero point, which we set at the median value of the host stellar mass in each band. \textbf{The choice of splitting the sample at the median mass ensures that equal weights are given above and below the median mass.} The parameter $\alpha$ is the slope of the luminosity-host mass correlation.

In this paper, we expand the polynomial up to second order ($N=2$), and the coefficients are denoted as $P0,\ P1, \rm{and} \ P2$. $\beta$ is the slope of the luminosity-color relation. We define model distance moduli to SN~Ia hosts in the $\rm \Lambda CDM$ cosmological framework as:  

\begin{eqnarray}\label{mumod}
%\begin{split}
% \mu_{model} (z_{hel},z_{cmb},H_0,q_0) &  = &
\mu_{\Lambda CDM}  = 5\log_{10} & & \bigg[\bigg(\frac{1+z_{hel}}{1+z_{cmb}} \bigg)\frac{cz_{cmb}}{H_0}\bigg(1+ \nonumber \\
 & & \frac{1-q_0}{2}z_{cmb}\bigg)\bigg] +25 
%\end{split}
 \end{eqnarray}

%\crb{This equation is incorrect. The right-most $z_{cmb}$ should be inside the second parenthesis, multiplying the $q_0$ term. Otherwise, there is no linear term in $z_{cmb}$.}

Here $q_0$ is the deceleration parameter. 
%For SNe~Ia with distances from calibrators, we set $\mu_{model}$ to the distances obtained directly via that method. 
The observed distance moduli ($\mu_{obs}$) for all SNe~Ia are calculated using Eqn.~\ref{muobs}. In calculating model distance moduli ($\mu_{model}$) we use Cepheid/TRGB/SBF-based distances when available, otherwise we use Eqn.~\ref{mumod} to predict distance. That is:

\begin{equation}
    \mu_{model} = 
    \begin{cases}
      %\mu \ (z_{hel},z_{cmb},H_0,q_0) & \rm SNe \ Ia\\
      \mu_{\Lambda CDM} & \rm SNe \ Ia\\
      \mu & \rm Cepheid/TRGB/SBF
    \end{cases} 
\end{equation}
%\crb{Given the discussion with Peter B., it's probably worth stating explicitly that we solve for $H_0$ and all the nuisance parameters simultaneously so that all errors and co-variances propagate to the final uncertainty in the Hubble constant.}\textbf{SU: added in next section}

We compare the observationally-determined distance modulus with the model distance modulus by defining a $\chi^2$ as: 
\begin{equation}\label{chi}
    \chi^2 =  \sum_{i} \frac{(\mu_{obs,i}-\mu_{model,i})^2}{\sigma^2_i+\sigma^2_{int}+\sigma^2_{pec,i}}
\end{equation}

Eqn.~\ref{chi} has three error terms. The first error term, $\sigma_i$, is the sum of the individual errors on observed quantities along with the covariance between peak magnitude and color, and the covariance between peak magnitude and color-stretch parameter. Ignoring the index, the first error term can be expanded as:
%\begin{equation}\label{sig}
%\begin{split}
 %   \sigma^2 = \sigma^2_{m_x} + (P1+2P2 \ s_{BV})^2 \sigma^2_{s_{BV}}  \\ +\beta^2 \ \sigma^2_{B-V}+  (2P1+4P2 \ s_{BV})^2 \ cov(m_x,s_{BV})\\ + \beta \ cov(m_x,B-V) +\ \alpha^2 \ \sigma^2_{\mathscr{M_*}}
  %  \end{split}
%\end{equation}

%\begin{equation}\label{sig}
%\begin{split}
 %   \sigma^2 = & \sigma_{m_x}^2 + (P1 + 2P2\ s_{BV})^2\sigma^2_{s_{BV}}\\
  %       & +\beta^2 \sigma^2_{B-V} - 2(P1 + 2P2\ s_{BV})\ cov(m_x,s_{BV}) \\
   %      &  + 2\beta(P1 + 2P2\ s_{BV})\ cov(s_{BV},B-V) \\
    %     &  -2\beta\ cov(m_x,B-V) + \alpha^2\sigma^2_M
%\end{split}    
%\end{equation}

\begin{equation}\label{sig}
    \begin{split}
        \sigma^2 = & \sigma_{m_x}^2 + (P1 + 2P2\ (s_{BV}-1))^2\sigma^2_{s_{BV}}\\
         & +\beta^2 \sigma^2_{B-V} - 2(P1 + 2P2\ (s_{BV}-1))\ cov(m_x,s_{BV}) \\
         &  + 2\beta(P1 + 2P2\ (s_{BV}-1))\ cov(s_{BV},B-V) \\
         &  -2\beta\ cov(m_x,B-V) + \alpha^2\sigma^2_M
    \end{split}
\end{equation}

%\crb{There's something not right here. The second term doesn't include any $\sigma$ factors. That can't be right.}
The second error term, $\sigma_{int}$, is the intrinsic random scatter that we add to the error budget as a free parameter to account for SN-to-SN variations not accounted for by $\sigma_i$ or $\sigma_{pec}$. The final error term $\sigma_{pec}$ is the error due to uncertainty in distance measurements resulting from galaxy peculiar velocities that scales with redshift. This term can be expressed as\footnote{One can obtain this term by taking the partial derivatives of the first order term of the distance equation, $$\mu(z) = 5log_{10}\bigg(\frac{cz}{H_0}\bigg)$$ }:
\begin{equation}
    \sigma_{pec} = 2.17\frac{V_{pec}}{cz_{cmb}}
\end{equation}

%{\bf [Where does this equation and the 2.17 come from? This comes out of the clear blue.]}

where $V_{pec}$ is a free parameter that represents the average peculiar velocity of SN Ia sample. \textbf{$V_{pec}$ is allowed to vary in each filter because the sample size and redshift distribution are different in each filter.\footnote{\textbf{Recently, \cite{peterson22} demonstrated that by separating the component of redshift due to peculiar velocity, the value of $H_0$ can change by $\sim 0.4 \ \rm km \ s^{-1} \ Mpc^{-1}$. We discuss this further in \S~\ref{h0all}.}}}
Note that $\sigma_{int}$ and $\sigma_{pec}$ are calculated for SNe~Ia that do not have distance measurements from calibrators. For calibrators, uncertainties from their distances are included.
%\crb{It might be clearer if you simply re-label $\sigma_{\mu}$ as $\sigma_{model}$ instead and that for the distant sample, this is dominated by peculiar velocities, whereas for the calibrating sample, we use the published uncertainties}.
 
To find the best values and uncertainties of all variables, we explore the parameter space with Markov Chain Monte Carlo (MCMC) techniques after converting the $\chi^2$ into a log-likelihood function

\begin{equation}
    \mathcal{L} = -0.5\big[\chi^2 +N log(2\pi\sigma^2)\big]
\end{equation}
where $\sigma$ includes all error terms mentioned in Eqn. \ref{chi}, and $N$ is the total number of data points. 
%\subsection{Tripp Calibration}

We use a Python package, \texttt{EMCEE} (\citealt{foreman13}) to explore the likelihood of each variable. EMCEE utilizes an affine invariant MCMC ensemble sampler proposed by \cite{goodman10}. The advantage of this sampler over traditional Metropolis-Hastings sampling is that it significantly reduces autocorrelation time. Another advantage of an affine invariant sampler is that we need to tune only two parameters to get the desired output: number of walkers and number of steps. 

%These two parameters are walkers and {\bf [missing verb here]} number of steps or samplers.

Briefly, we start by assigning initial maximum likelihood values of variables to the walkers. The walkers then start wandering and explore the full posterior distribution. After an initial run, we inspect the samplers for their performance. A good way to do this is to look at the time series of variables in the chain and compute the autocorrelation time, $\tau$. We then set the number of steps to greater than $50\tau$ as suggested by the online documentation\footnote{  \href{https://emcee.readthedocs.io/en/stable/tutorials/autocorr/}{https://emcee.readthedocs.io/en/stable/tutorials/autocorr/}}. When the chains are sufficiently burnt-in (e.g., they forget their initial start point), we can safely throw away some steps that are a few times higher than the burnt-in steps. Another criterion of good sampling is the acceptance fraction, $a_f$. This is the fraction of steps that are accepted after the sampling is done. The suggested value of $a_f$ is between $0.2-0.5$ (\citealt{gelman1996}).  

We run \texttt{EMCEE} with 80 walkers. We sample 3000 steps and throw away the first 500. In each run, we obtain $a_f \sim 0.46$.
%We calculate $\tau\sim 100$ and $a_f \sim 0.45$ for the optimal sampling. 
One can visualize the output of two-dimensional and one-dimensional posterior probability distributions in a corner plot, and check for any unwanted multi-modal distribution. From the marginalized distributions we take the 16th, 50th, and 84th percentiles to obtain best-fit values and uncertainties.

%\crb{emcee does not use Metropolis-Hastings. M-H is a very inefficient sampling algorithm and emcee is one of several improved methods. Specifically, emcee uses the ``Affine Invariant Markov chain Monte Carlo Ensemble sampler''. You might want to read up on it and maybe give a very brief explanation, since you refer to ``walkers''. You should state  how you compute the ``best-fit'' values and uncertainties of the parameters from the chains, and perhaps how you determine that you've run the MCMC for long enough. }

Previously, \cite{burns18} employed the `No U-Turn Sampler' using the \texttt{STAN} data modeling language. We can therefore check the consistency between the two sampling methods. We set $\rm H_0 = 72 \ km \ s^{-1} \ Mpc^{-1}$ and the deceleration parameter $\rm q_0 = \Omega_m /2 -\Omega_{\Lambda}= -0.53$ as priors as used in \citet{burns18}.  We compare the luminosity calibration between this work and \citet{burns18} using the Tripp method in the $B$-band. This comparison is shown in Table~\ref{comp} for the Cepheid calibration. Tripp nuisance variables for both CSP-I and the combined CSP-I $\& $-II samples are shown. We confirm that for the CSP-I sample, results from both \texttt{EMCEE} (this work) and \texttt{STAN} (\citealt{burns18}) agree.

We note that both $\sigma_{int}$ and $V_{pec}$ have increased for the combined sample in comparison to CSP-I. This is due to the addition of more low-redshift SNe~Ia in the combined sample. For example, for $z<0.01$, there are 14 SNe~Ia in CSP-I, and 13 SNe~Ia in CSP-II. Therefore, in the combined sample, there is almost a two-fold increase of low-redshift SNe~Ia. To prove the effect of low-redshift SNe~Ia on $V_{pec}$, we calculate $V_{pec}$ by removing $z<0.01$ SNe~Ia. With this restricted sample $V_{pec}$ is $\rm 243 \ km \ s^{-1}$ in CSP-I, and $\rm 251 \ km \ s^{-1}$ in the combined sample. The values of $\sigma_{int}$ in these cases are 0.14 mag and 0.18 mag, respectively. \textbf{The $\beta$ parameter is also steeper in the combined sample than CSP-I sample. This could be due to the fact that in the combined sample, we have more star-forming galaxies that produce more dust.}

%\crb{If you really are allowing $H_0$ to vary, then that should be included in Table 3. The increased $V_{pec}$ still bothers me. We should try to figure out which data are pushing this to a higher value. It must be a low-z objects... What happens to $V_{pec}$ when excluding low-z objects. We could see if you include $V_{pec}$ in tables 8-10, which might be worth doing anyway.}

\begin{table*}
    \centering
    \caption{Comparing the Cepheid calibration of CSP SNe~Ia using the Tripp method. We include $1\sigma$ statistical uncertainties in the parentheses.}
    \begin{tabular}{l|c|c|c}
    \hline
    \hline
    Parameter & \cite{burns18} & This Work & This Work\\
    &(CSP-I)& (CSP-I) & (CSP-I \& II)\\
    \hline
    $ P0  \ (\rm mag)$ & -19.18~(06)   & -19.18~(02) & -19.09~(01)\\
    $ P1  \ (\rm mag)$ & -0.89~(11)   & -0.87~(12) & -1.22~(08) \\
    $ P2  \ (\rm mag)$ & -0.02~(30)   & -0.07~(30) & -1.45~(30)\\
    $ \beta$ & 2.65~(08)   & 2.76~(10) & 3.06~(08)\\
    $ \alpha \ \rm (mag/dex)$ & -0.06~(03)   & -0.08~(04)& -0.02~(01) \\
    $ \sigma_{int}  \ (mag)$ & 0.13   & 0.13 & 0.17\\
    $ V_{pec} \ (\rm km \ s^{-1})$ &  310  & 297 & 442\\
    \hline
    \end{tabular}
    \label{comp}
\end{table*}

\subsection{Measurement of $H_0$}\label{h0val}
\subsubsection{Using Calibrators Separately}\label{h0ind}
The principal objective of this paper is to determine $H_0$ using various distance calibrators: Cepheids, TRGB, and SBF. In this section, we calculate $H_0$ using various calibrators separately. In the next section (\S~\ref{h0all}), we combine all calibrators to obtain global values of $H_0$. We follow the procedure described in \S ~\ref{standard}, but now include $H_0$ as a variable. We note that we solve for $H_0$ and all other nuisance parameters simultaneously so that all errors and co-variances propagate to the final uncertainty in $H_0$.

\begin{figure*}[htbp]
    \centering
    \includegraphics[width=\textwidth]{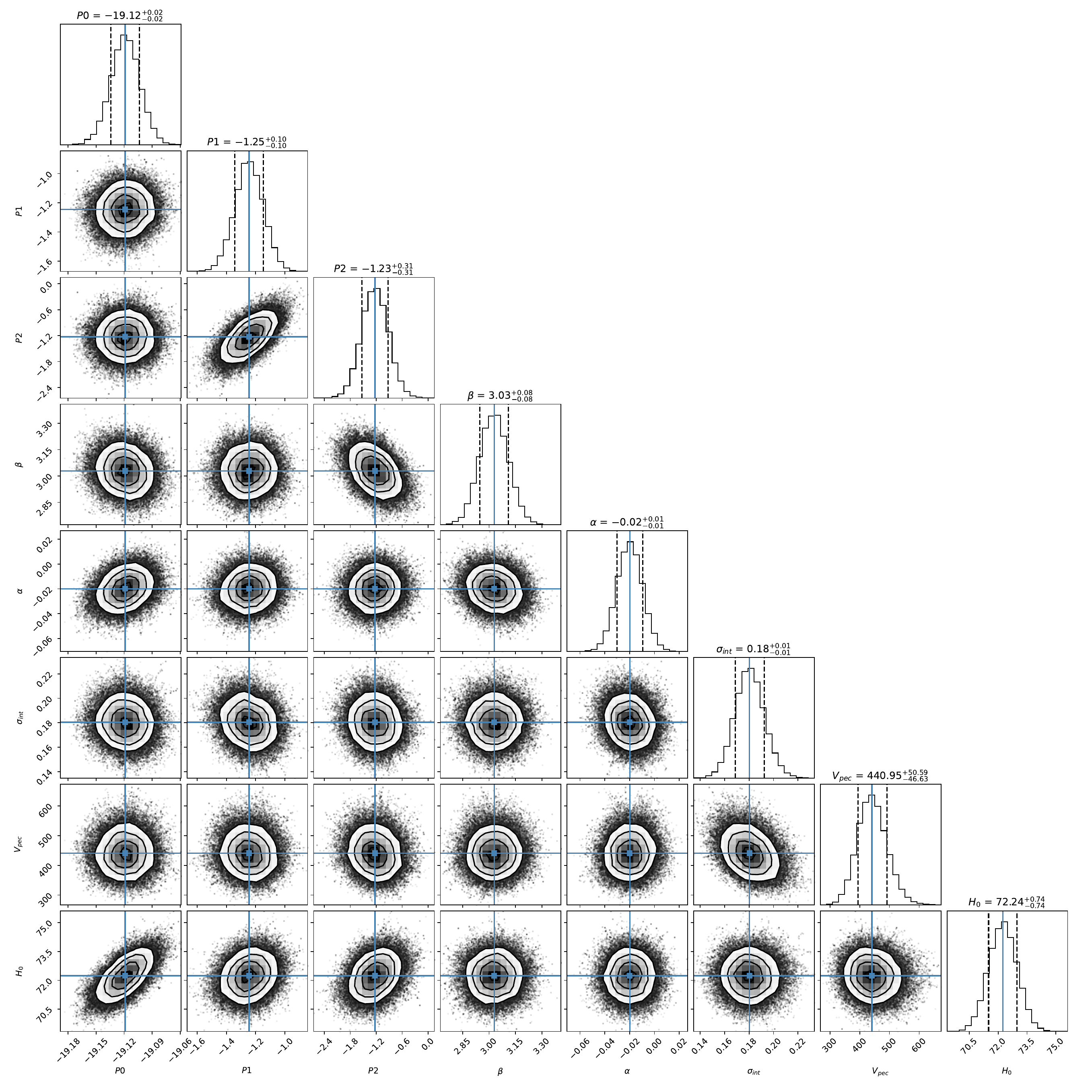}
    \caption{Posterior distribution of MCMC fitting parameters in determining $H_0$ using the $B$-band peak magnitude based on a Cepheid calibration. See Table~\ref{comp} for the units of parameters.}
    \label{fig:mcmcH0}
\end{figure*}

We present $H_0$ and other nuisance variables using different calibration methods across $uBgVriYJH$ bands in Table~\ref{tab:all}. Figure~\ref{fig:mcmcH0} represents a corner plot of posterior probability distributions of variables from \texttt{EMCEE} output using the Cepheid calibration. Using the $B$-band light-curve fit, we get $\rm H_0=72.24\pm 0.74 \ km \ s^{-1} \ Mpc^{-1}$ from the Cepheid calibration
%\footnote{By adding a host mass offset of 0.3 dex that we discussed in \S \ref{sec:data} we get $\rm H_0=72.37 \pm 0.70 \ km \ s^{-1} \ Mpc^{-1}$. This is because the effect of host mass is insignificant.} 
and $\rm H_0=70.31 \pm 0.70 \ km \ s^{-1} \ Mpc^{-1}$ from the TRGB calibration. When we use the SBF calibration, we obtain different values depending on the compilation used. We obtain $\rm H_0=69.52 \pm 0.93 \ km \ s^{-1} \ Mpc^{-1}$ using the SBF compilation from \cite{khetan21} and $\rm H_0=77.11 \pm 1.11 \ km \ s^{-1} \ Mpc^{-1}$ using the compilation from \cite{jensen21}. If we combine these two compilations, we obtain $\rm H_0=72.62 \pm 0.90 \ km \ s^{-1} \ Mpc^{-1}$. Throughout this paper, we present statistical errors unless otherwise
stated. We define statistical errors as the 1-$\sigma$ confidence interval of the
marginalized parameter in question. For $H_0$, this includes the uncertainty in all 
nuisance parameters in Eqn.~\ref{muobs}. We discuss systematic errors due to the variation in the SN~Ia distance scales among various calibrators and due to SN~Ia sample selection effects in \S~\ref{comcal} and in \S~\ref{sel}, respectively. 

When we compare $H_0$ using Cepheid and TRGB calibrations, we find them to be consistent with previously published studies (See Table~\ref{tab:h0}). When we compare $H_0$ using the SBF calibration from \cite{khetan21}, we also obtain a consistent result. Similarly, we obtain consistent $H_0$ using the SBF calibrators from \cite{jensen21}, with a recent study by \cite{garnavich22}, who also used SBF calibrators from \cite{jensen21}. However, these SBF-calibrated $H_0$ values differ depending on which SBF sample is used: \cite{khetan21} or \cite{jensen21}. The difference is significant ($> 5 \sigma$) when we consider statistical error only, but reduced to $< 2 \sigma$, when we consider systematic errors of $\rm \sim 3.0 \ km \ s^{-1} \ Mpc^{-1}$.

It is currently not clear why we are observing significant differences in $H_0$ from the two SBF compilations. As noted earlier, the average distance moduli of 22 SBF hosts that we use in this work from \cite{jensen21} is 1.5 mag fainter than that of 24 SBF hosts from \cite{khetan21}. In the rest of this study (from Table~\ref{tab:91} onward), we concatenate SBF distances from \cite{jensen21}  and \cite{khetan21}, and use the combined sample as the SBF sample\footnote{For the seven hosts that are common between \cite{jensen21} and \cite{khetan21}, we use distances from \cite{jensen21}, since \cite{jensen21} provide the most recent measurements of them.}.

We also note from Table~\ref{tab:all} that SN~Ia absolute magnitudes ($P0$ values) are different among the calibration methods. $H_0$ decreases with increasing SN~Ia luminosity, which is a direct consequence of Eqn.~\ref{mumod}. These two terms are also dependent on each other (See Figure~\ref{fig:mcmcH0}). In the absence of any distance calibration, they are fully degenerate. It is the systematic differences in the calibrator distances that drive differences in $H_0$. We compare various calibrators in \S \ref{comcal}.

Next, we investigate how $H_0$ varies when SNe~Ia are observed across various photometric bandpasses. In the Cepheid calibration, $H_0$ is consistent among $BVri$ bands, but gives larger values in $ugYJH$ bands\textbf{\footnote{We note that \cite{galbany22} used publicly available CSP-I data to calculate $H_0$ utilizing Cepheid calibration in $YJH$ bands. Their results are consistent with ours within systematic uncertainties.}}. In the TRGB calibration, $H_0$ is different ($\sim 2 \sigma$) between optical ($uBgVri$) and near-infrared ($YJH$) bands. Previously, \cite{freedman19} found smaller value of $H_0$ in the $H$ band compare to the $B$ band, using the CSP-I SN~Ia sample. We also obtain similar result when we use the CSP-I sample. It is the addition of SNe~Ia from CSP-II that  changes the trend in $H_0$ (see Figure~\ref{fig:h0trgb}). In the combined SBF calibration, we find smaller $H_0$ in the $J$ and $H$ bands compared to rest of the bands. 

A conspicuous trend in the TRGB-calibrated values of $H_0$ can be seen in Figure \ref{fig:h0trgb}. While the combined CSP-I and II sample always yields a value for $H_0$ between the CSP-I and CSP-II subsamples (as one would expect), CSP-I tends to give a higher value of $H_0$ than CSP-II in the optical, but the opposite is true in the NIR. This raises the concern that there is a systematic error in the NIR photometry that has been introduced when switching our NIR camera (RetroCam) from the Swope telescope to the du Pont. Yet there does not appear to be any such systematic in the photometry of the local sequence photometry (see Suntzeff et al., in preparation). Examining the distribution of residuals reveals that they are significantly non-Gaussian in the NIR, with a bright ``tail" (see Figure \ref{fig:hist_trgb_opt_NIR}). This tail is seen in both the CSP-I and CSP-II subsamples, indicating it is not likely that a systematic offset is responsible for the different values of $H_0$. Fitting with a two-component Gaussian indicates that nearly 20\% of the combined CSP-I and CSP-II sample originate from this brighter tail in the NIR, yet no such tail is seen in the optical. These overly bright residuals do not correlate with any of the corrections we apply ($s_{BV}$, $B-V$ color, host mass, or red-shift). Interestingly, the mode of the residuals in the optical appear to line up with the mode of residuals in the NIR for the same $H_0$, as shown in Figure \ref{fig:hist_trgb_opt_NIR}. This could indicate the existence of an overly-NIR-bright subpopulation of SNe~Ia and warrants further investigation.

We also calculate $H_0$ and other variables by excluding 91T and 91bg-like SNe~Ia\footnote{We note that not all SNe~Ia have spectroscopic information at the right phase for sub-typing.}. We show our results in Table~\ref{tab:91}. Although they are $\emph{bona fide}$ SNe~Ia, they have peculiar spectra, and may have separate progenitor channels.
Moreover, 91T-like SNe~Ia \citep{filippenko92b,phillips92,ruiz92} are relatively over-luminous \citep{boone21,phillips22,Yang22}, and 91bg-like events are sub-luminous \citep{filippenko92a,leibundgut93}.
%Moreover, 91T-like SNe~Ia (\citealt{phillips92}) are found to be relatively over-luminous (Phillips et al. in preparation), and 91bg-like to be sub-luminous (\citealt{filippenko92}).
In both Tables~\ref{tab:all} and \ref{tab:91}, we see slightly larger values of $H_0$ when the TRGB calibration is used in near-infrared ($YJH$) bands. This scenario is reversed in the SBF calibration. $H_0$ in the Cepheid calibration remains consistent across the bands.

\begin{figure}
    \centering
    \includegraphics[width=\columnwidth]{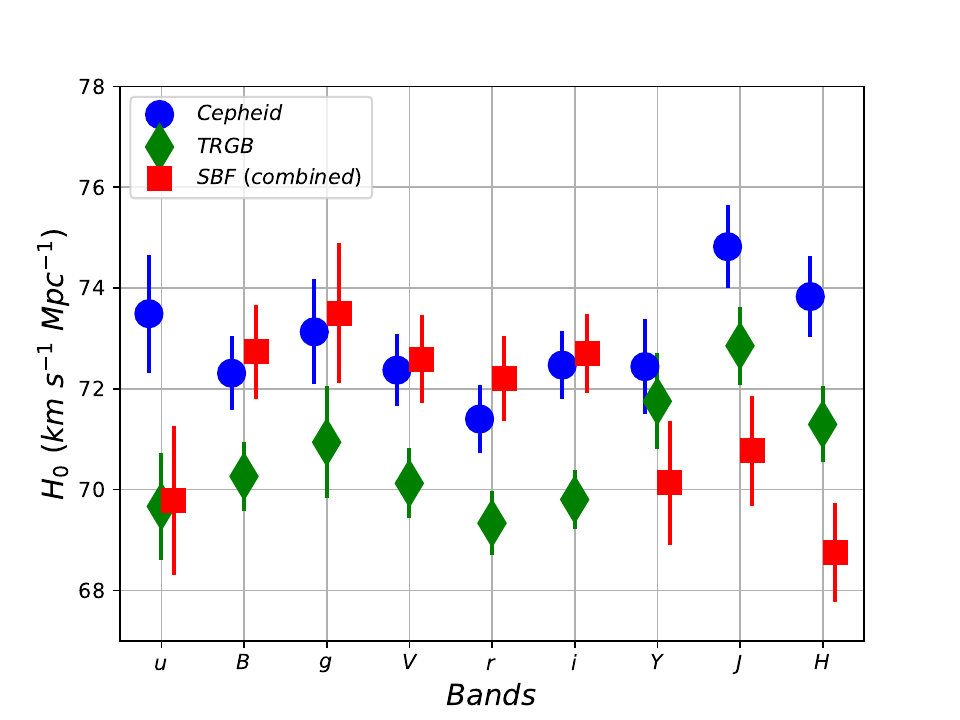}
    \caption{$H_0$ across various bands. Values of $H_0$ are taken from Table~\ref{tab:all}. See text for discussion.}
    \label{fig:h0bands}
\end{figure}

\begin{figure}
    \centering
    \includegraphics[width=\columnwidth]{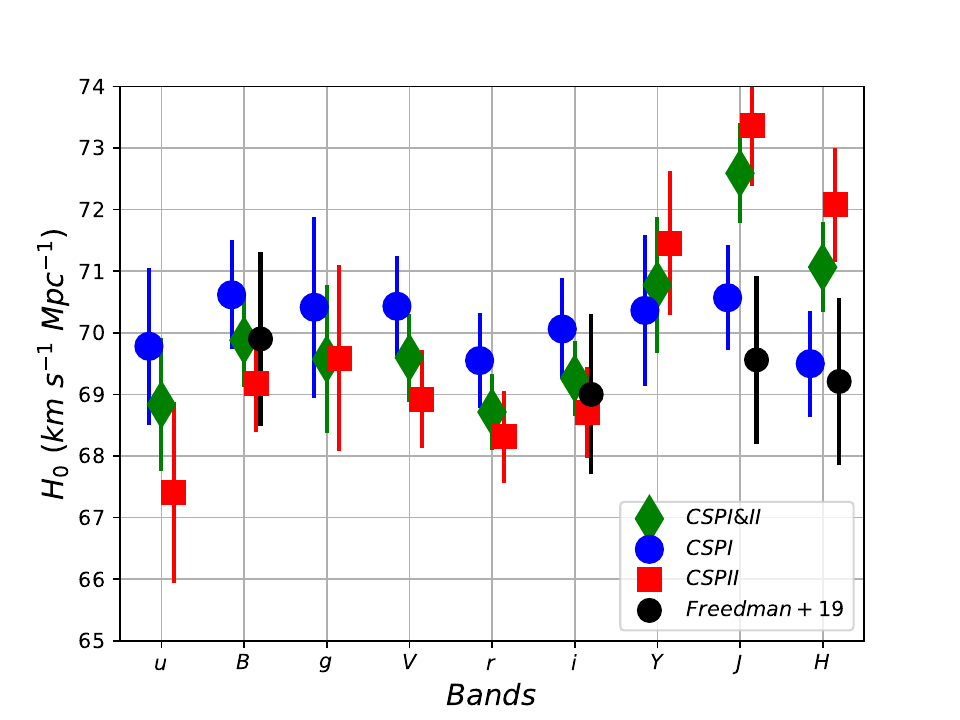}
    \caption{$H_0$ across various bands in TRGB calibration. See text for discussion.}
    \label{fig:h0trgb}
\end{figure}

\begin{figure}
    \centering
    \includegraphics[width=\columnwidth]{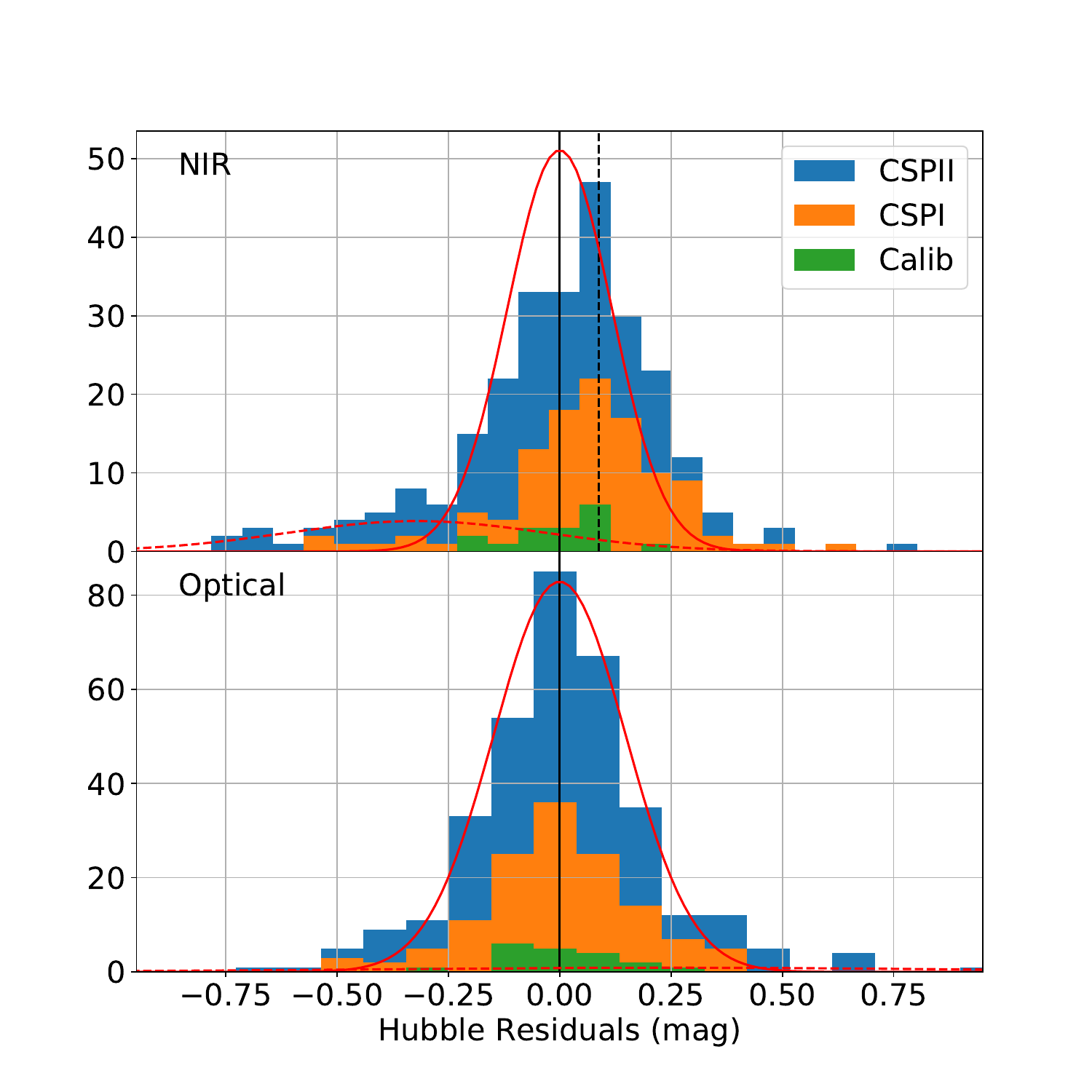}
    \caption{The Hubble diagram residuals in the optical (bottom panel) and NIR (top panel). The histograms are stacked with the green corresponding to the TRGB calibrations, the orange corresponding to the CSP-I sample, and the blue corresponding to the CSP-II. A two-component Gaussian fit is show for the NIR residuals. \textbf{The vertical dashed line shows the location of zero-residuals in the NIR if we force $H_0$ to the value that we obtain from optical.}}
    \label{fig:hist_trgb_opt_NIR}
\end{figure}

\begin{table*}
    \centering
    \caption{$H_0$ and all other nuisance parameters for the full sample in various bands. The number of calibrators used in each band is mentioned in Table~\ref{tab:sample}. For $H_0$, if the uncertainty is greater than one $\rm km \ s^{-1} \ Mpc^{-1}$, we explicitly include decimals in quoting uncertainties. For all other cases, they should be read as the first two digits after decimals. We keep this convention throughout the paper.}
    \begin{tabular}{c|c|c|c|c|c|c|c|c|c}
    \hline
    \hline
   Calibration & Band & $H_0$ & $\sigma_{int}$ & $V_{pec}$ & $P0$ & $P1$& $P2$ & $\alpha$ & $\beta$\\
    && $(\rm km \ s^{-1} \ Mpc^{-1})$ & (mag) & ($\rm km \ s^{-1}$) & (mag) & (mag) & (mag) & $\rm (mag/dex)$ & \\
    \hline 
 & $ u $ & 73.61 (1.07) & 0.23 (02) & 479 (62) & -18.721 (021) & -1.89 (14) & -1.89 (48) & -0.06 (01) & 4.30 (13) \\
& $ B $ & 72.24 (0.74) & 0.18 (01) & 440 (48) & -19.119 (015) & -1.25 (09) & -1.23 (30) & -0.02 (01) & 3.03 (08) \\
& $ g $ & 72.93 (0.94) & 0.15 (01) & 471 (46) & -19.156 (023) & -1.21 (11) & -1.18 (33) & -0.03 (01) & 2.59 (09) \\
& $ V $ & 72.39 (0.66) & 0.19 (01) & 463 (48) & -19.116 (014) & -1.28 (09) & -1.40 (31) & -0.02 (01) & 2.08 (07) \\
Cepheid& $ r $ & 71.37 (0.65) & 0.17 (01) & 442 (44) & -19.070 (015) & -1.14 (08) & -0.83 (28) & -0.03 (00) & 1.57 (06) \\
& $ i $ & 72.41 (0.63) & 0.15 (01) & 422 (41) & -18.462 (014) & -0.74 (08) & -0.54 (25) & -0.03 (00) & 1.11 (06) \\
& $ Y $ & 72.48 (0.80) & 0.12 (01) & 360 (37) & -18.476 (021) & -0.40 (07) & 0.66 (23) & -0.03 (00) & 0.49 (06) \\
& $ J $ & 74.75 (0.81) & 0.17 (01) & 370 (47) & -18.616 (017) & -0.68 (1) & 0.12 (30) & -0.03 (01) & 0.51 (08) \\
& $ H $ & 73.78 (0.71) & 0.14 (01) & 369 (45) & -18.386 (015) & -0.41 (09) & 0.45 (30) & -0.03 (01) & 0.23 (08) \\
  \hline 
 & $ u $ & 69.67 (1.06) & 0.23 (02) & 446 (62) & -18.844 (025) & -1.47 (14) & -0.20 (40) & -0.07 (02) & 3.87 (10) \\
& $ B $ & 70.31 (0.70) & 0.18 (01) & 431 (49) & -19.179 (015) & -1.08 (09) & -0.53 (28) & -0.00 (01) & 2.87 (06) \\
& $ g $ & 71.00 (1.15) & 0.15 (01) & 464 (49) & -19.216 (032) & -1.08 (11) & -0.64 (33) & -0.01 (01) & 2.50 (09) \\
& $ V $ & 70.15 (0.64) & 0.18 (01) & 452 (48) & -19.186 (014) & -1.08 (09) & -0.59 (27) & -0.01 (01) & 1.91 (06) \\
TRGB & $ r $ & 69.26 (0.64) & 0.17 (01) & 441 (44) & -19.137 (015) & -0.96 (08) & -0.11 (25) & -0.01 (00) & 1.43 (05) \\
& $ i $ & 69.76 (0.62) & 0.15 (01) & 429 (44) & -18.540 (014) & -0.56 (07) & 0.04 (22) & -0.01 (00) & 0.97 (05) \\
& $ Y $ & 71.65 (0.99) & 0.12 (01) & 348 (38) & -18.502 (028) & -0.33 (07) & 0.80 (24) & -0.03 (00) & 0.50 (06) \\
& $ J $ & 72.92 (0.81) & 0.18 (01) & 332 (47) & -18.674 (016) & -0.55 (10) & 0.77 (29) & -0.04 (01) & 0.30 (06) \\
& $ H $ & 71.45 (0.74) & 0.15 (01) & 340 (42) & -18.467 (016) & -0.29 (09) & 0.85 (26) & -0.05 (01) & 0.19 (06) \\
\hline        
& $ u $ & 66.24 (1.50) & 0.22 (01) & 466 (54) & -18.949 (045) & -1.39 (17) & -0.27 (49) & -0.02 (02) & 4.01 (14) \\
& $ B $ & 69.52 (0.93) & 0.18 (01) & 441 (40) & -19.214 (026) & -0.90 (09) & -0.06 (29) & -0.01 (01) & 2.85 (08) \\
& $ g $ & 68.86 (1.75) & 0.14 (01) & 464 (48) & -19.269 (052) & -0.92 (12) & -0.56 (34) & 0.00 (01) & 2.51 (09) \\
& $ V $ & 69.63 (0.91) & 0.18 (01) & 453 (42) & -19.211 (024) & -0.90 (09) & -0.22 (29) & -0.01 (01) & 1.91 (08) \\
SBF\footnote{\citealt{khetan21}}& $ r $ & 69.40 (0.91) & 0.17 (01) & 408 (39) & -19.139 (025) & -0.90 (08) & -0.10 (26) & -0.01 (01) & 1.49 (07) \\
& $ i $ & 70.03 (0.89) & 0.15 (01) & 394 (37) & -18.538 (025) & -0.59 (07) & -0.06 (23) & -0.02 (01) & 1.04 (06) \\
& $ Y $ & 64.11 (1.66) & 0.12 (01) & 346 (35) & -18.745 (056) & -0.34 (08) & 0.85 (25) & -0.02 (01) & 0.48 (06) \\
& $ J $ & 67.54 (1.17) & 0.17 (01) & 372 (39) & -18.838 (033) & -0.61 (11) & 0.41 (32) & -0.03 (01) & 0.41 (09) \\
& $ H $ & 64.99 (1.02) & 0.14 (01) & 386 (39) & -18.665 (03) & -0.37 (10) & 0.63 (30) & -0.03 (01) & 0.21 (08) \\
\hline
& $ u $ & 73.04 (1.64) & 0.22 (02) & 480 (62) & -18.724 (047) & -1.34 (16) & -0.24 (48) & -0.01 (02) & 3.93 (14) \\
& $ B $ & 77.11 (1.11) & 0.17 (01) & 462 (49) & -18.963 (028) & -0.84 (08) & -0.68 (29) & 0.02 (01) & 2.95 (08) \\
& $ g $ & 76.71 (1.76) & 0.14 (01) & 487 (47) & -19.035 (049) & -0.95 (12) & -0.68 (35) & -0.00 (01) & 2.50 (09) \\
& $ V $ & 77.56 (1.08) & 0.18 (01) & 484 (48) & -18.949 (028) & -0.85 (09) & -0.84 (28) & 0.02 (01) & 1.99 (08) \\
SBF\footnote{\citealt{jensen21}}& $ r $ & 76.37 (0.98) & 0.16 (01) & 461 (45) & -18.913 (025) & -0.81 (08) & -0.38 (26) & 0.01 (01) & 1.53 (07) \\
& $ i $ & 76.70 (1.01) & 0.14 (01) & 446 (43) & -18.327 (026) & -0.58 (07) & -0.40 (25) & -0.00 (01) & 1.07 (06) \\
& $ Y $ & 73.90 (1.64) & 0.12 (01) & 367 (37) & -18.428 (048) & -0.36 (08) & 0.64 (24) & -0.02 (01) & 0.48 (06) \\
& $ J $ & 75.46 (1.47) & 0.17 (01) & 374 (46) & -18.594 (04) & -0.61 (10) & 0.24 (30) & -0.02 (01) & 0.48 (08) \\
& $ H $ & 76.08 (1.39) & 0.14 (01) & 375 (46) & -18.315 (038) & -0.33 (10) & 0.50 (30) & -0.01 (01) & 0.26 (07) \\
\hline
& $ u $ & 69.81 (1.44) & 0.22 (02) & 474 (62) & -18.843 (04) & -1.36 (16) & -0.14 (46) & -0.02 (02) & 4.04 (14) \\
& $ B $ & 72.62 (0.90) & 0.18 (01) & 449 (49) & -19.115 (022) & -0.87 (08) & -0.04 (28) & -0.00 (01) & 2.84 (08) \\
& $ g $ & 73.32 (1.39) & 0.14 (01) & 481 (51) & -19.138 (039) & -0.94 (12) & -0.62 (36) & -0.00 (01) & 2.52 (09) \\
& $ V $ & 72.87 (0.85) & 0.18 (01) & 470 (50) & -19.112 (02) & -0.84 (08) & -0.12 (28) & -0.01 (01) & 1.90 (07) \\
SBF & $ r $ & 72.12 (0.83) & 0.17 (01) & 447 (45) & -19.056 (02) & -0.85 (08) & 0.04 (26) & -0.01 (01) & 1.46 (07) \\
(combined)& $ i $ & 72.65 (0.81) & 0.15 (01) & 433 (45) & -18.460 (02) & -0.59 (07) & -0.01 (22) & -0.02 (00) & 1.04 (06) \\
& $ Y $ & 70.12 (1.32) & 0.12 (01) & 363 (37) & -18.547 (040) & -0.35 (08) & 0.70 (24) & -0.03 (00) & 0.50 (06) \\
& $ J $ & 70.69 (1.11) & 0.17 (01) & 362 (47) & -18.747 (029) & -0.63 (11) & 0.47 (31) & -0.03 (01) & 0.44 (08) \\
& $ H $ & 68.58 (0.99) & 0.14 (01) & 361 (46) & -18.560 (026) & -0.41 (10) & 0.88 (29) & -0.03 (01) & 0.20 (07) \\
\hline 
    \end{tabular}
    
    \label{tab:all}
\end{table*}

\begin{table*}
    \centering
    \caption{$H_0$ and all other nuisance parameters. Here, we exclude 91T and 91bg-like SNe~Ia.}
    \begin{tabular}{c|c|c|c|c|c|c|c|c|c}
    \hline
    \hline
    Calibration & Band & $H_0$ & $\sigma_{int}$ & $V_{pec}$ & $P0$ & $P1$& $P2$ & $\alpha$ & $\beta$\\
    && $(\rm km \ s^{-1} \ Mpc^{-1})$ & (mag) & ($\rm km \ s^{-1}$) & (mag) & (mag) & (mag) & $\rm (mag/dex)$ & \\
    \hline 

 & $ u $ & 72.59 (1.10) & 0.19 (02) & 501 (66) & -18.717 (024) & -1.88 (15) & -2.85 (57) & -0.07 (01) & 4.33 (14) \\
& $ B $ & 71.67 (0.68) & 0.15 (01) & 456 (48) & -19.117 (016) & -1.22 (10) & -1.39 (37) & -0.03 (01) & 3.01 (08) \\
& $ g $ & 72.61 (0.89) & 0.12 (01) & 473 (51) & -19.151 (024) & -1.27 (12) & -1.70 (43) & -0.03 (01) & 2.55 (09) \\
& $ V $ & 71.67 (0.69) & 0.15 (01) & 471 (52) & -19.114 (015) & -1.26 (09) & -1.71 (33) & -0.03 (01) & 2.06 (08) \\
Cepheid & $ r $ & 70.75 (0.63) & 0.14 (01) & 451 (46) & -19.070 (015) & -1.10 (09) & -0.93 (32) & -0.03 (00) & 1.53 (06) \\
& $ i $ & 71.88 (0.60) & 0.12 (01) & 442 (41) & -18.462 (015) & -0.59 (08) & -0.35 (30) & -0.03 (00) & 1.11 (06) \\
& $ Y $ & 71.99 (0.77) & 0.10 (01) & 370 (38) & -18.475 (022) & -0.23 (08) & 0.83 (27) & -0.03 (00) & 0.51 (06) \\
& $ J $ & 73.80 (0.72) & 0.14 (01) & 365 (44) & -18.614 (016) & -0.47 (10) & 0.10 (38) & -0.03 (01) & 0.52 (07) \\
& $ H $ & 73.57 (0.70) & 0.11 (01) & 387 (44) & -18.378 (016) & -0.22 (11) & 0.51 (41) & -0.03 (01) & 0.23 (07) \\
\hline
 & $ u $ & 69.26 (1.08) & 0.21 (02) & 453 (65) & -18.839 (025) & -1.47 (18) & -0.48 (57) & -0.09 (02) & 3.87 (11) \\
& $ B $ & 69.87 (0.69) & 0.16 (01) & 436 (49) & -19.171 (016) & -1.09 (10) & -0.68 (30) & -0.01 (01) & 2.86 (07) \\
& $ g $ & 70.73 (1.16) & 0.13 (01) & 462 (51) & -19.216 (033) & -1.14 (12) & -0.77 (43) & -0.02 (01) & 2.45 (09) \\
& $ V $ & 69.91 (0.66) & 0.16 (01) & 458 (50) & -19.167 (016) & -1.21 (09) & -1.15 (26) & -0.01 (01) & 1.86 (06) \\
TRGB & $ r $ & 68.93 (0.63) & 0.15 (01) & 449 (47) & -19.136 (015) & -0.91 (09) & 0.01 (33) & -0.01 (01) & 1.43 (05) \\
& $ i $ & 69.41 (0.61) & 0.12 (01) & 443 (45) & -18.534 (015) & -0.44 (07) & 0.25 (23) & -0.01 (00) & 0.99 (05) \\
& $ Y $ & 71.48 (1.02) & 0.10 (01) & 360 (37) & -18.494 (029) & -0.16 (07) & 0.99 (24) & -0.03 (00) & 0.53 (06) \\
& $ J $ & 72.29 (0.74) & 0.14 (01) & 335 (42) & -18.665 (017) & -0.34 (10) & 0.92 (34) & -0.04 (01) & 0.33 (06) \\
& $ H $ & 71.08 (0.74) & 0.12 (01) & 363 (44) & -18.463 (016) & -0.05 (10) & 1.18 (29) & -0.04 (01) & 0.16 (06) \\
\hline
& $ u $ & 69.75 (1.51) & 0.21 (02) & 477 (61) & -18.825 (046) & -1.22 (21) & 0.04 (59) & -0.03 (02) & 4.06 (14) \\
& $ B $ & 72.74 (0.90) & 0.16 (01) & 454 (50) & -19.090 (023) & -0.92 (13) & -0.53 (49) & -0.01 (01) & 2.90 (09) \\
& $ g $ & 73.94 (1.54) & 0.13 (01) & 474 (47) & -19.109 (045) & -0.99 (14) & -0.94 (39) & -0.01 (01) & 2.51 (09) \\
& $ V $ & 72.86 (0.87) & 0.16 (01) & 478 (53) & -19.080 (026) & -1.20 (12) & -1.59 (36) & -0.02 (01) & 1.96 (1) \\
SBF & $ r $ & 72.03 (0.86) & 0.15 (01) & 459 (48) & -19.048 (022) & -0.79 (10) & 0.17 (40) & -0.02 (01) & 1.47 (07) \\
(combined) & $ i $ & 72.65 (0.78) & 0.12 (01) & 453 (43) & -18.446 (021) & -0.46 (07) & 0.22 (25) & -0.03 (00) & 1.04 (06) \\
& $ Y $ & 70.92 (1.37) & 0.10 (01) & 377 (39) & -18.506 (041) & -0.17 (08) & 0.84 (25) & -0.03 (01) & 0.52 (06) \\
& $ J $ & 71.15 (1.10) & 0.14 (01) & 363 (41) & -18.710 (029) & -0.35 (12) & 0.86 (34) & -0.04 (01) & 0.47 (08) \\
& $ H $ & 69.36 (1.04) & 0.11 (01) & 390 (47) & -18.526 (028) & -0.09 (11) & 1.44 (29) & -0.03 (01) & 0.21 (08) \\

\hline        
    \end{tabular}
    
    \label{tab:91}
\end{table*}

The use of nine $uBgVriYJH$ bands from optical to near-infrared provides us an opportunity to measure inter-band scatter in the values of $H_0$, which can be considered as an additional systematic uncertainty. We show inter-band scatter in Table~\ref{tab:bandcut} and find that the inter-band scatter is $\sim 0.03 \ \rm km \ s^{-1} \ Mpc^{-1}$, which is much smaller than the statistical error in $H_0$.

Finally, in measuring $H_0$, we have used host stellar mass as an additional correction factor in the SN~Ia luminosity calibration. If we do not use this correction, $H_0$ does not change significantly. We show this in Table~\ref{tab:hostffect}.

\begin{table}[]
    \centering
    \caption{Effect of host mass correction in determining $H_0$. We do not find significant differences in $H_0$ whether host mass correction is applied or not. Shown here are the values of $H_0$ from $B$-band light-curve fits and using various distance calibration.}
    \begin{tabular}{lcc}
    \hline
    \hline
          &$H_0$ ($\rm km \ s^{-1} \ Mpc^{-1}$)&\\
         \cmidrule{2-3}
         Calibration&Linear correction &No correction \\
         \hline
         Cepheid &72.24~(0.74)&72.37~(0.71)\\
         TRGB &70.31~(0.70)& 70.25~(0.71)\\
         
         SBF &72.62~(0.90)& 72.45~(0.94)\\
         
         \hline
    \end{tabular}
    
    \label{tab:hostffect}
\end{table}

\subsubsection{Combining All Calibrators}\label{h0all}
In this section, we combine all calibrators from Cepheid, TRGB, and SBF to derive $H_0$. We utilize $B$ and $H$ band data to represent optical and near-infrared results, respectively. Combining all calibrators also allows us to determine systematic error that may originate from the variation in distance scales among different calibrators (see \S~\ref{comcal}). We follow the same procedure as in \S~\ref{h0ind} and present $H_0$ in Table~\ref{tab:h0all}. 

\textbf{It has been common in SN~Ia cosmology analyses since Joint Light Curve Analysis (JLA; \citealt{betoule14}) to correct SN Ia redshifts using estimates of peculiar velocities to account for local, large scale structures, with estimates most commonly derived from the 2M++ survey (\citealt{carrick15}). These have been shown to improve the overall fit ($\chi^2$ and dispersion) in large SN~Ia samples (\citealt{peterson23}).  Here, starting with the combined analysis, we include these velocity corrections calculated from \cite{carrick15}. The mean sample velocity correction for the CSP survey is $\rm 90 \ km \ sec^{-1}$ and the standard deviation of the corrections is $\rm 320 \ km \ sec^{-1}$. At the sample mean, $\rm cz=10,800 \ km \ sec^{-1}$, this is equivalent to a mean correction of $\rm 5log((cz+90)/cz)=0.017$ mag and a standard deviation of the corrections of 0.064 mag. To assess the significance of these corrections to the Hubble diagram, we can compare the SN $\sigma_{int}$ with the corrections to $\sigma_{int}$ with randomized application of corrections drawn from a gaussian of the same width as the corrections. For the $B$-band analysis, which is the largest sample, the improvement would occur by chance 1\% of the time, i.e., $\sim$ 99\% confidence in the usefulness of the corrections. For the smaller $H$-band analysis, it is 90\% confidence. The impact to $H_0$ is to increase it by $\rm 0.55\ km \ s^{-1} \ Mpc^{-1}$ for $B$ and $\rm 0.6 \ km \ s^{-1} \ Mpc^{-1}$ for $H$.
}

From the $B$ and $H$ bands we obtain $\rm H_0=71.76 \pm 0.58 \ km \ s^{-1} \ Mpc^{-1}$ and $\rm H_0=73.22 \pm 0.68 \ km \ s^{-1} \ Mpc^{-1}$. We consider these two values of $H_0$ as agnostic values of $H_0$ from this work, since we do not prefer one calibration method over others. We discuss and calculate systematic uncertainties among various distance scales by performing an analysis in \S\ref{comcal}. 

\begin{table*}
    \centering
    \caption{Determination of $H_0$ from the combined Cepheid, TRGB, and SBF distance calibrators.}
    \begin{tabular}{c|c|c|c|c|c|c|c|c}
    \hline
    \hline
     Band & $H_0$ & $\sigma_{int}$ & $V_{pec}$ & $P0$ & $P1$& $P2$ & $\alpha$ & $\beta$\\
    & $(\rm km \ s^{-1} \ Mpc^{-1})$ & (mag) & ($\rm km \ s^{-1}$) & (mag) & (mag) & (mag) & $\rm (mag/dex)$ & \\
    \hline

    $ B $ & 71.76 (0.58) & 0.17 (01) & 466 (56) & -19.138 (01) & -1.48 (09) & -1.76 (26) & -0.02 (00) & 2.90 (06) \\
 $ H $ & 73.22 (0.68) & 0.16 (01) & 373 (48) & -18.451 (011) & -0.18 (09) & 1.13 (26) & -0.07 (00) & 0.23 (06) \\
\hline
 \end{tabular}
    
    \label{tab:h0all}
\end{table*}

\section{Discussion}\label{sec:dis}

\subsection{Systematic Difference among Various Calibrators}\label{comcal}
We see in Table~\ref{tab:all} that $H_0$ and $P0$ are different when they are obtained using different calibration methods. In this context, we can address the question of how absolute magnitudes of SNe~Ia would vary between the calibrating SNe~Ia samples. To answer this, we perform a MCMC analysis where we use all three calibrations simultaneously, and assign a separate absolute magnitude for each case (e.g. $P0_{Cepheid}$, $P0_{TRGB}$, and $P0_{SBF}$). The absolute magnitude of a distant SN~Ia is $P0_{dist}$. In performing the MCMC sampling, we draw $P0s$ of calibrating SNe~Ia from a normal distribution that has a mean of $P0_{dist}$ and a standard deviation, $\sigma_{cal}$ that we add as an additional fitting parameter. The advantage of fitting simultaneously, rather than fitting each calibrator separately, is that all other nuisance parameters are solved consistently, and their uncertainties are propagated correctly. 
%Furthermore, the value of $\sigma_{cal}$ encapsulates all the systematic uncertainties present in the distance ladders below the SNe~Ia. 
We show this analysis in Table~\ref{tab:allcal}. 

\textbf{The complete uncertainty in $H_0$ is given as the error in the posterior and including marginalization over $\sigma_{cal}$. We use the error in $P0_{dist}$ to calculate the systematic error in $H_0$. To do this, we first remove the statistical error in $P0$ (Table~\ref{tab:h0all}) from the total error in $P0_{dist}$ (Table~\ref{tab:allcal}) to get the systematic error. Since $\rm \mu \sim 5\times log_{10}(cz/H_0)$, $\rm \sigma_{H_0}/H_0 \sim \sigma_{P0}/2.17$. Therefore, $\sigma_{H_0}$ corresponds to $\rm 1.19 \ km \ s^{-1} \ Mpc^{-1}$ in $B$-band, and $\rm 1.28 \ km \ s^{-1} \ Mpc^{-1}$ in $H$-band. These values of $\sigma_{H_0}$ are the systematic uncertainties as determined from the variation of distance scales among the calibrators.}

Adding systematic errors to $H_0$, 
%from this section and from \S~\ref{sel}, 
we can write $\rm H_0=71.76 \pm 0.58 \ (stat) \pm 1.19 \ (sys) \ km \ s^{-1} \ Mpc^{-1}$ from SNe~Ia $B$-band luminosity, and  $\rm H_0=73.22 \pm 0.68 \ (stat) \pm 1.28 \ (sys) \ km \ s^{-1} \ Mpc^{-1}$ from SNe~Ia $H$-band luminosity. These results provide important insight on the differences in $H_0$ that we see from various calibrators. The most recent result from the SH0Es program (\citealt{riess22}), that uses Cepheid calibrators, reports $\rm H_0=73.04 \ \pm 1.04 \ (total) \ km \ s^{-1} \ Mpc^{-1}$. The CCHP program (\citealt{freedman21}), that uses TRGB calibrators, reports $\rm H_0=69.80 \pm 0.60 \ (stat) \pm 1.6 \ (sys) \ km \ s^{-1} \ Mpc^{-1}$. %Clearly, there is a difference of $\rm 3.24 \ km \ s^{-1} \ Mpc^{-1}$ at $\sim 2 \sigma$ level between these two findings. 
Moreover, using SBF calibrators, \cite{garnavich22} reports $\rm H_0=74.60 \pm 0.90 \ (stat) \pm 2.7 \ (sys) \ km \ s^{-1} \ Mpc^{-1}$ and \cite{khetan21} reports $\rm H_0=70.50 \pm 2.37 \ (stat) \pm 3.38 \ (sys) \ km \ s^{-1} \ Mpc^{-1}$. \textbf{If we combine Cepheids, TRGB and the space-based SBF from \cite{jensen21} (i.e., excluding the ground-based SBF from \citealt{khetan21}) we find $\rm H_0=72.21 \pm 0.60 \ (stat) \ km \ s^{-1} \ Mpc^{-1}$ for $B$-band.} It is evident from our analysis that the differences in $H_0$ between previously published studies can be explained simply by the systematic differences in the SN~Ia distance scales among various calibrators. %Moreover, our result  removes the existing tension in $H_0$ that exist between the late-time $H_0$ from various distant calibrators and the early-time $H_0$ from the CMB.

\begin{figure*}
    \centering
    
    \includegraphics[width=\textwidth]{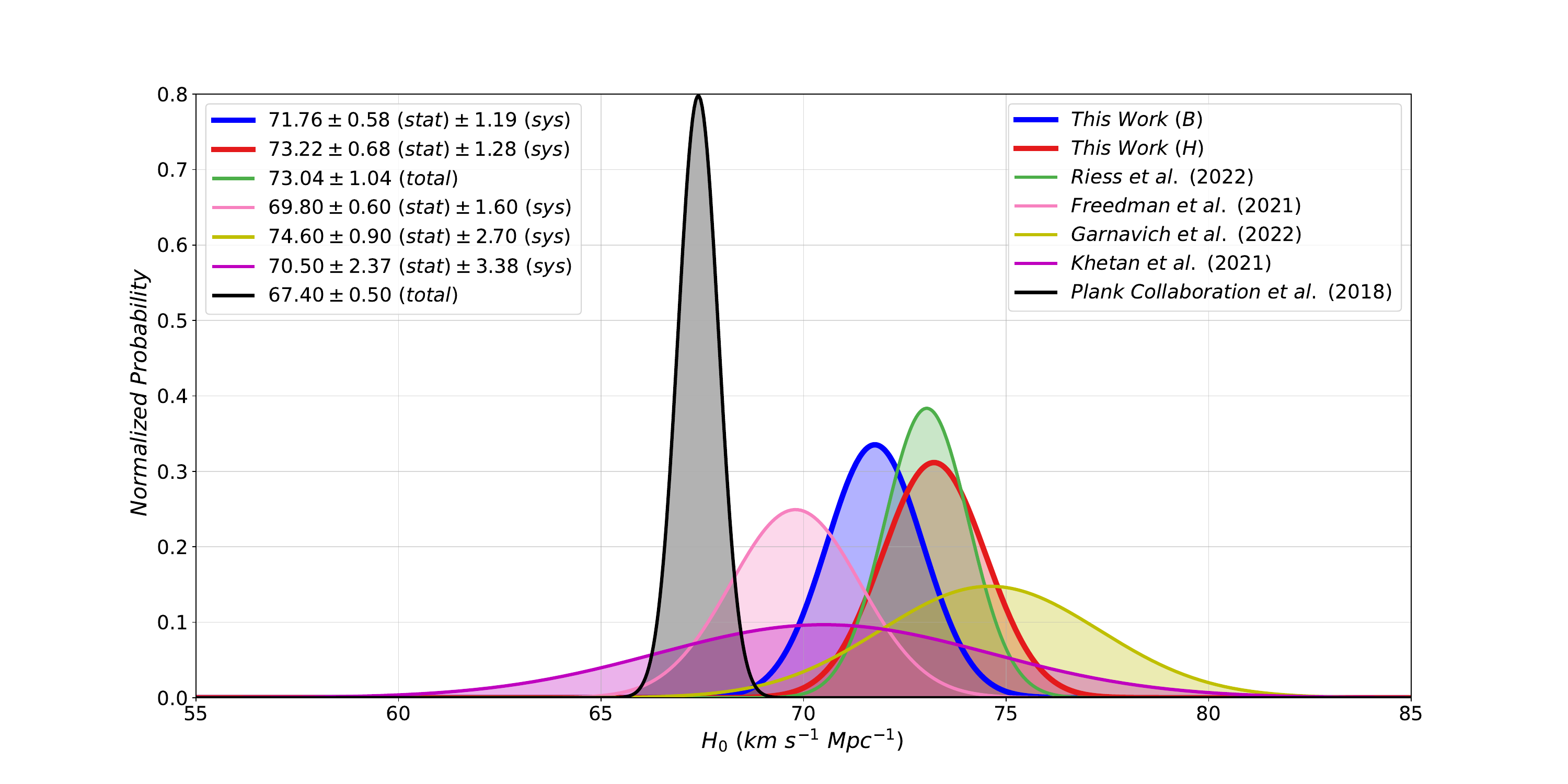}
    \caption{Normalized probability distributions of $H_0$. We show $H_0$ for $B$ and $H$ bands that we obtain from combining Cepheid, TRGB, and SBF distance calibrations. We also show most recent $H_0$ measurements from \cite{riess22}, \cite{freedman21}, \cite{garnavich22}, \cite{khetan21}, and \cite{planck18}. Combined statistical and systematic errors are used to construct these distributions. It is easy to see the effect of systematic uncertainties that are dominating in each case. We discuss this in detail in \S\ref{comcal}. Note that the two $H_0$ values from this work overlap with all SN~Ia based published $H_0$ values shown here.}
    \label{fig:pdfh0}
\end{figure*}

It is not immediately clear what could cause this level of disagreement between different calibration methods. Therefore, we will not give preference to one method over others. Our measurements of $H_0$ with added systematic uncertainties can be considered as representative values of $H_0$ with the current state of uncertainty in the SN~Ia distance scale. We visualize these numbers in Figure~\ref{fig:pdfh0}.

%This is an additional systematic corresponding to 5\% in $H_0$ that is comparable with the statistical error. This systematic increases in $H$-band and exceeds the statistical error.

\begin{table*}
    \centering
    \caption{Absolute magnitude ($M$) of various calibrating SNe~I as discussed in \S \ref{comcal}. We convert error in the absolute magnitude ($\sigma_{P0}$) of distant SNe~Ia to corresponding $\sigma_{H_0}$. The values of $\sigma_{H_0}$ represent the systematic uncertainty in $H_0$ that originates from the variation in the SN~Ia luminosity among various distance calibrators.}
    \begin{tabular}{lcccc}
    \hline
    \hline
     &&Values&\\
    \cmidrule{2-5}
    Parameter& $M_B$  &$\sigma_{H_0} (B)$& $M_H$ &$\sigma_{H_0} (H)$\\
    &(mag)&$\rm \ (km \ s^{-1} \ Mpc^{-1})$&(mag)&$\rm (\ km \ s^{-1} \ Mpc^{-1}$)\\
    
    \hline 
      $P0_{dist}$   & -19.147~(038)&&-18.435~(040)& \\
      
       $P0_{Cepheid}$  & -19.138~(037) &&-18.400~(040)&\\
       $P0_{TRGB}$  & -19.165~(047)& 1.19 &-18.476~(050)&1.28\\
       $P0_{SBF}$ & -19.105~(036)&&-18.462~(044)&\\
       \hline 
       
    \end{tabular}
    
    \label{tab:allcal}
\end{table*}

\subsection{SN~Ia Hubble Residual and  Intrinsic Luminosity Scatter}\label{sec:hd}

\begin{figure*}[htbp]
    \centering
    \includegraphics[width=\textwidth]{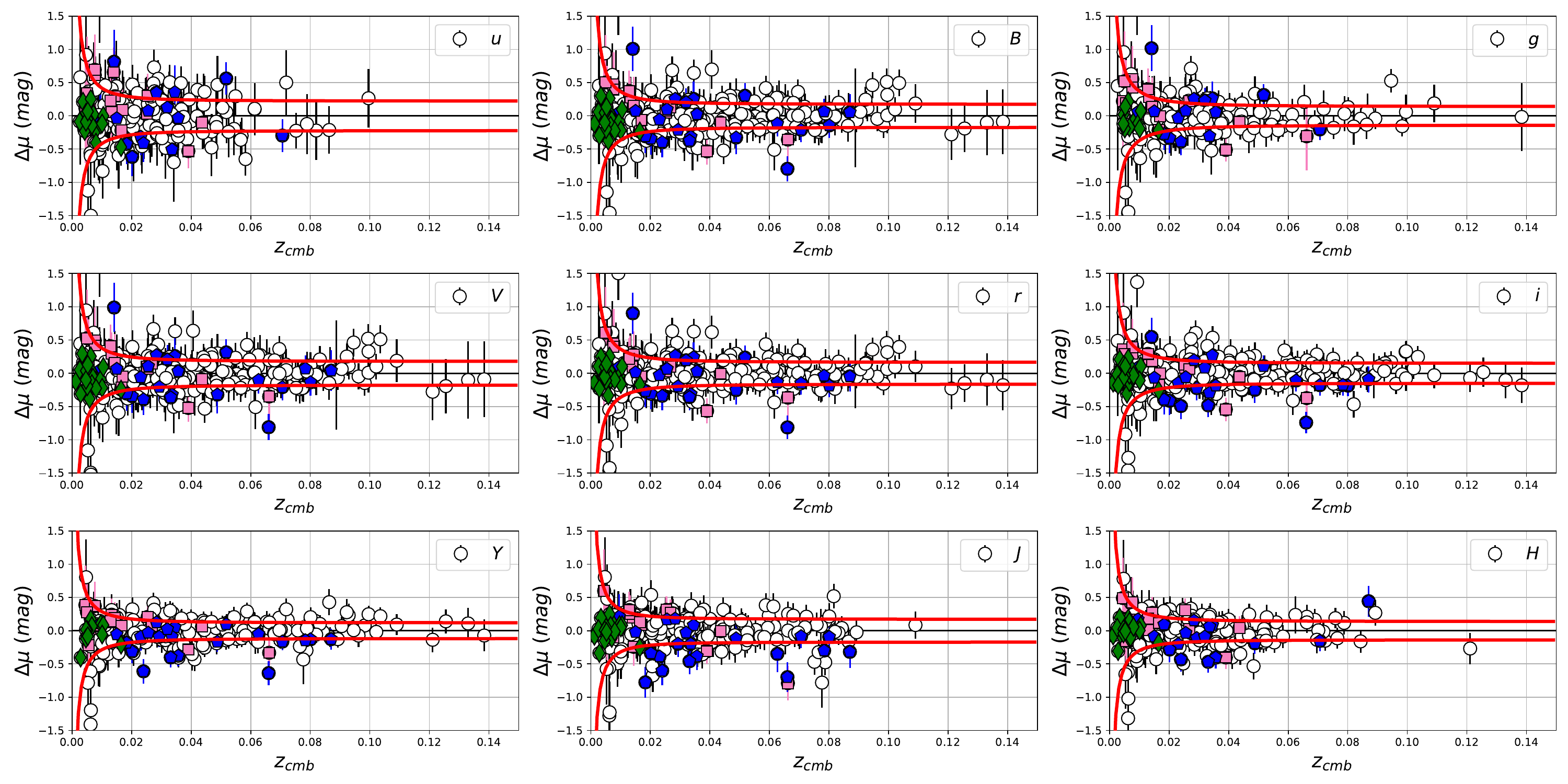}
    \caption{Hubble residuals in different bands with the Cepheid calibration. The red curves are combined errors due to peculiar velocity ($V_{pec}$) and intrinsic scatter ($\sigma_{int}$). Blue points are 91T-like, and pink squares are 91bg-like objects. Green diamonds are Cepheid calibrators.}
    \label{fig:hd}
\end{figure*}

In this section, we investigate the precision of SNe~Ia distances as a function of the observed photometric band by investigating the Hubble residuals. Hubble residuals ($\Delta \mu$) are the differences between the observed and the predicted distance moduli in the sense that a negative $\Delta \mu$ indicates an over-luminous SN~Ia compared to its expected brightness, after corrections and distance predicted by the Hubble-Lema{\^i}tre law. Hubble residuals when plotted against redshifts show how well the model fits the data. In Figure~\ref{fig:hd} we show Hubble residuals in $uBgVriYJH$ bands when Cepheid calibration is used in determining $H_0$.

\begin{figure}
    \centering
    \includegraphics[width=\columnwidth]{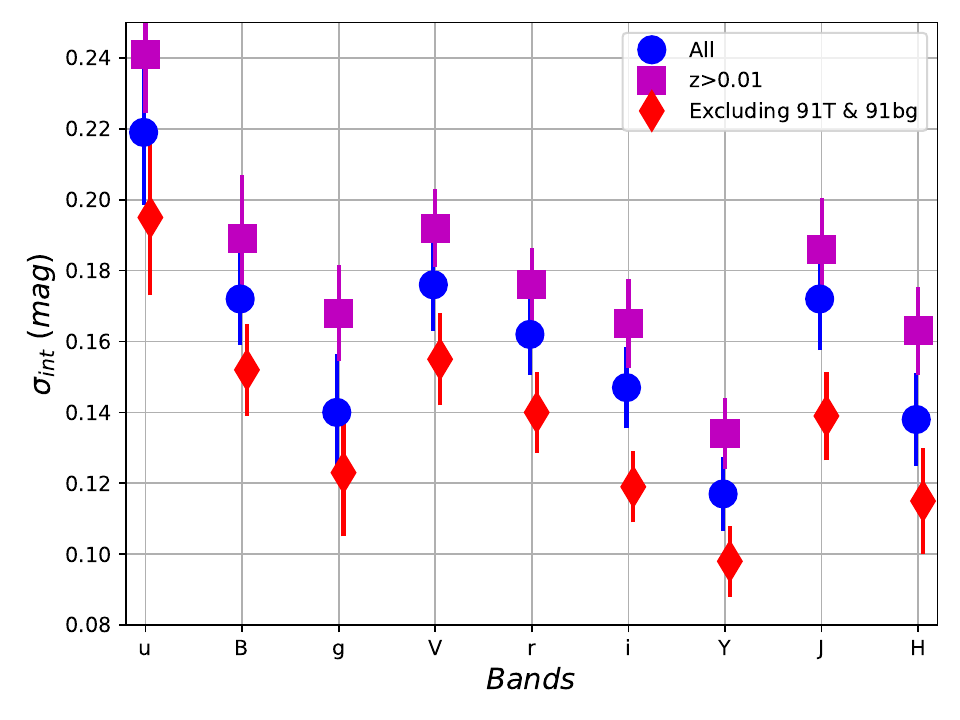}
    \caption{Variation of $\sigma_{int}$ as a function of wavelength from optical to near-infrared in Cepheid calibration. We notice that $\sigma_{int}$ decreases if 91T-like and 91bg-like SNe~Ia are excluded. Including only distant SNe~Ia ($z>0.01$) increases $\sigma_{int}$ (see \S \ref{sel}).}
    \label{fig:sigma}
\end{figure}

A parameter we are interested to study is the intrinsic luminosity scatter of SNe~Ia, $\sigma_{int}$, introduced earlier. This term describes how well we can standardize SN~Ia distances. It will be interesting to see how $\sigma_{int}$ varies between optical and near-infrared bands, and at which band it has the smallest value. Smaller values of $\sigma_{int}$ indicate better calibration. We present $\sigma_{int}$ for the whole sample in Table~\ref{tab:all}. We also show the same, but excluding 91T and 91bg-like SNe~Ia, in Table~\ref{tab:91}. Figure~\ref{fig:sigma} shows $\sigma_{int}$ as a function of wavelength for various cases in Cepheid calibration.

It is easy to see that $\sigma_{int}$ is reduced in all bands when excluding 91T and 91bg-like objects, but this reduction is not significant.  In both cases, we find that SNe~Ia show the smallest value of $\sigma_{int}$ in the $Y$-band. This can be visually inspected in Figure~\ref{fig:hd}, where we see the scatter in SN~Ia distances is the smallest in the $Y$-band. Red curves represent the combined uncertainty due to $V_{pec}$ and $\sigma_{int}$. The effect of $V_{pec}$ becomes insignificant for redshifts greater than $\sim 0.02$. 

It is interesting to note that, among the three near-infrared bands ($YJH$), $Y$ provides the best calibration, and $\sigma_{int}$ increases as one moves to longer wavelengths. While the number of available calibrators in $Y$-band is smaller (see Table~\ref{tab:sample}), \cite{phillips12} interprets this as being due to a combined effect of higher signal-to-noise and relative insensitivity to color effects (both reddening from dust and intrinsic color variations).
%We also note that in $Y$-band K-corrections have the smallest uncertainty (\citealt{boldt14}).  

%%%%%%%%%%%
\subsection{Effect of Various Cuts}\label{sel}

So far, we have studied $H_0$ and $\sigma_{int}$ both from our full sample and when excluding 91T and 91bg-like SNe~Ia. In this section, we apply various cuts and perform additional calculations for a number of specific cases. We exclude:
low-redshift objects ($z<0.01$), faster decliners ($s_{BV}<0.5$), redder objects ($B-V>0.5$  mag), and objects for which the earliest observation is later than five days post maximum ($t_0>5$ day). We also combine all these cuts. Values of $H_0$ and $\sigma_{int}$ using these cuts are shown in Table~\ref{tab:intcut} for Cepheid calibration, in Table~\ref{tab:intcuttrgb} for TRGB calibration, and in Table~\ref{tab:intcutsbf} for SBF calibration. In Table~\ref{tab:intcutall} we show the same combining all three calibrators in $B$ and $H$ bands.

It is interesting to note that (when comparing with Table~\ref{tab:all}) excluding low-redshift SNe~Ia increases (with low significance) $\sigma_{int}$ in all bands (see Figure~\ref{fig:sigma}). Note that in this case $V_{pec}$ decreases. It is because they both contribute to the denominator of the $\chi^2$. In all other cuts, we do not see appreciable changes in $\sigma_{int}$. When we combine all cuts, the $z>0.01$ cut dominates the change in $\sigma_{int}$. As we have seen before, $\sigma_{int}$ is the smallest in the $Y$ band, especially for $t_0<5$ days, where we find $\sigma_{int} = 0.11\pm 0.01$ mag. 

%\crb{Curious that $\sigma_{int}$ does this. The immediate questions are:  1) is the change significant? and 2) What happens to $V_{pec}$? Can you include uncertainties for $\sigma_{int}$ and the values (and uncertainties) for $V_{pec}$ in tables 8,9, and 10? I can imagine that if $V_{pec}$ goes down, $\sigma_{int}$ could go up, since they both contribute to the denominator in your $\chi^2$}

It has been suggested that SNe~Ia in the near-infrared could be better standard candles since the effect of dust is reduced. A few studies (e.g. \citealt{kevin04}, \citealt{barone12}, \citealt{arturo19}, \citealt{peterson23}, and Do et al. in preparation) present Hubble residual dispersion and $\sigma_{int}$ of SNe~Ia in near-infrared bands, but do not compare the same SNe~Ia with the optical bands. Also, these studies are made with either a much smaller sample or with a number of selection criteria. Our results do not show a monotonic decrease in $\sigma_{int}$ from optical to near-infrared, but we do find the smallest $\sigma_{int}$ in the $Y$-band, and the second-smallest in the $H$-band. 

We recall that in \S~\ref{h0val}, we found values of $H_0$ in $YJH$ bands to be higher in TRGB calibration and lower in SBF calibration. Here, we see that, except for $z>0.01$ cut, TRGB gives consistent values of $H_0$ across various bands. However, these cuts do not make SBF-based $H_0$ consistent between optical and near-infrared.

\begin{table*}
    \centering
    \caption{$H_0$ and all other nuisance parameters in Cepheid calibration when applying various cuts.}
    \begin{tabular}{l|c|c|c|c|c|c|c|c|c}
    \hline
    \hline
   Selection &  Band & $H_0$ & $\sigma_{int}$ & $V_{pec}$ & $P0$ & $P1$& $P2$ & $\alpha$ & $\beta$\\
    && $(\rm km \ s^{-1} \ Mpc^{-1})$ & (mag) & ($\rm km \ s^{-1}$) & (mag) & (mag) & (mag) & $\rm (mag/dex)$ & \\
    \hline 
   & $ u $ & 74.02 (1.13) & 0.25 (01) & 184 (122) & -18.705 (022) & -1.87 (14) & -2.13 (44) & -0.06 (01) & 4.38 (13) \\
& $ B $ & 72.60 (0.73) & 0.19 (01) & 189 (97) & -19.110 (015) & -1.22 (09) & -1.18 (28) & -0.02 (01) & 3.03 (07) \\
& $ g $ & 73.40 (0.97) & 0.17 (01) & 268 (88) & -19.140 (023) & -1.18 (10) & -1.19 (31) & -0.03 (01) & 2.61 (09) \\
& $ V $ & 72.75 (0.70) & 0.19 (01) & 222 (95) & -19.106 (014) & -1.25 (09) & -1.45 (27) & -0.02 (00) & 2.10 (07) \\
$z>0.01$& $ r $ & 71.75 (0.64) & 0.18 (01) & 252 (74) & -19.062 (014) & -1.11 (08) & -0.80 (25) & -0.03 (00) & 1.57 (07) \\
& $ i $ & 72.89 (0.62) & 0.17 (01) & 239 (73) & -18.452 (014) & -0.72 (07) & -0.43 (24) & -0.03 (00) & 1.11 (06) \\
& $ Y $ & 72.82 (0.78) & 0.13 (01) & 186 (76) & -18.469 (020) & -0.38 (07) & 0.65 (22) & -0.03 (00) & 0.53 (06) \\
& $ J $ & 74.95 (0.72) & 0.19 (01) & 202 (112) & -18.611 (015) & -0.68 (09) & 0.08 (29) & -0.03 (01) & 0.56 (08) \\
& $ H $ & 73.97 (0.73) & 0.16 (01) & 184 (98) & -18.377 (014) & -0.38 (09) & 0.44 (28) & -0.03 (01) & 0.27 (07) \\
\hline      
& $ u $ & 73.64 (1.15) & 0.22 (02) & 497 (66) & -18.698 (021) & -1.90 (13) & -2.57 (53) & -0.06 (01) & 4.30 (14) \\
& $ B $ & 72.48 (0.74) & 0.17 (01) & 457 (51) & -19.108 (015) & -1.24 (09) & -1.33 (36) & -0.02 (01) & 3.03 (08) \\
& $ g $ & 73.21 (0.99) & 0.14 (01) & 486 (50) & -19.139 (024) & -1.20 (10) & -1.45 (41) & -0.03 (01) & 2.58 (10) \\
& $ V $ & 72.59 (0.75) & 0.17 (01) & 472 (50) & -19.105 (015) & -1.28 (08) & -1.57 (34) & -0.02 (01) & 2.08 (08) \\
$s_{BV}>0.5$ & $ r $ & 71.60 (0.66) & 0.16 (01) & 461 (45) & -19.061 (015) & -1.13 (08) & -0.79 (35) & -0.03 (00) & 1.55 (07) \\
& $ i $ & 72.68 (0.65) & 0.15 (01) & 441 (44) & -18.453 (014) & -0.73 (08) & -0.50 (30) & -0.03 (00) & 1.09 (06) \\
& $ Y $ & 72.47 (0.82) & 0.12 (01) & 371 (41) & -18.464 (020) & -0.40 (07) & 0.37 (31) & -0.02 (00) & 0.47 (06) \\
& $ J $ & 74.28 (0.76) & 0.16 (01) & 391 (51) & -18.605 (015) & -0.74 (09) & -0.83 (37) & -0.02 (01) & 0.46 (08) \\
& $ H $ & 73.66 (0.77) & 0.13 (01) & 382 (50) & -18.374 (015) & -0.46 (10) & -0.17 (46) & -0.02 (01) & 0.22 (07) \\
\hline 
& $ u $ & 73.96 (1.09) & 0.23 (02) & 432 (64) & -18.699 (021) & -1.92 (13) & -2.41 (57) & -0.06 (01) & 4.24 (16) \\
& $ B $ & 72.60 (0.72) & 0.18 (01) & 372 (50) & -19.107 (015) & -1.23 (09) & -1.01 (37) & -0.01 (01) & 2.86 (11) \\
& $ g $ & 73.30 (0.94) & 0.15 (01) & 390 (46) & -19.140 (023) & -1.20 (09) & -0.98 (41) & -0.02 (01) & 2.36 (14) \\
& $ V $ & 72.75 (0.72) & 0.18 (01) & 369 (46) & -19.103 (015) & -1.25 (08) & -1.17 (36) & -0.02 (01) & 1.92 (10) \\
$B-V<0.5$& $ r $ & 71.90 (0.64) & 0.17 (01) & 365 (43) & -19.057 (014) & -1.10 (08) & -0.34 (33) & -0.02 (00) & 1.33 (09) \\
(mag)& $ i $ & 72.84 (0.66) & 0.15 (01) & 367 (44) & -18.448 (014) & -0.72 (07) & -0.32 (31) & -0.02 (00) & 0.96 (10) \\
& $ Y $ & 72.39 (0.84) & 0.12 (01) & 343 (41) & -18.463 (022) & -0.41 (07) & 0.20 (31) & -0.02 (00) & 0.46 (10) \\
& $ J $ & 74.31 (0.81) & 0.16 (01) & 368 (53) & -18.607 (016) & -0.76 (09) & -1.03 (38) & -0.03 (01) & 0.57 (12) \\
& $ H $ & 73.60 (0.76) & 0.13 (01) & 367 (49) & -18.375 (015) & -0.47 (09) & -0.46 (41) & -0.03 (01) & 0.29 (11) \\
\hline
& $ u $ & 73.70 (1.15) & 0.22 (02) & 494 (67) & -18.708 (021) & -1.88 (14) & -2.24 (44) & -0.06 (01) & 4.42 (13) \\
& $ B $ & 72.27 (0.73) & 0.17 (01) & 443 (52) & -19.112 (015) & -1.25 (10) & -1.27 (30) & -0.02 (01) & 3.05 (08) \\
& $ g $ & 73.12 (0.98) & 0.14 (01) & 471 (48) & -19.145 (023) & -1.20 (11) & -1.25 (34) & -0.03 (01) & 2.63 (10) \\
& $ V $ & 72.49 (0.71) & 0.18 (01) & 459 (48) & -19.107 (015) & -1.29 (09) & -1.47 (29) & -0.02 (01) & 2.10 (08) \\
$t_0<5$& $ r $ & 71.52 (0.66) & 0.17 (01) & 444 (47) & -19.061 (014) & -1.14 (08) & -0.81 (28) & -0.03 (00) & 1.57 (07) \\
(day)& $ i $ & 72.72 (0.66) & 0.15 (01) & 435 (45) & -18.455 (014) & -0.73 (08) & -0.41 (24) & -0.03 (01) & 1.11 (07) \\
& $ Y $ & 72.67 (0.85) & 0.11 (01) & 379 (46) & -18.466 (023) & -0.37 (08) & 0.77 (25) & -0.02 (00) & 0.51 (08) \\
& $ J $ & 74.72 (0.82) & 0.17 (01) & 371 (54) & -18.609 (016) & -0.70 (10) & 0.26 (33) & -0.02 (01) & 0.48 (08) \\
& $ H $ & 74.17 (0.74) & 0.13 (01) & 374 (50) & -18.374 (015) & -0.38 (10) & 0.59 (33) & -0.02 (01) & 0.19 (08) \\
\hline 
& $ u $ & 73.49 (1.19) & 0.24 (02) & 191 (122) & -18.700 (022) & -1.92 (13) & -2.59 (55) & -0.06 (01) & 4.35 (17) \\
& $ B $ & 72.55 (0.76) & 0.19 (01) & 147 (97) & -19.106 (016) & -1.25 (08) & -1.16 (39) & -0.01 (01) & 2.93 (11) \\
& $ g $ & 73.04 (0.94) & 0.17 (01) & 213 (89) & -19.140 (022) & -1.21 (10) & -1.15 (42) & -0.03 (01) & 2.46 (14) \\
& $ V $ & 72.62 (0.73) & 0.20 (01) & 148 (93) & -19.102 (015) & -1.27 (09) & -1.33 (36) & -0.01 (01) & 1.96 (11) \\
All cuts& $ r $ & 71.77 (0.65) & 0.18 (01) & 194 (90) & -19.058 (014) & -1.11 (08) & -0.48 (36) & -0.02 (00) & 1.40 (10) \\
& $ i $ & 72.73 (0.66) & 0.17 (01) & 207 (92) & -18.452 (014) & -0.75 (08) & -0.40 (33) & -0.03 (00) & 1.01 (09) \\
& $ Y $ & 72.52 (0.83) & 0.14 (01) & 186 (94) & -18.464 (022) & -0.39 (08) & 0.36 (33) & -0.02 (00) & 0.51 (10) \\
& $ J $ & 73.76 (0.85) & 0.18 (01) & 195 (116) & -18.609 (016) & -0.76 (10) & -0.95 (48) & -0.03 (01) & 0.63 (12) \\
& $ H $ & 73.22 (0.75) & 0.15 (01) & 193 (101) & -18.373 (015) & -0.45 (09) & -0.58 (41) & -0.02 (01) & 0.21 (11) \\
\hline 

    \end{tabular}
    
    \label{tab:intcut}
\end{table*}

% commenting out other tables

\begin{table*}
    \centering
    \caption{$H_0$ and all other nuisance parameters in TRGB calibration when applying various cuts.}
    \begin{tabular}{l|c|c|c|c|c|c|c|c|c}
    \hline
    \hline
   Selection &  Band & $H_0$ & $\sigma_{int}$ & $V_{pec}$ & $P0$ & $P1$& $P2$ & $\alpha$ & $\beta$\\
    && $(\rm km \ s^{-1} \ Mpc^{-1})$ & (mag) & ($\rm km \ s^{-1}$) & (mag) & (mag) & (mag) & $\rm (mag/dex)$ & \\
    \hline 

& $ u $ & 69.00 (1.00) & 0.25 (01) & 193 (120) & -18.863 (025) & -1.46 (14) & -0.31 (44) & -0.07 (02) & 3.89 (11) \\
& $ B $ & 70.21 (0.67) & 0.19 (01) & 200 (94) & -19.178 (016) & -1.11 (09) & -0.67 (25) & -0.00 (01) & 2.87 (06) \\
& $ g $ & 71.09 (1.25) & 0.17 (01) & 251 (89) & -19.209 (033) & -1.06 (10) & -0.66 (32) & -0.01 (01) & 2.51 (09) \\
& $ V $ & 70.12 (0.65) & 0.20 (01) & 225 (100) & -19.178 (015) & -1.19 (08) & -0.98 (23) & -0.01 (01) & 1.90 (06) \\
$z>0.01$& $ r $ & 69.31 (0.61) & 0.18 (01) & 242 (79) & -19.137 (015) & -0.94 (09) & -0.10 (25) & -0.01 (01) & 1.43 (05) \\
& $ i $ & 69.77 (0.58) & 0.17 (01) & 237 (79) & -18.542 (014) & -0.56 (07) & 0.07 (21) & -0.01 (00) & 0.98 (05) \\
& $ Y $ & 72.13 (1.04) & 0.13 (01) & 175 (73) & -18.491 (029) & -0.32 (07) & 0.76 (21) & -0.03 (01) & 0.54 (06) \\
& $ J $ & 72.29 (0.75) & 0.19 (01) & 149 (103) & -18.688 (016) & -0.56 (10) & 0.69 (27) & -0.04 (01) & 0.34 (06) \\
& $ H $ & 70.80 (0.71) & 0.17 (01) & 153 (95) & -18.482 (015) & -0.25 (09) & 0.89 (27) & -0.05 (01) & 0.21 (06) \\
\hline
& $ u $ & 68.82 (1.03) & 0.23 (02) & 463 (63) & -18.862 (027) & -1.53 (15) & -0.55 (53) & -0.07 (02) & 3.87 (11) \\
& $ B $ & 70.22 (0.69) & 0.17 (01) & 442 (49) & -19.176 (016) & -1.12 (09) & -0.69 (31) & -0.00 (01) & 2.85 (07) \\
& $ g $ & 70.94 (1.23) & 0.14 (01) & 473 (50) & -19.217 (035) & -1.08 (11) & -0.59 (40) & -0.01 (01) & 2.51 (09) \\
& $ V $ & 69.92 (0.62) & 0.18 (01) & 462 (51) & -19.177 (015) & -1.21 (09) & -1.11 (27) & -0.01 (01) & 1.87 (06) \\
$s_{B-V}>0.5$& $ r $ & 69.18 (0.64) & 0.17 (01) & 452 (46) & -19.142 (015) & -0.94 (08) & 0.04 (28) & -0.01 (01) & 1.45 (06) \\
& $ i $ & 69.70 (0.59) & 0.15 (01) & 439 (45) & -18.541 (014) & -0.56 (07) & 0.03 (24) & -0.01 (00) & 0.97 (05) \\
& $ Y $ & 72.03 (0.99) & 0.12 (01) & 364 (40) & -18.482 (03) & -0.33 (07) & 0.57 (28) & -0.02 (01) & 0.48 (06) \\
& $ J $ & 72.24 (0.74) & 0.17 (01) & 345 (49) & -18.672 (017) & -0.65 (10) & -0.01 (33) & -0.04 (01) & 0.24 (06) \\
& $ H $ & 70.80 (0.71) & 0.14 (01) & 356 (48) & -18.480 (015) & -0.29 (11) & 0.73 (37) & -0.05 (01) & 0.17 (06) \\
\hline 
& $ u $ & 69.25 (1.02) & 0.23 (01) & 389 (57) & -18.844 (025) & -1.51 (14) & -0.38 (47) & -0.06 (02) & 3.68 (12) \\
& $ B $ & 70.52 (0.66) & 0.18 (01) & 359 (46) & -19.167 (016) & -1.11 (08) & -0.51 (28) & 0.00 (01) & 2.68 (08) \\
& $ g $ & 70.17 (1.17) & 0.15 (01) & 374 (46) & -19.236 (033) & -1.04 (10) & -0.10 (37) & -0.01 (01) & 2.17 (13) \\
& $ V $ & 70.29 (0.64) & 0.18 (01) & 358 (46) & -19.169 (015) & -1.19 (08) & -0.87 (24) & -0.00 (01) & 1.68 (07) \\
$B-V<0.5$ & $ r $ & 69.61 (0.61) & 0.17 (01) & 350 (42) & -19.128 (014) & -0.92 (08) & 0.20 (25) & 0.00 (00) & 1.25 (07) \\
(mag)& $ i $ & 69.94 (0.62) & 0.15 (01) & 361 (45) & -18.530 (014) & -0.56 (07) & 0.14 (22) & -0.00 (00) & 0.81 (07) \\
& $ Y $ & 72.10 (1.07) & 0.12 (01) & 332 (41) & -18.478 (029) & -0.33 (07) & 0.47 (27) & -0.02 (01) & 0.49 (10) \\
& $ J $ & 72.32 (0.71) & 0.18 (01) & 320 (51) & -18.671 (016) & -0.65 (10) & 0.04 (33) & -0.04 (01) & 0.18 (08) \\
& $ H $ & 70.82 (0.71) & 0.15 (01) & 332 (45) & -18.479 (016) & -0.31 (12) & 0.59 (42) & -0.05 (01) & 0.15 (08) \\
\hline 

& $ u $ & 68.60 (1.01) & 0.23 (02) & 450 (65) & -18.867 (024) & -1.51 (15) & -0.40 (45) & -0.08 (02) & 3.93 (11) \\
& $ B $ & 70.09 (0.68) & 0.18 (01) & 421 (51) & -19.178 (015) & -1.13 (09) & -0.67 (26) & -0.01 (01) & 2.87 (06) \\
& $ g $ & 70.89 (1.22) & 0.15 (01) & 453 (48) & -19.212 (032) & -1.08 (11) & -0.62 (34) & -0.02 (01) & 2.51 (10) \\
& $ V $ & 69.84 (0.63) & 0.18 (01) & 450 (51) & -19.179 (015) & -1.23 (09) & -1.00 (24) & -0.01 (01) & 1.88 (06) \\
$t_0<5$ & $ r $ & 69.14 (0.64) & 0.17 (01) & 438 (46) & -19.138 (014) & -0.96 (09) & -0.08 (27) & -0.01 (01) & 1.43 (06) \\
(day)& $ i $ & 69.69 (0.59) & 0.15 (01) & 435 (48) & -18.542 (014) & -0.56 (08) & 0.14 (22) & -0.01 (01) & 0.98 (06) \\
& $ Y $ & 71.46 (1.03) & 0.11 (01) & 369 (44) & -18.502 (028) & -0.28 (08) & 0.89 (24) & -0.02 (01) & 0.54 (07) \\
& $ J $ & 71.96 (0.79) & 0.17 (01) & 341 (52) & -18.687 (017) & -0.55 (11) & 0.87 (30) & -0.03 (01) & 0.29 (07) \\
& $ H $ & 70.92 (0.74) & 0.14 (01) & 354 (47) & -18.480 (016) & -0.21 (10) & 1.10 (29) & -0.05 (01) & 0.14 (06) \\
\hline
& $ u $ & 68.69 (1.09) & 0.23 (02) & 179 (125) & -18.856 (031) & -1.26 (21) & 0.71 (91) & -0.03 (02) & 3.69 (15) \\
& $ B $ & 70.32 (0.68) & 0.16 (01) & 225 (93) & -19.173 (016) & -1.16 (10) & -0.26 (44) & 0.00 (01) & 2.66 (1) \\
& $ g $ & 68.89 (1.19) & 0.14 (01) & 211 (74) & -19.290 (036) & -0.98 (11) & 1.32 (46) & 0.00 (01) & 1.60 (25) \\
& $ V $ & 70.03 (0.68) & 0.17 (01) & 222 (89) & -19.169 (016) & -1.27 (08) & -0.85 (27) & 0.00 (01) & 1.62 (09) \\
All cuts& $ r $ & 69.14 (0.65) & 0.15 (01) & 267 (71) & -19.140 (015) & -1.00 (08) & 0.70 (30) & 0.00 (00) & 1.25 (09) \\
& $ i $ & 69.69 (0.65) & 0.14 (01) & 293 (85) & -18.534 (016) & -0.59 (08) & 0.43 (26) & 0.00 (00) & 0.78 (09) \\
& $ Y $ & 71.21 (1.14) & 0.13 (01) & 166 (103) & -18.497 (034) & -0.31 (08) & 0.64 (36) & -0.02 (01) & 0.65 (22) \\
& $ J $ & 72.06 (0.84) & 0.18 (01) & 116 (93) & -18.672 (017) & -0.57 (13) & 0.53 (47) & -0.03 (01) & 0.13 (09) \\
& $ H $ & 70.99 (0.85) & 0.15 (02) & 261 (130) & -18.476 (017) & -0.27 (12) & 0.84 (44) & -0.05 (01) & 0.10 (07) \\
\hline 
 \end{tabular}
    
    \label{tab:intcuttrgb}
\end{table*}

\begin{table*}
    \centering
    \caption{$H_0$ and all other nuisance parameters in combined SBF calibration when applying various cuts.}
    \begin{tabular}{l|c|c|c|c|c|c|c|c|c}
    \hline
    \hline
   Selection &  Band & $H_0$ & $\sigma_{int}$ & $V_{pec}$ & $P0$ & $P1$& $P2$ & $\alpha$ & $\beta$\\
    && $(\rm km \ s^{-1} \ Mpc^{-1})$ & (mag) & ($\rm km \ s^{-1}$) & (mag) & (mag) & (mag) & $\rm (mag/dex)$ & \\
    \hline 
& $ u $ & 69.71 (1.35) & 0.25 (01) & 180 (123) & -18.839 (040) & -1.36 (16) & -0.26 (47) & -0.01 (02) & 4.07 (14) \\
& $ B $ & 72.73 (0.82) & 0.19 (01) & 215 (101) & -19.106 (020) & -0.98 (10) & -0.60 (35) & -0.00 (01) & 2.93 (08) \\
& $ g $ & 73.33 (1.38) & 0.17 (01) & 305 (85) & -19.136 (040) & -0.92 (11) & -0.62 (33) & -0.00 (01) & 2.54 (09) \\
& $ V $ & 72.55 (0.79) & 0.19 (01) & 265 (92) & -19.098 (020) & -1.19 (09) & -1.42 (26) & -0.01 (01) & 2.01 (07) \\
$z>0.01$& $ r $ & 72.09 (0.75) & 0.18 (01) & 262 (81) & -19.057 (02) & -0.87 (08) & -0.10 (28) & -0.01 (01) & 1.49 (07) \\
& $ i $ & 72.39 (0.75) & 0.16 (01) & 255 (77) & -18.468 (019) & -0.60 (07) & -0.05 (23) & -0.02 (01) & 1.04 (06) \\
& $ Y $ & 69.66 (1.21) & 0.13 (01) & 181 (77) & -18.562 (038) & -0.35 (07) & 0.68 (21) & -0.03 (01) & 0.53 (06) \\
& $ J $ & 70.10 (1.05) & 0.19 (01) & 180 (112) & -18.761 (028) & -0.63 (11) & 0.39 (31) & -0.04 (01) & 0.49 (08) \\
& $ H $ & 68.31 (0.99) & 0.16 (01) & 173 (101) & -18.563 (026) & -0.34 (11) & 0.94 (3) & -0.03 (01) & 0.23 (08) \\
\hline

& $ u $ & 69.18 (1.47) & 0.22 (02) & 489 (67) & -18.861 (045) & -1.39 (17) & -0.16 (50) & -0.02 (02) & 4.07 (14) \\
& $ B $ & 72.64 (0.89) & 0.17 (01) & 464 (53) & -19.102 (023) & -1.00 (11) & -0.73 (39) & -0.00 (01) & 2.92 (08) \\
& $ g $ & 73.62 (1.55) & 0.14 (01) & 498 (49) & -19.124 (046) & -0.95 (12) & -0.80 (39) & -0.00 (01) & 2.53 (09) \\
& $ V $ & 72.63 (0.82) & 0.17 (01) & 479 (51) & -19.089 (021) & -1.22 (09) & -1.60 (26) & -0.01 (01) & 1.98 (08) \\
$s_{BV}>0.5$& $ r $ & 71.84 (0.80) & 0.16 (01) & 469 (47) & -19.064 (021) & -0.85 (09) & 0.05 (34) & -0.01 (01) & 1.49 (07) \\
& $ i $ & 72.35 (0.76) & 0.15 (01) & 449 (46) & -18.466 (021) & -0.60 (08) & -0.07 (26) & -0.02 (01) & 1.04 (06) \\
& $ Y $ & 70.66 (1.40) & 0.12 (01) & 381 (41) & -18.521 (045) & -0.36 (08) & 0.40 (29) & -0.02 (01) & 0.48 (06) \\
& $ J $ & 70.50 (1.06) & 0.17 (01) & 383 (49) & -18.739 (03) & -0.68 (12) & 0.00 (35) & -0.03 (01) & 0.41 (08) \\
& $ H $ & 68.04 (0.99) & 0.14 (01) & 374 (48) & -18.575 (029) & -0.34 (11) & 1.14 (30) & -0.03 (01) & 0.20 (08) \\
\hline
& $ u $ & 68.77 (1.58) & 0.23 (01) & 410 (60) & -18.869 (047) & -1.36 (16) & 0.26 (52) & -0.01 (02) & 3.80 (18) \\
& $ B $ & 72.07 (0.86) & 0.18 (01) & 390 (50) & -19.134 (025) & -0.79 (12) & 0.78 (55) & 0.01 (01) & 2.45 (14) \\
& $ g $ & 72.76 (1.47) & 0.15 (01) & 396 (47) & -19.150 (045) & -0.92 (11) & -0.23 (39) & 0.01 (01) & 2.24 (13) \\
& $ V $ & 71.72 (0.84) & 0.19 (01) & 392 (53) & -19.155 (022) & -0.53 (08) & 1.92 (25) & 0.01 (01) & 1.19 (11) \\
$B-V<0.5$  &  $ r $ & 71.46 (0.78) & 0.17 (01) & 366 (44) & -19.076 (021) & -0.82 (08) & 0.63 (27) & -0.00 (01) & 1.14 (10) \\
(mag)& $ i $ & 72.06 (0.75) & 0.15 (01) & 378 (46) & -18.474 (020) & -0.59 (07) & 0.23 (25) & -0.01 (01) & 0.84 (08) \\
& $ Y $ & 71.05 (1.47) & 0.12 (01) & 354 (43) & -18.505 (047) & -0.35 (07) & 0.28 (31) & -0.02 (00) & 0.48 (10) \\
& $ J $ & 70.88 (1.13) & 0.17 (01) & 360 (52) & -18.733 (032) & -0.70 (11) & -0.16 (39) & -0.04 (01) & 0.51 (13) \\
& $ H $ & 68.16 (1.00) & 0.14 (01) & 357 (49) & -18.573 (029) & -0.36 (11) & 1.02 (36) & -0.04 (01) & 0.25 (13) \\
\hline
& $ u $ & 69.38 (1.44) & 0.22 (02) & 476 (67) & -18.847 (043) & -1.40 (17) & -0.40 (47) & -0.02 (02) & 4.14 (14) \\
& $ B $ & 72.43 (0.85) & 0.17 (01) & 449 (50) & -19.109 (022) & -1.01 (11) & -0.66 (39) & -0.00 (01) & 2.93 (09) \\
& $ g $ & 73.11 (1.40) & 0.14 (01) & 480 (50) & -19.139 (041) & -0.94 (12) & -0.66 (36) & -0.01 (01) & 2.55 (10) \\
& $ V $ & 72.22 (0.84) & 0.17 (01) & 471 (53) & -19.101 (021) & -1.22 (09) & -1.49 (25) & -0.01 (01) & 2.02 (08) \\
$t_0<5$ & $ r $ & 71.77 (0.77) & 0.17 (01) & 448 (48) & -19.060 (020) & -0.88 (09) & -0.05 (32) & -0.01 (01) & 1.48 (08) \\
(day)& $ i $ & 72.12 (0.78) & 0.15 (01) & 441 (47) & -18.472 (021) & -0.60 (08) & 0.01 (25) & -0.02 (01) & 1.04 (07) \\
& $ Y $ & 69.50 (1.33) & 0.11 (01) & 388 (43) & -18.560 (041) & -0.31 (09) & 0.77 (26) & -0.02 (01) & 0.53 (08) \\
& $ J $ & 69.11 (1.08) & 0.17 (01) & 373 (55) & -18.781 (029) & -0.63 (12) & 0.67 (32) & -0.02 (01) & 0.41 (09) \\
& $ H $ & 68.10 (0.96) & 0.14 (01) & 369 (51) & -18.575 (027) & -0.32 (12) & 1.24 (27) & -0.03 (01) & 0.15 (07) \\
\hline
& $ u $ & 68.85 (1.51) & 0.25 (01) & 167 (117) & -18.861 (047) & -1.38 (17) & -0.00 (54) & -0.01 (02) & 3.92 (19) \\
& $ B $ & 72.15 (0.88) & 0.20 (01) & 167 (99) & -19.125 (024) & -0.86 (13) & 0.30 (59) & 0.01 (01) & 2.56 (14) \\
& $ g $ & 72.67 (1.49) & 0.17 (01) & 223 (94) & -19.147 (045) & -0.96 (11) & -0.45 (40) & -0.00 (01) & 2.33 (14) \\
& $ V $ & 72.21 (0.84) & 0.19 (01) & 173 (97) & -19.101 (023) & -1.18 (10) & -1.21 (33) & -0.00 (01) & 1.76 (12) \\
All cuts& $ r $ & 71.35 (0.77) & 0.18 (01) & 209 (92) & -19.078 (021) & -0.83 (08) & 0.56 (30) & -0.00 (01) & 1.16 (10) \\
& $ i $ & 71.93 (0.78) & 0.17 (01) & 217 (95) & -18.477 (022) & -0.61 (07) & 0.18 (27) & -0.01 (01) & 0.88 (1) \\
& $ Y $ & 70.80 (1.46) & 0.14 (01) & 181 (96) & -18.512 (046) & -0.33 (08) & 0.34 (33) & -0.02 (01) & 0.52 (11) \\
& $ J $ & 69.86 (1.19) & 0.19 (01) & 179 (113) & -18.748 (034) & -0.69 (13) & 0.02 (44) & -0.03 (01) & 0.54 (15) \\
& $ H $ & 67.90 (1.01) & 0.16 (01) & 169 (108) & -18.573 (028) & -0.33 (12) & 1.13 (35) & -0.03 (01) & 0.21 (12) \\

\hline 
 \end{tabular}
    
    \label{tab:intcutsbf}
\end{table*}

\begin{table*}
    \centering
    \caption{$H_0$ and all other nuisance parameters in combined calibration when applying various cuts.}
    \begin{tabular}{l|c|c|c|c|c|c|c|c|c}
    \hline
    \hline
   Selection &  Band & $H_0$ & $\sigma_{int}$ & $V_{pec}$ & $P0$ & $P1$& $P2$ & $\alpha$ & $\beta$\\
    && $(\rm km \ s^{-1} \ Mpc^{-1})$ & (mag) & ($\rm km \ s^{-1}$) & (mag) & (mag) & (mag) & $\rm (mag/dex)$ & \\
    \hline 
$z>0.01$& $ B $ & 71.75 (0.64) & 0.19 (01) & 276 (110) & -19.139 (011) & -1.47 (08) & -1.71 (24) & -0.02 (00) & 2.90 (06) \\
& $ H $ & 73.14 (0.64) & 0.17 (01) & 311 (88) & -18.452 (011) & -0.19 (09) & 1.06 (27) & -0.07 (00) & 0.25 (06) \\

\hline
$s_{BV}>0.5$ & $ B $ & 71.73 (0.59) & 0.17 (01) & 470 (56) & -19.133 (011) & -1.52 (09) & -1.99 (28) & -0.02 (00) & 2.88 (06) \\
& $ H $ & 73.31 (0.67) & 0.15 (01) & 396 (55) & -18.448 (012) & -0.18 (09) & 1.10 (29) & -0.07 (00) & 0.24 (06) \\

\hline
$(B-V)<0.5$ & $ B $ & 71.94 (0.58) & 0.17 (01) & 395 (52) & -19.131 (01) & -1.46 (09) & -1.68 (28) & -0.02 (00) & 2.75 (07) \\
& $ H $ & 73.32 (0.62) & 0.15 (01) & 378 (59) & -18.448 (011) & -0.19 (09) & 1.08 (30) & -0.07 (00) & 0.21 (08) \\

\hline
$t_0<5$ & $ B $ & 71.58 (0.58) & 0.17 (01) & 478 (55) & -19.137 (010) & -1.50 (09) & -1.78 (28) & -0.02 (00) & 2.91 (06) \\
& $ H $ & 73.38 (0.74) & 0.16 (02) & 415 (57) & -18.446 (010) & -0.12 (09) & 1.30 (27) & -0.07 (00) & 0.22 (06) \\

\hline
All cuts & $ B $ & 71.79 (0.59) & 0.19 (01) & 240 (107) & -19.133 (011) & -1.49 (09) & -1.78 (29) & -0.02 (00) & 2.77 (07) \\
& $ H $ & 73.35 (0.75) & 0.16 (02) & 396 (88) & -18.447 (011) & -0.14 (10) & 1.16 (32) & -0.07 (00) & 0.22 (08) \\
\hline 
 \end{tabular}
    
    \label{tab:intcutall}
\end{table*}

Exploring the values of $H_0$ by applying various cuts, we can investigate the scatter that is introduced in $H_0$ measurements due to SNe~Ia sample selection. To compute this scatter, we gather all $H_0$ measurements in a given band for a particular calibration, and calculate the standard deviation. We show these scatters in Table~\ref{tab:cut} for $B$ and $H$ band.

We find comparable scatter between $B$ and $H$ bands in the  Cepheid and SBF calibrations. In the TRGB calibration, we find larger scatter in $H$ band compared to $B$ band. Also, $H_0$ is the least sensitive to SNe~Ia sample selection in the TRGB calibration in the optical ($0.14\ \rm km \ s^{-1} \ Mpc^{-1}$), and most sensitive in the combined calibration ($0.60\ \rm km \ s^{-1} \ Mpc^{-1}$) followed by the Cepheid calibration ($0.46\ \rm km \ s^{-1} \ Mpc^{-1}$) in the near-infrared.   Finally, we investigate inter-band scatter for various criteria. They are shown in Table~\ref{tab:bandcut}. 

%In this case, scatters in $H_0$ range between $1\sim 2 \ \rm km \ s^{-1} \ Mpc^{-1}$. We consider these scatters ad additional systematic uncertainties and add them in quadrature with what we obtain in \S~\ref{comcal}.

\begin{table}[]
    \centering
    \caption{Scatter in $H_0$ from various sample cuts as seen in Tables~\ref{tab:intcut}, \ref{tab:intcuttrgb}, and \ref{tab:intcutsbf}.}
    \begin{tabular}{lccc}
    \hline
    \hline
    Calibration & &Scatter in $H_0$ $(\rm km \ s^{-1} \ Mpc^{-1})$\\
    \cmidrule{2-4}
    && $B$ &$H$\\
    \hline
         Cepheid && 0.35 & 0.46 \\
         TRGB && 0.14&0.33  \\
         SBF && 0.22&0.20\\
         \hline
         Combined && 0.17&0.60 \\
         \hline
    \end{tabular}
    
    \label{tab:cut}
\end{table}

\begin{table*}[]
    \centering
    \caption{Inter-band scatter of $H_0$ for various cuts.}
    \begin{tabular}{lccccccc}
    \hline
    \hline
     &&&&Scatter in $H_0$ $(\rm km \ s^{-1} \ Mpc^{-1})$&&&\\
    \cmidrule{2-8}
    Calibration&Full & No 91T/91bg & $z>0.01$ & $s_{BV}>0.5$ & $B-V<0.5$ mag & $t_0<5$ day& All cuts  \\
    \cmidrule{1-8}
    Cepheid & 1.10 & 1.07 & 1.01 & 1.01 & 0.93 & 1.11 & 0.89\\
    TRGB & 1.24 & 1.19 & 1.20 & 1.21 & 1.09 & 1.20 & 1.01\\
    SBF & 1.42 & 2.07 & 2.08 & 2.04 & 2.03 & 2.19 & 2.11\\
    \hline 
    \end{tabular}
    
    \label{tab:bandcut}
\end{table*}

\subsection{SN~Ia-Host Correlation}
\subsubsection{SN~Ia Luminosity-Host Mass }\label{sec:lum}

A number of works have studied the correlations between the properties of SNe~Ia and their hosts (e.g., \citealp{uddin20, ponder21, johansson21, uddin17b, sullivan10, lampeitl10a, neill09, childress13a, rigault13, roman18, peter20} and references therein). The most significant outcome is how the decline-rate and color-corrected luminosity of SNe~Ia varies with the stellar mass of their host galaxies. In contrast to most other studies, \cite{uddin20}, using CSP-I data, showed this correlation in nine bandpasses ($uBgVriYJH$), and found evidence that the correlation between SN~Ia luminosity and host mass is insignificant except for the $u$-band. They also found that the Hubble residual offsets ($\Delta_{HR}$) do not vary strongly from optical to near-infrared wavelengths. Here, we define $\Delta_{HR}$ as the difference of mean $\Delta \mu$ of SNe~Ia between massive hosts and low-mass hosts, often called the mass-step. In this work, CSP-I data have been refit with the latest \texttt{SNooPy} version, and we obtain a different set of $\Delta_{HR}$ values, which weakens correlations and mass-steps\footnote{For example, in \cite{uddin20}, $\Delta_{HR}$ in the $u$-band was $0.147\pm 0.044$ mag for CSP-I data, compared to $0.090\pm 0.088$ mag for the same sample in this work.}. 

\begin{figure*}[htbp]
    \centering
    \includegraphics[width=\textwidth]{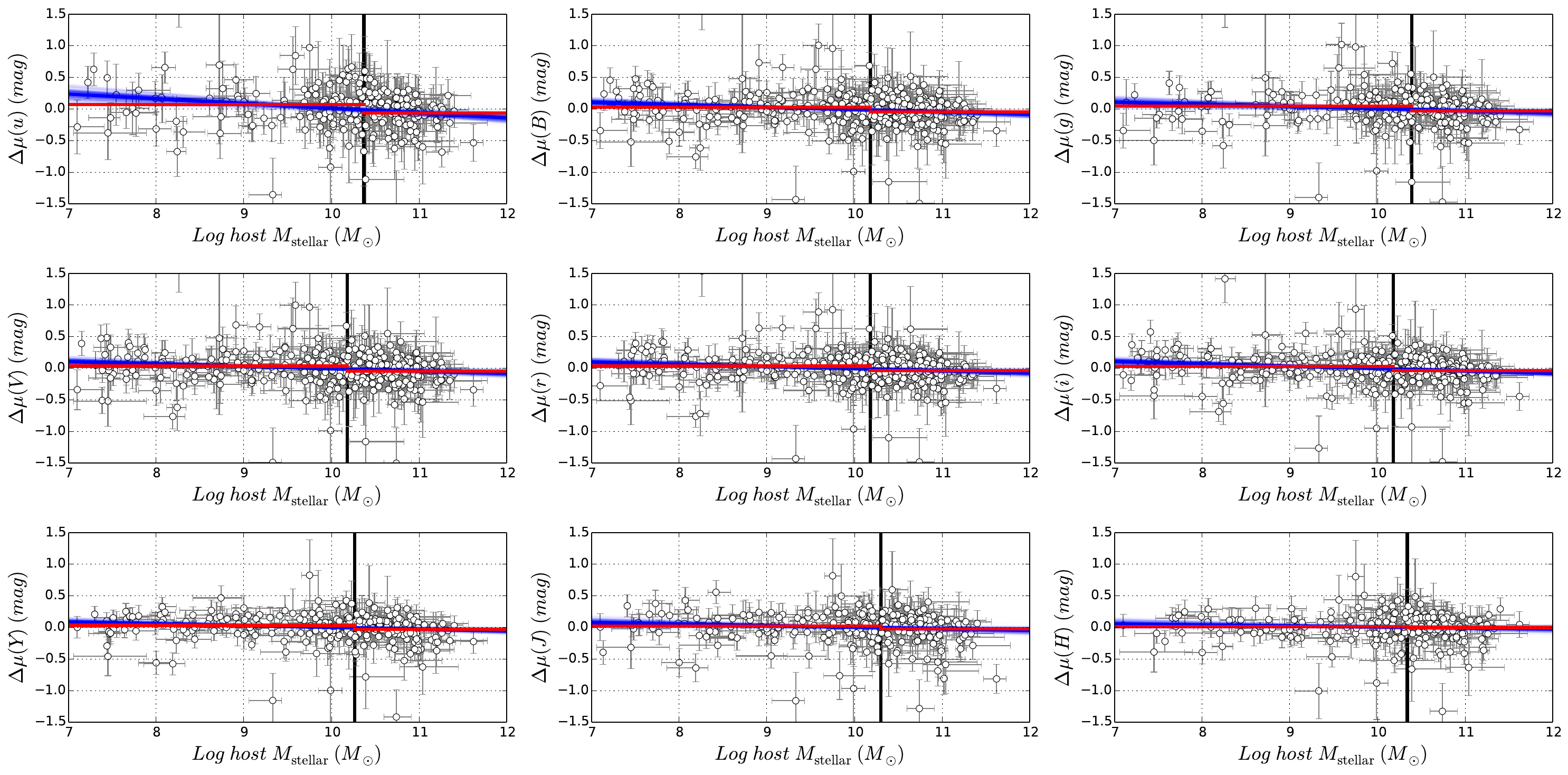}
    \caption{Hubble residuals ($\Delta_\mu$) as a function of host stellar mass in different bands. The black solid line is the median point of host stellar mass where we split the sample to calculate $\Delta_{HR}$. Red lines are the weighted mean Hubble residuals. Blue thick lines are the best-fit slopes of correlations, and the blue shaded lines are the $1\sigma$ dispersion.}
    \label{fig:masscorr}
\end{figure*}

\begin{figure}[htbp]
    \centering
    \includegraphics[width=\columnwidth]{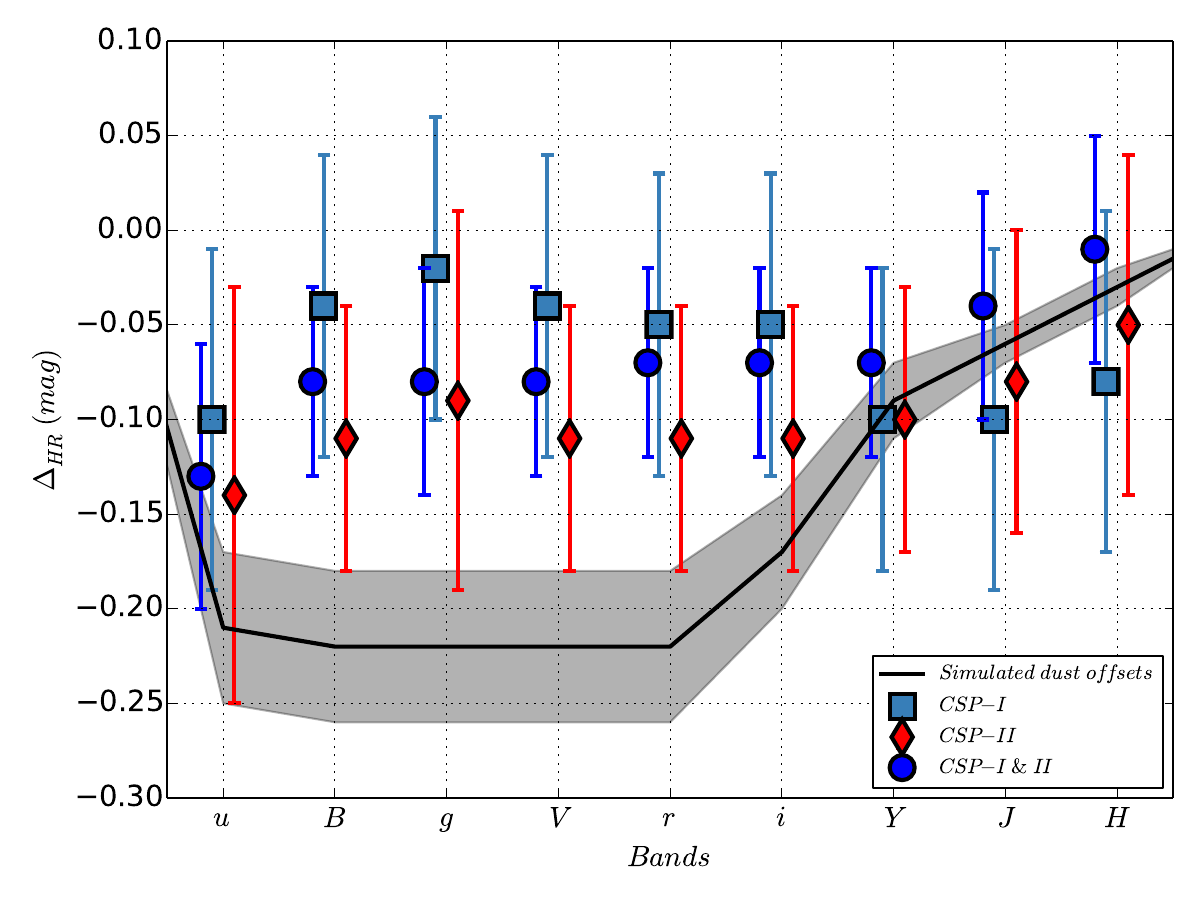}
    \caption{Simulated and observed Hubble residual offsets ($\Delta_{HR}$) as a function of wavelength. This figure is similar to Figure~13 in \cite{uddin20}, but now including the CSP-II sample. Simulated dust offset calculation is based on \cite{bs20}. We do not find $\Delta_{HR}$ to vary significantly from optical to near-infrared wavelengths. See \cite{uddin20} for the construction of the simulated dust offset. See text for discussion.}
    \label{fig:offsim}
\end{figure}

We find similar results using the CSP-II and the combined CSP-I \& II samples. We show the correlation between $\Delta \mu$ and host galaxy mass from this work in Figure~\ref{fig:masscorr}. Slopes and $\Delta_{HR}$ are given in Table~\ref{tab:slopes}. We also reproduce Figure~13 of \cite{uddin20} with more data from this work in Figure~\ref{fig:offsim}, where we show the behavior of $\Delta_{HR}$ across different bands. 

%With added data from CSP-II, we still find no significant correlations between SN~Ia Hubble residuals and host stellar mass in terms of slope and $\Delta_{HR}$, except for $u$-band, where the slope is $-0.066\pm 0.021$ mag/dex, and the mass-step is $-0.151\pm0.069$ mag. We also find no systematic variation of slope or $\Delta_{HR}$ with wavelength. Indeed, our results are consistent with there being no systematic correlation of $\Delta_{HR}$ with host galaxy mass.

With added data from CSP-II, we still find no significant  $\Delta_{HR}$, except for $u$-band, where the mass-step is $-0.132\pm0.071$ mag ($\sim 2 \sigma$). We also find no systematic variation of slope or $\Delta_{HR}$ with wavelength. Indeed, our results are consistent with there being no systematic correlation of $\Delta_{HR}$ with host galaxy mass.

On the other hand, the correlation between SN~Ia Hubble residuals and host stellar mass as measured by the slopes between them show moderate significance ($2 - 3 \ \sigma$) in all bands except for the $H$ band. In the $u$-band, the slope is maximum with a value of $-0.075\pm 0.020$ mag/dex.

While the origin of this correlation in not well established, \cite{bs20} predicted a wavelength dependence of $\Delta_{HR}$ with a progressive decrease when moving from the optical to near-infrared (see Figure~\ref{fig:offsim}). Their prediction was based on different dust distributions (different $R_V$) in massive and low-mass hosts, and they concluded that dust is responsible for the observed mass-step. In contrast to their prediction, our results indicate that dust may not be responsible for $\Delta_{HR}$, since we do not see variation of $\Delta_{HR}$ with wavelength. A study published by \cite{ponder21} is consistent with our findings that dust is not responsible for the observed mass-step, but the study by \cite{johansson21} is compatible with the findings in \cite{bs20}, i.e., it suggests that allowing for diversity in the extinction laws among host galaxies alleviates the need for a mass-step.

We note that SNe~Ia in CSP-I were obtained from targeted surveys, and their hosts are mostly massive. Therefore, the initial disagreement with literature results could have been due to the lack of low-mass hosts in CSP-I. In CSP-II, SNe~Ia came mostly from untargeted searches, and we have a mixture of massive and low-mass hosts. However, the disagreement still holds with the addition of more low-mass hosts.

Previously, \cite{childress14} argued that the mass-step originates from the differences in progenitor age distributions. Distinguishing SNe~Ia into prompt and tardy groups, their empirical formula predicts a redshift evolution of mass-step. But that prediction was not supported by the analysis made in \cite{uddin17b}.

While it is debatable that dust or SN~Ia progenitor age distributions explain a possible mass-step, the effect of metallicity should be investigated. There is a hint in our study that this may be the case. We find that the mass-step is relatively strong in $u$-band, and expected to be stronger in the ultraviolet (UV) wavelengths, if the variation in metallicity drives this correlation (\citealp{lentz00, brown14}). Studying this correlation using Hubble residuals calculated from UV may reveal the effect of metallicity, which will be investigated in a future work.
%We note that, \cite{hayden13} concluded metallicity of hosts are responsible for the observed correlation.

%\crb{This, to me, is one of the more interesting parts of this paper. This paragraph definitely needs to be fleshed out more. Start with a summary of what \cite{bs20} claimed, how this must therefore be a function of wavelength, how your previous results tentatively contradicted this, but we didn't have sufficient low-mass objects to tell for sure, but now with CSP-II, we've got what we need.}

\begin{table*}
    \centering
    \caption{Slope of SN~Ia luminosity-host mass correlation and Hubble residual offset in various bands.}
    \begin{tabular}{ccccc}
    \hline
     \hline
     &  &&Correlations&\\
    \cmidrule{3-5}
    Sample&Filter& Slope  & $\Delta_{HR}$  & Median Host Mass \\
    & &(mag/dex)&(mag)&$\rm Log~(M_0/M_{\odot}$)\\
   
    \hline 
& $ u $ &-0.082 (0.050) & -0.096 (0.109) & 10.50 \\
& $ B $ &-0.034 (0.036) & -0.040 (0.094) & 10.54 \\
& $ g $ &-0.022 (0.036) & -0.021 (0.090) & 10.54 \\
& $ V $ &-0.041 (0.038) & -0.040 (0.095) & 10.54 \\
CSP-I& $ r $ &-0.039 (0.033) & -0.054 (0.089) & 10.54 \\
& $ i $ &-0.034 (0.029) & -0.052 (0.086) & 10.54 \\
& $ Y $ &-0.061 (0.027) & -0.100 (0.083) & 10.55 \\
& $ J $ &-0.073 (0.033) & -0.096 (0.096) & 10.58 \\
& $ H $ &-0.024 (0.029) & -0.077 (0.092) & 10.59 \\
\hline
& $ u $ &-0.078 (0.022) & -0.138 (0.104) & 10.17 \\
& $ B $ &-0.043 (0.013) & -0.109 (0.067) & 9.95 \\
& $ g $ &-0.042 (0.017) & -0.090 (0.089) & 10.13 \\
& $ V $ &-0.044 (0.013) & -0.110 (0.068) & 9.95 \\
CSP-II& $ r $ &-0.042 (0.012) & -0.107 (0.065) & 9.95 \\
& $ i $ &-0.046 (0.010) & -0.106 (0.062) & 9.95 \\
& $ Y $ &-0.039 (0.010) & -0.095 (0.063) & 9.96 \\
& $ J $ &-0.042 (0.015) & -0.085 (0.075) & 10.02 \\
& $ H $ &-0.032 (0.015) & -0.052 (0.078) & 10.11 \\
\hline
& $ u $ &-0.075 (0.020) & -0.132 (0.071) & 10.37 \\
& $ B $ &-0.037 (0.012) & -0.076 (0.051) & 10.18 \\
& $ g $ &-0.034 (0.013) & -0.077 (0.062) & 10.39 \\
& $ V $ &-0.037 (0.012) & -0.076 (0.051) & 10.18 \\
CSP-I \& II& $ r $ &-0.035 (0.011) & -0.073 (0.049) & 10.18 \\
& $ i $ &-0.037 (0.010) & -0.065 (0.048) & 10.18 \\
& $ Y $ &-0.028 (0.009) & -0.065 (0.050) & 10.27 \\
& $ J $ &-0.025 (0.013) & -0.038 (0.056) & 10.30 \\
& $ H $ &-0.015 (0.012) & -0.014 (0.057) & 10.34 \\
 \hline
         
    \end{tabular}
    
    \label{tab:slopes}
\end{table*}

%\begin{figure*}[htbp]
 %   \centering
  %  \includegraphics[width=\textwidth]{CSPIIMassCorr.pdf}
   % \caption{Same as Figure~\ref{fig:masscorr} but for CSP-II only.}
    %\label{fig:csp2masscorr}
%\end{figure*}

\subsubsection{SN~Ia Luminosity-Projected Distance}\label{sep}

\begin{figure*}[htbp]
\begin{center}
\includegraphics[width=\textwidth]{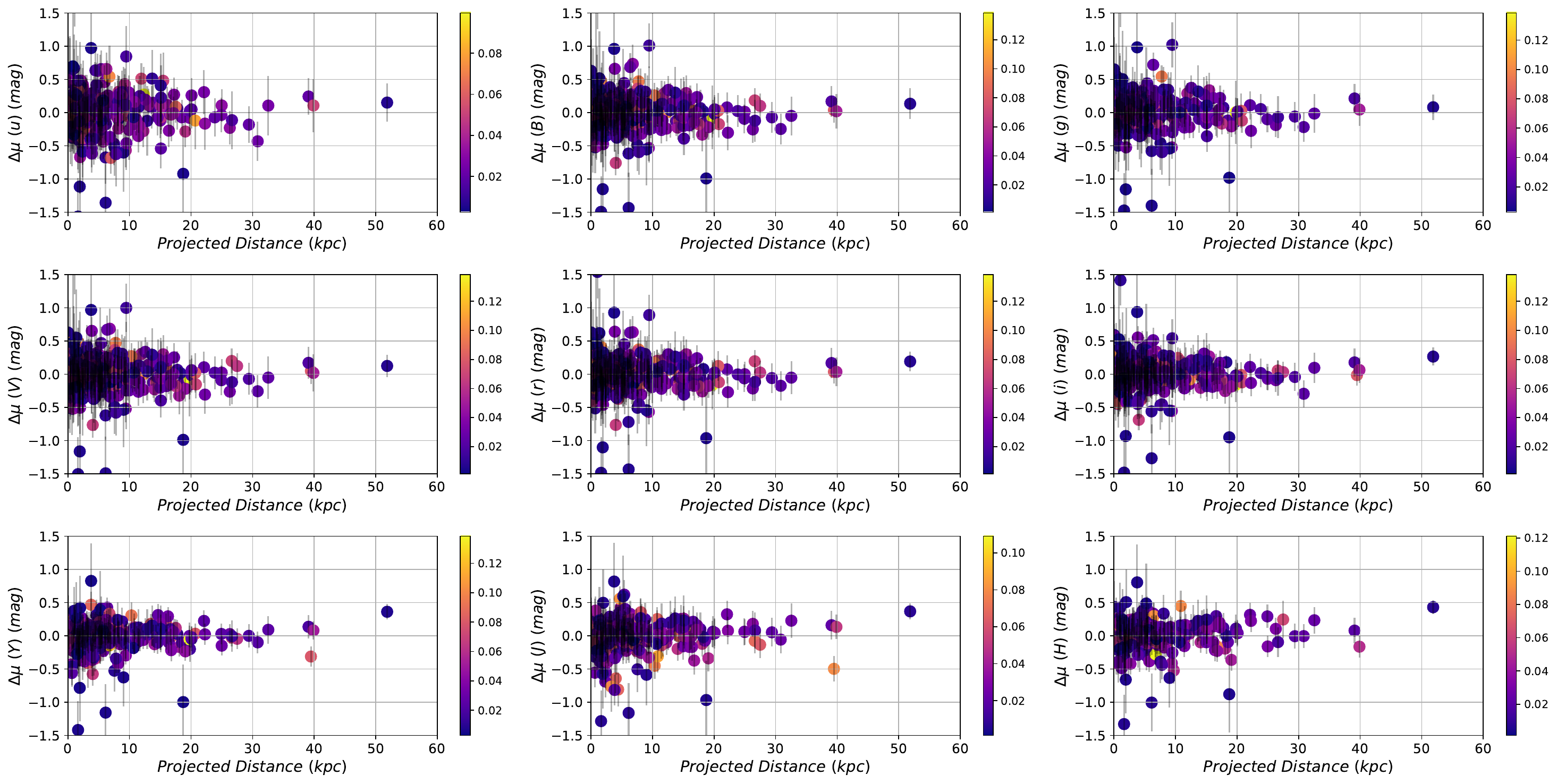} 
\caption{Distribution of $\Delta \mu$ with projected distance from host centers in the $uBgVriYJH$ bands. In all cases, the dispersion of $\Delta \mu$ is smaller for SNe~Ia, which explode beyond 10 kpc from their host centers. Color represents redshift.}
\label{sepall}
\end{center}
\end{figure*}

Using the CSP-I sample, \cite{uddin20} showed that SNe~Ia that explode beyond 10 kpc from host centers have a smaller scatter in their Hubble residuals than those that explode within 10 kpc. We extend this study by adding CSP-II data. In Figure~\ref{sepall} we show the distribution of Hubble residuals according to the projected distance for the full CSP sample in various bands. We calculate the dispersion in Hubble residuals of SNe~Ia inside and outside 10 kpc from host centers and list them in Table~\ref{scd}. Here, we also find a similar result with a larger sample, namely that SNe~Ia beyond 10 kpc have a smaller dispersion in their Hubble residuals. Previously, \cite{wang97} performed a similar study where they showed that SNe~Ia have 3-4 times less scatter in their observed peak brightness when they explode beyond 7.5 kpc from host centers. To see if the redshift affects this trend, we color code Figure~\ref{sepall} with redshift and find that there is no difference in redshift distributions between these two samples.

\begin{table}[htp]
\caption{Dispersion in Hubble residuals ($\Delta_{\mu}$) of SNe~Ia that are within 10 kpc from their host centers, and those that are beyond 10 kpc in various bands. On average, SNe~Ia exploding beyond 10 kpc from host centers have $\sim0.1$ mag smaller scatter. SNe~Ia are also labeled with respective redshifts.}
\begin{center}
\begin{tabular}{cccc}
\hline
\hline
Band & &Dispersion (mag)&\\
\cmidrule{2-4}
&   $ <10 \ kpc$  & $ >10 \ kpc$ & $Difference$ \\
\hline
$ u $ & 0.388 & 0.253 & 0.135\\
$ B $ & 0.312 & 0.195 & 0.117\\
$ g $ & 0.335 & 0.197 & 0.138\\
$ V $ & 0.313 & 0.196 & 0.118\\
$ r $ & 0.295 & 0.181 & 0.114\\
$ i $ & 0.273 & 0.171 & 0.102\\
$ Y $ & 0.243 & 0.179 & 0.064\\
$ J $ & 0.275 & 0.216 & 0.060\\
$ H $ & 0.251 & 0.205 & 0.046\\
\hline
\end{tabular}
\end{center}
\label{scd}
\end{table}%

\cite{uddin20} performed a Monte Carlo simulation and showed that the observed differences in the standard deviations of these two subgroups of SNe~Ia are statistically significant. 
%observed differences in the standard deviations of these two subgroups of SNe~Ia did not happen by chance, meaning they are intrinsically different. 
Here we repeat the same analysis and find that the mean of the differences in the scatter is $0.13\pm0.05$ mag.

To make a statistically robust conclusion, we performed a \texttt{Kolmogorov$–$Smirnov (K-S)} test to compare these two populations of SNe~Ia, with one population inside and the other population outside 10 kpc of host centers. The null hypothesis of the non-parametric \texttt{K-S} is that the two distributions are drawn from the same parent distribution. One can reject the null hypothesis if:

\begin{equation}\label{dstat}
D_{n,m} \ > \ c(\alpha) \ \sqrt{\frac{n+m}{n\times m}} \ \end{equation}
where $D_{n,m}$ is the D-statistic for two samples of size $n$ and $m$. This quantifies the maximum difference between two cumulative distribution functions. For a given confidence level, $\alpha$ the quantity $c(\alpha)$ is defined as:
\begin{equation}
    c(\alpha) =\rm  \sqrt{-ln\bigg(\frac{\alpha}{2}\bigg)\times \frac{1}{2}}
\end{equation}

\begin{figure*}% [hpbt] what you need
    \centering
     
        \includegraphics[width=\textwidth]{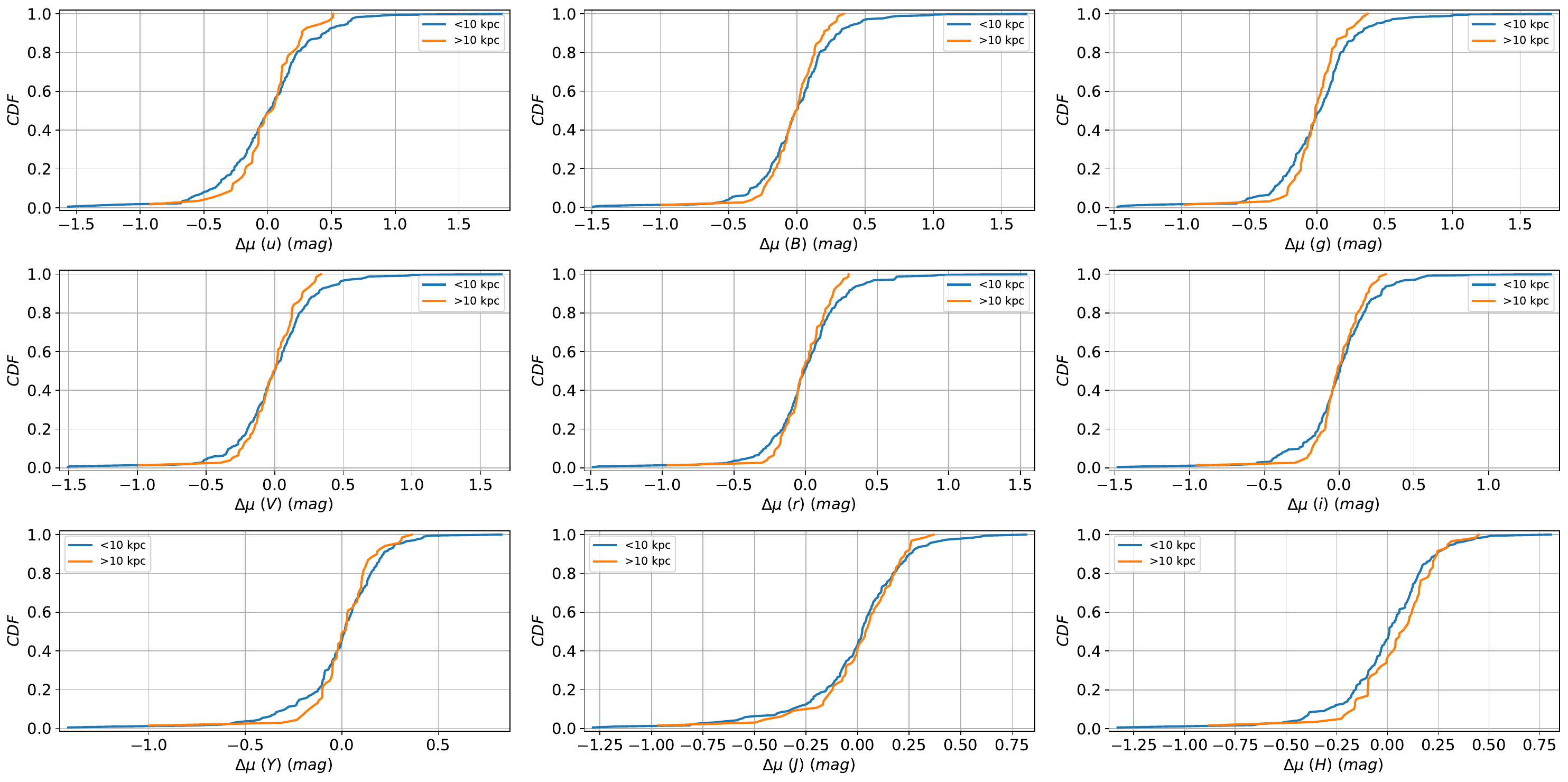}

    \caption{CDFs of Hubble residuals of two samples, shown for all bands. Maximum differences between the distributions can be read from Table~\ref{tab:kstest}.}\label{fig:cdf}
\end{figure*}

We present \texttt{K-S} test results in Table~\ref{tab:kstest}. First, we calculate the D-statistic in each band. Then we calculate the term $c(\alpha)$ for a range of confidence levels ($\alpha$), and select the $\alpha$ as soon as the null hypothesis rejection condition of Eqn.~\ref{dstat} is met. Cumulative distribution functions (CDF) for these two samples are shown in Figure~\ref{fig:cdf}.

\begin{figure}
    \centering
    \includegraphics[width=\columnwidth]{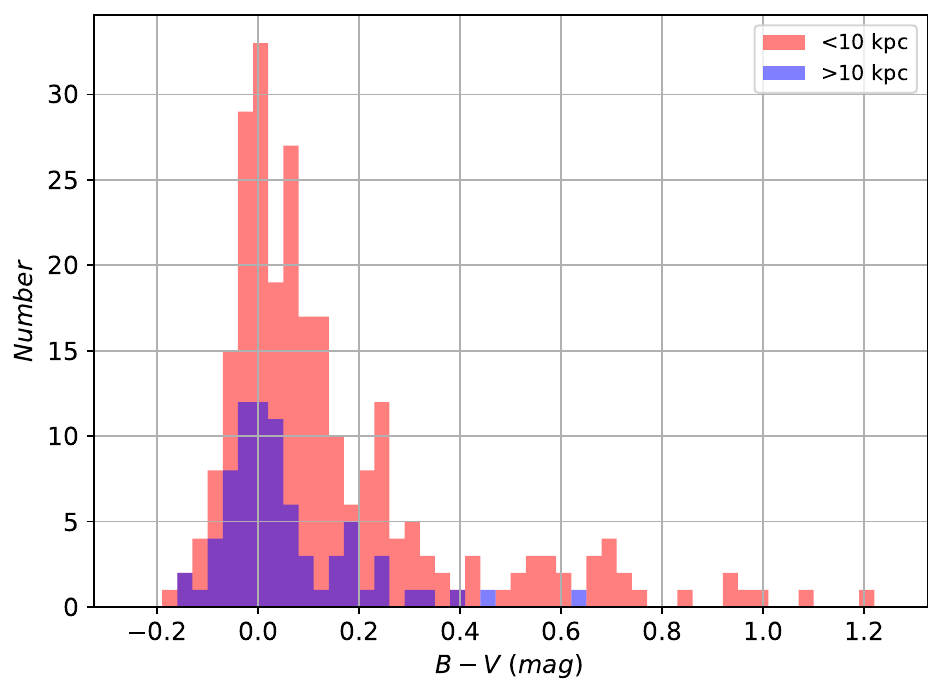}
    \caption{Distributions of SN~Ia color for the inner and the outer samples. We see that SNe~Ia exploding inside 10 kpc from host centers have a broader $(B-V)$ color range compare to those exploding beyond 10 kpc from host centers.}
    \label{fig:bvdist}
\end{figure}

It is not clear why we see smaller dispersion in SNe~Ia luminosity beyond a distance of 10 kpc from their hosts. We speculate that it could be due to the decrease of dust as we move from the center of a galaxy to its outskirts. From Table~\ref{scd} we find that the differences in dispersion are  smaller in the near-infrared bands than those in the optical. We also notice that (see Figure~\ref{fig:bvdist}) SNe~Ia that explode within 10 kpc from their host exhibit a wider range in $(B-V)$ color compare to those that explode beyond 10 kpc. The larger scatter closer to host centers may also originate from the additional error due to host-galaxy subtraction, for which further investigations are needed.

\begin{table}[]
    \centering
    \caption{Two sample \texttt{K-S} statistics. In each band, we calculate the D-statistic, sample weighted $c(\alpha)$ for the confidence level ($\alpha$) at which the condition in Eqn.~\ref{dstat} is met.}
    \begin{tabular}{cccc}
    \hline
    \hline 
    Band &&K-S Statistics&\\
    \cmidrule{2-4}
    & $D_{n,m}$& $c(\alpha) \ \sqrt{\frac{n+m}{n\times m}}$  & $\alpha$ (\%)\\
    \hline 
$ u $  & 0.188 & 0.185 &89 \\
$ B $  & 0.102 & 0.101 &40 \\
$ g $  & 0.168 & 0.166 &84 \\
$ V $  & 0.109 & 0.109 &51 \\
$ r $  & 0.095 & 0.095 &30 \\
$ i $  & 0.126 & 0.126 &69 \\
$ Y $  & 0.127 & 0.126 &61 \\
$ J $  & 0.143 & 0.142 &72 \\
$ H $  & 0.159 & 0.158 &77 \\
\hline
    \end{tabular}
    
    \label{tab:kstest}
\end{table}

%%%%%%%%%%

\FloatBarrier
\section{Conclusion}\label{sec:con}
We have presented an analysis of a combined SNe~Ia sample from CSP-I and CSP-II. We calculated the value of the Hubble constant, $H_0$, using Cepheid, TRGB, and SBF calibrations applied to this combined sample. %Using the distances of 322 SNe~Ia observed in the $B$-band and 
Combining all calibration methods, we have derived a global value of Hubble constant as $\rm H_0= 71.76 \pm 0.58$ (stat) $\pm 1.19$ (sys) $\rm km \ s^{-1} \ Mpc^{-1}$ from $B$-band data and %. Similar calculation from $H$-band data provides 
$\rm H_0= 73.22 \pm 0.68$ (stat) $\pm 1.28$ (sys) $\rm km \ s^{-1} \ Mpc^{-1}$ from $H$-band data. We have derived systematic errors by assigning equal weights to various calibration methods. %Our analysis minimizes the tension between the late-time $H_0$ and the early-time $H_0$ from the CMB. 
We consider these $H_0$ values to be representative measurements for the current time, given the established disagreements observed in recent $H_0$ studies. 
%Additionally, our analysis addresses the existing tension between late-time $H_0$ from various distant calibrators and the early-time $H_0$ from CMB. Unsurprisingly, the dominant systematic comes from $\sigma_{cal}$, the calibration of the SN~Ia distance scale. 
%It is clear from this that increased numbers of nearby SNe~Ia are not what is required to make progress, but rather finding the sources of the systematic errors that plague the Cepheid, TRGB, and/or SBF distance methods. 

%"If we assign equal weight to the Cepheid, TRGB, and SBF calibrators we can derive the systematic errors required for consistency in the first rung of the distance ladder, resulting in an increased systematic error in H_0.  As a result, the tension between the late-time H0 we derive by combining the various distant calibrators and the early-time H0 from the CMB is minimal."

%Systematic uncertainties in current $H_0$ measurements are within a few $\rm km \ s^{-1} \ Mpc^{-1}$ compared to statistical uncertainties, that are $\rm \sim 1 \ km \ s^{-1} \ Mpc^{-1}.$ While future studies need to focus on improving systematic uncertainties, discovering more local and distant calibrators, and addressing the velocity effect due to local mass concentration may aid in reconciling different $H_0$ measurements that we see in present-day analyses.   

Using these calibrators separately, although $H_0$ using the Cepheid and TRGB calibrations are consistent with previous measurements, we found a significant difference in $H_0$ calculated from two different SBF calibrator compilations. This difference in $H_0$ may originate from how the SBF distances are measured. In $YJH$ bands, the values of $H_0$ are larger in the TRGB calibration, and smaller in the SBF calibration in comparison to optical bands ($uBgVri$). Excluding 91T and 91bg-like SNe~Ia does not change $H_0$ significantly, but $\sigma_{int}$ is reduced. We have found that $\sigma_{int}$ is the smallest in $Y$-band, and there is no gradual improvement of $\sigma_{int}$ from the optical to the near-infrared.

We have revisited the correlation between the luminosity of SNe~Ia and their host mass. We did not find any significant correlations from the optical to the NIR, except for $u$-band, where a $\sim 3 \sigma$ negative slope is detected with a $\sim 2 \sigma$ difference in Hubble residuals at the median host mass. Dust has been hypothesized as the source of this correlation, but we do not see the corresponding decrease in this correlation with increasing wavelength. Likewise, the progenitor age distribution, that theoretically explains this correlation, predicts a redshift evolution of the mass-step, which is not evident from observed data. We suspect that metallicity may drive this correlation, and therefore, studies should be made in the UV wavelengths where the metallicity effect is more prominent due to increased line opacity.

Finally, we have confirmed our previous findings that SNe~Ia exploding more than 10 kpc from their host centers have less scatter in their Hubble residuals than those closer to their host centers. While the exact reason for this finding is unknown, dust may play a role, since we find a reduction of the differences in scatter from optical to near-infrared wavelengths.

\acknowledgments
We thank Adam Riess for providing feedback on this paper and for discussion on the combined analysis and peculiar velocity correction. We also thank Nandita Khetan for useful discussions on the SBF calibration. The work of the $Carnegie \ Supernova \ Project$ has been supported by the National Science Foundation (NSF) under grants AST0306969, AST0607438, AST1008343, AST1613426, AST1613472, AST613455., and also by the Danish Agency for Science and Technology and Innovation through a Sapere Aude Level 2 grant (PI Stritzinger). S. U. acknowledges financial support from the Distinguished Professor Chair funds of N. Suntzeff. N. B. S. and K. K. gratefully acknowledge the support from the George P. and Cynthia Woods Mitchell Institute for Fundamental Physics and Astronomy. We also thank the Mitchell Foundation for their support of the Cooks Branch Workshop on Supernovae. M. S. acknowledges grants from the Villum FONDEN No. 24349 and the Independent Research Fund Denmark (IRFD No. 8021-00170B). C. G. acknowledges support from research grant by the VILLUM FONDEN No. 25501. L. G. acknowledges financial support from the Spanish Ministerio de Ciencia e Innovaci\'on (MCIN), the Agencia Estatal de Investigaci\'on (AEI) 10.13039/501100011033, and the European Social Fund (ESF) "Investing in your future" under the 2019 Ram\'on y Cajal program RYC2019-027683-I and the PID2020-115253GA-I00 HOSTFLOWS project, from Centro Superior de Investigaciones Cient\'ificas (CSIC) under the PIE project 20215AT016, and the program Unidad de Excelencia Mar\'ia de Maeztu CEX2020-001058-M. B. J. S. is supported by NSF grants AST-1920392 and AST-1911074. We acknowledge the Las Campanas Observatory for the outstanding support during our observing runs, and to the Carnegie Observatories  Time Allocation Committee for generous time allocations.


%\vspace{5mm}
%\facilities{HST(STIS), Swift(XRT and UVOT), AAVSO, CTIO:1.3m,


%\software{astropy \citep{2013A&A...558A..33A},  
 %       SExtractor \citep{1996A&AS..117..393B}
  %        }


\newpage
\appendix

\section{Data}\label{alldata}
In this section, we present data used in this study. First, we present data for CSP SNe~Ia in the $B$-band. Data for all other bands are available online\footnote{\href{https://github.com/syeduddin/h0csp}{https://github.com/syeduddin/h0csp}}. We also present similar tables for various calibrators. A description of various columns are given in Table~\ref{tab:explanation}.

\setcounter{table}{0}
\renewcommand{\thetable}{\Alph{section}\arabic{table}}

\begin{table}[]
    \centering
    \caption{Description of various columns used to present data.}
    \begin{tabular}{l|l}
    \hline
    \hline
        Parameter & Description  \\
        \hline
        $Name$ & SN Ia name\\
        $Host$ & Host galaxy name\\
        $z_{cmb}$ & CMB redshift\\
        $z_{hel}$ & Heliocentric redshift\\
        $B_{max}$ & Peak $B$-band magnitude\\
        $eB_{max}$ & Error in $B_{max}$\\
        $s_{BV}$ & Color-stretch parameter\\
        $es_{BV}$ & Error in $s_{BV}$\\
        $(B-V)$ & $(B-V)$ color\\
        $e(B-V)$ & Error in $(B-V)$\\
        $cov(B_{max}, \ s_{BV})$ & Covariance between $B_{max}$ and $s_{BV}$ \\
        $cov(B_{max}, \ B-V)$ & Covariance between $B_{max}$ and $B-V$ \\
        %$cov(s_{BV}, \ B-V)$ & Covariance between $s_{BV}$ and $B-V$ \\
        $log \ (M_*)$ & Host galaxy stellar mass in unit of $M_{\odot}$ \\
        $Sample$ & CSP-I or CSP-II (for Table~\ref{csptable})\\
        $Subtype$ & Ia, 91T, or 91bg (for Table~\ref{csptable})\\
        $\mu$ & Distance modulus from calibrators (for Tables~\ref{cephtable}, \ \ref{trgbtable}, \ \ref{irsbftable}, and \ref{sbftable})\\
        $e\mu$ & Error in  $\mu$ (for Tables~\ref{cephtable}, \ \ref{trgbtable}, \ \ref{irsbftable}, and \ref{sbftable})\\
    
        \hline
    \end{tabular}
    
    \label{tab:explanation}
\end{table}

\begin{longrotatetable}
\thispagestyle{plain}

\begin{deluxetable*}{llllllllllllllll}
\tablecaption{CSP SN Ia light-curve fitting with \texttt{SNooPy} and host galaxy stellar mass. This is a partial table. Full version for various bands are available in github page.\label{csptable}}
\tablewidth{600pt}
\tabletypesize{\scriptsize}
\tablehead{
\colhead{$Name$} & \colhead{$Host$} & \colhead{$z_{cmb}$} & 
\colhead{$z_{hel}$} & \colhead{$B_{max}$} & 
\colhead{$eB_{max}$} & \colhead{$s_{BV}$} & 
\colhead{$es_{BV}$} & \colhead{$(B-V)$} & 
\colhead{$e(B-V)$} & \colhead{$cov(B_{max}, \ s_{BV})$} & \colhead{$cov(B_{max}, \ B-V)$} & \colhead{$log \ (M_*$)} & \colhead{$Sample$}&\colhead{$Subtype$} \\ 
&  & &
 & \colhead{(mag)} & 
\colhead{(mag)} & & 
 & \colhead{(mag)} & 
\colhead{(mag)} &  &   & \colhead{$(M_{\odot})$} & & \\
} 
\startdata
ASAS14ad & KUG 1237+183 & 0.0274 & 0.0264 & 16.244 & 0.005 & 1.01 & 0.018 & -0.011 & 0.01118 & 1e-05 & -2e-05 & 8.61 & CSPII & Ia \\
ASAS14hp & 2MASX J21303015-7038489 & 0.0387 & 0.03889 & 16.547 & 0.003 & 1.074 & 0.01 & -0.049 & 0.005 & -1e-05 & -1e-05 & 9.09 & CSPII & Ia \\
ASAS14hr & 2MASX J01504127-1431032 & 0.0327 & 0.0336 & 17.021 & 0.009 & 0.802 & 0.01 & 0.061 & 0.01345 & 4e-05 & -8e-05 & 10.34 & CSPII & Ia \\
ASAS14hu & ESO 058- G 012 & 0.0219 & 0.02159 & 15.431 & 0.004 & 1.052 & 0.007 & -0.045 & 0.0064 & -1e-05 & -2e-05 & 9.93 & CSPII & Ia \\
ASAS14jc & 2MASX J07353554-6246099 & 0.0118 & 0.01132 & 15.245 & 0.006 & 0.915 & 0.005 & 0.437 & 0.01 & 1e-05 & -4e-05 & 9.46 & CSPII & Ia \\
ASAS14jg & 2MASX J23331223-6034201 & 0.0143 & 0.01482 & 14.732 & 0.035 & 1.285 & 0.022 & 0.049 & 0.05941 & -2e-05 & -0.00122 & 8.76 & CSPII & Ia \\
ASAS14jz & GALEXASC J184443.33-524819.2 & 0.0156 & 0.01579 & 14.663 & 0.004 & 0.989 & 0.006 & -0.041 & 0.00566 & 1e-05 & -2e-05 & 8.23 & CSPII & Ia \\
ASAS14kd & 2MASX J22532475+0447583 & 0.0231 & 0.0243 & 16.01 & 0.005 & 1.169 & 0.007 & 0.234 & 0.00707 & 2e-05 & -2e-05 & 10.14 & CSPII & Ia-91T \\
ASAS14kq & 2MASX J23451480-2947009 & 0.0326 & 0.03358 & 16.612 & 0.009 & 1.147 & 0.011 & -0.021 & 0.01204 & 6e-05 & -8e-05 & 8.91 & CSPII & Ia \\
ASAS14lo & UGC 06837 & 0.021 & 0.01992 & 15.876 & 0.036 & 0.952 & 0.018 & 0.127 & 0.0608 & -1e-05 & -0.0013 & 9.69 & CSPII & Ia \\
ASAS14lp & NGC 4666 & 0.0062 & 0.0051 & 11.914 & 0.004 & 1.029 & 0.004 & 0.243 & 0.0064 & 0.0 & -2e-05 & 10.74 & CSPII & Ia \\
ASAS14lq & 2MASX J22571481-2058014 & 0.0251 & 0.02617 & 16.013 & 0.047 & 0.999 & 0.024 & 0.02 & 0.0794 & -5e-05 & -0.00221 & 10.65 & CSPII & Ia \\
ASAS14lt & IC 0299 & 0.0315 & 0.03202 & 16.368 & 0.01 & 0.944 & 0.008 & -0.033 & 0.0164 & 2e-05 & -0.0001 & 10.64 & CSPII & Ia \\
ASAS14lw & GALEXASC J010647.95-465904.1 & 0.0203 & 0.02089 & 15.602 & 0.006 & 1.25 & 0.007 & -0.012 & 0.00922 & 2e-05 & -4e-05 & 6.35 & CSPII & Ia \\
ASAS14me & ESO 113- G 047 & 0.0174 & 0.0178 & 15.074 & 0.004 & 1.078 & 0.005 & 0.007 & 0.00566 & -0.0 & -2e-05 & 9.05 & CSPII & Ia \\
ASAS14mf & GALEXASC J000454.54-322615.3 & 0.0302 & 0.03108 & 16.511 & 0.007 & 0.984 & 0.005 & -0.012 & 0.01063 & 1e-05 & -5e-05 & 8.93 & CSPII & Ia \\
ASAS14mw & AM 0139-655 NED02 & 0.0271 & 0.02739 & 15.879 & 0.003 & 1.063 & 0.007 & -0.034 & 0.005 & -1e-05 & -1e-05 & 10.68 & CSPII & Ia \\
ASAS14my & NGC 3774 & 0.0217 & 0.0205 & 15.553 & 0.006 & 0.923 & 0.005 & 0.015 & 0.01 & 1e-05 & -4e-05 & 10.31 & CSPII & Ia \\
ASAS15aj & NGC 3449 & 0.012 & 0.01091 & 14.577 & 0.006 & 0.831 & 0.007 & 0.103 & 0.00922 & 1e-05 & -4e-05 & 11.2 & CSPII & Ia \\
ASAS15al & GALEXASC J045749.46-213526.3 & 0.0338 & 0.03378 & 16.841 & 0.036 & 1.077 & 0.028 & 0.055 & 0.0592 & -3e-05 & -0.0013 & 8.07 & CSPII & Ia \\
ASAS15as & SDSS J093916.69+062551.1 & 0.0298 & 0.02868 & 16.179 & 0.005 & 1.076 & 0.014 & 0.029 & 0.00781 & -2e-05 & -2e-05 & 7.66 & CSPII & Ia \\
ASAS15ba & SDSS J140455.12+085514.0 & 0.024 & 0.02312 & 15.975 & 0.004 & 0.967 & 0.004 & 0.008 & 0.0064 & 0.0 & -2e-05 & 7.99 & CSPII & Ia \\
ASAS15be & GALEXASC J025245.83-341850.6 & 0.0214 & 0.02188 & 15.611 & 0.005 & 1.134 & 0.006 & 0.019 & 0.0064 & 2e-05 & -2e-05 & 7.8 & CSPII & Ia \\
ASAS15bm & LCRS B150313.2-052600 & 0.0214 & 0.02079 & 15.625 & 0.007 & 0.991 & 0.005 & 0.127 & 0.0099 & 1e-05 & -5e-05 & 10.02 & CSPII & Ia \\
ASAS15cb & VCC 1810 & 0.0411 & 0.04001 & 16.953 & 0.047 & 1.086 & 0.053 & 0.107 & 0.0794 & -5e-05 & -0.00221 & 10.28 & CSPII & Ia \\
ASAS15cd & CGCG 064-017 & 0.0354 & 0.03429 & 16.693 & 0.006 & 1.003 & 0.005 & -0.021 & 0.00922 & 1e-05 & -4e-05 & 9.95 & CSPII & Ia \\
ASAS15da & 2MASX J05235106-2442201 & 0.0488 & 0.0487 & 17.679 & 0.044 & 0.853 & 0.025 & -0.132 & 0.07603 & -2e-05 & -0.00194 & 10.4 & CSPII & Ia \\
ASAS15db & NGC 5996 & 0.0114 & 0.01099 & 14.602 & 0.004 & 0.955 & 0.003 & 0.086 & 0.0064 & 0.0 & -2e-05 & 10.11 & CSPII & Ia \\
ASAS15dd & CGCG 107-031 & 0.0247 & 0.02436 & 16.065 & 0.006 & 0.849 & 0.006 & -0.007 & 0.00922 & 1e-05 & -4e-05 & 10.41 & CSPII & Ia \\
ASAS15eb & ESO 561- G 012 & 0.0173 & 0.01647 & 15.339 & 0.01 & 0.821 & 0.018 & -0.129 & 0.01562 & 4e-05 & -0.0001 & 10.96 & CSPII & Ia \\
ASAS15fr & 2MASX J09202045-0738229 & 0.0345 & 0.03341 & 16.816 & 0.039 & 0.905 & 0.021 & -0.133 & 0.0658 & -2e-05 & -0.00152 & 9.98 & CSPII & Ia-91T \\
ASAS15ga & NGC 4866 & 0.0077 & 0.00663 & 15.068 & 0.037 & 0.496 & 0.03 & 0.431 & 0.04958 & 0.00074 & -0.00137 & 10.38 & CSPII & Ia-91bg \\
ASAS15go & 2MASX J06113048-1629085 & 0.0193 & 0.01891 & 15.847 & 0.004 & 1.071 & 0.012 & 0.224 & 0.0064 & -1e-05 & -2e-05 & 10.18 & CSPII & Ia \\
ASAS15gr & ESO 366- G 015 & 0.0248 & 0.02428 & 15.867 & 0.004 & 1.035 & 0.007 & -0.036 & 0.0064 & -0.0 & -2e-05 & 8.94 & CSPII & Ia \\
ASAS15hf & ESO 375- G 041 & 0.0072 & 0.00617 & 13.996 & 0.006 & 0.943 & 0.005 & 0.009 & 0.01 & 1e-05 & -4e-05 & 9.56 & CSPII & Ia \\
\enddata
\end{deluxetable*}
\end{longrotatetable}

\begin{longrotatetable}

\begin{deluxetable*}{llllllllllllllll}
\tablecaption{Cepheid calibration SN Ia light-curve fitting with \texttt{SNooPy}, host galaxy stellar mass, and distance modulus.\label{cephtable}}
\tablewidth{700pt}
\tabletypesize{\scriptsize}
\tablehead{
\colhead{$Name$} & \colhead{$Host$} & \colhead{$z_{cmb}$} & 
\colhead{$z_{hel}$} & \colhead{$B_{max}$} & 
\colhead{$eB_{max}$} & \colhead{$s_{BV}$} & 
\colhead{$es_{BV}$} & \colhead{$(B-V)$} & 
\colhead{$e(B-V)$} & \colhead{$cov(B_{max}, \ s_{BV})$} & \colhead{$cov(B_{max}, \ B-V)$}  & \colhead{$log \ (M_*$)} & \colhead{$\mu$}&\colhead{$e\mu$} \\ 
 &  &  &
 & \colhead{(mag)} & 
\colhead{(mag)} & & 
 & \colhead{(mag)} & 
\colhead{(mag)} &  &   & \colhead{$(M_{\odot})$} & \colhead{(mag)}&\colhead{(mag)} \\ 
} 
\startdata
SN1981B & N4536 & 0.0072 & 0.00603 & 11.979 & 0.005 & 0.941 & 0.019 & 0.068 & 0.00781 & 1e-05 & 2e-05 & 10.47 & 30.835 & 0.05 \\
SN1990N & N4639 & 0.0041 & 0.003 & 12.714 & 0.006 & 1.1 & 0.006 & 0.066 & 0.00849 & 1e-05 & 4e-05 & 10.14 & 31.812 & 0.084 \\
SN1994ae & N3370 & 0.0051 & 0.004 & 13.142 & 0.005 & 1.043 & 0.006 & 0.06 & 0.00707 & 2e-05 & 2e-05 & 9.69 & 32.12 & 0.051 \\
SN1995al & N3021 & 0.0059 & 0.005 & 13.343 & 0.01 & 1.043 & 0.017 & 0.138 & 0.01345 & 5e-05 & 0.0001 & 9.87 & 32.464 & 0.158 \\
SN1998aq & N3982 & 0.0043 & 0.0037 & 12.357 & 0.006 & 0.968 & 0.004 & -0.104 & 0.00781 & 1e-05 & 4e-05 & 10.02 & 31.722 & 0.071 \\
SN2001el & N1448 & 0.0037 & 0.0039 & 12.822 & 0.008 & 0.959 & 0.005 & 0.144 & 0.01 & 2e-05 & 6e-05 & 10.34 & 31.287 & 0.037 \\
SN2002fk & N1309 & 0.0066 & 0.00712 & 13.202 & 0.006 & 0.977 & 0.004 & -0.091 & 0.00849 & 1e-05 & 4e-05 & 9.94 & 32.541 & 0.059 \\
SN2003du & U9391 & 0.0067 & 0.00638 & 13.478 & 0.005 & 1.002 & 0.004 & -0.073 & 0.0064 & 1e-05 & 2e-05 & 8.6 & 32.848 & 0.067 \\
SN2005W & N0691 & 0.008 & 0.00888 & 14.208 & 0.006 & 0.938 & 0.006 & 0.171 & 0.00781 & 1e-05 & 4e-05 & 9.9 & 32.83 & 0.109 \\
SN2005cf & N5917 & 0.007 & 0.00646 & 13.246 & 0.008 & 0.972 & 0.005 & 0.005 & 0.01063 & 1e-05 & 6e-05 & 9.65 & 32.363 & 0.12 \\
SN2006D & M1337 & 0.0096 & 0.00852 & 14.194 & 0.005 & 0.839 & 0.003 & 0.098 & 0.0064 & 0.0 & 2e-05 & 9.76 & 32.92 & 0.123 \\
SN2006bh & N7329 & 0.0105 & 0.01084 & 14.369 & 0.004 & 0.828 & 0.003 & -0.01 & 0.00566 & 0.0 & 2e-05 & 10.43 & 33.246 & 0.117 \\
SN2007A & N0105 & 0.0165 & 0.01763 & 15.718 & 0.008 & 1.047 & 0.02 & 0.201 & 0.01 & 8e-05 & 6e-05 & 10.57 & 34.527 & 0.25 \\
SN2007af & N5584 & 0.0063 & 0.00546 & 13.184 & 0.005 & 0.949 & 0.002 & 0.065 & 0.00707 & 0.0 & 2e-05 & 9.82 & 31.772 & 0.052 \\
SN2007sr & N4038 & 0.0067 & 0.00547 & 12.794 & 0.047 & 0.988 & 0.017 & 0.145 & 0.0586 & -1e-05 & 0.00221 & 10.05 & 31.603 & 0.116 \\
SN2009Y & N5728 & 0.0101 & 0.00935 & 13.992 & 0.008 & 1.143 & 0.01 & 0.146 & 0.01204 & -0.0 & 6e-05 & 10.76 & 33.094 & 0.205 \\
SN2009ig & N1015 & 0.008 & 0.0088 & 13.453 & 0.007 & 1.061 & 0.014 & 0.102 & 0.00922 & 2e-05 & 5e-05 & 10.35 & 32.563 & 0.074 \\
SN2011by & N3972 & 0.0034 & 0.00284 & 12.855 & 0.007 & 0.968 & 0.003 & -0.006 & 0.00922 & 1e-05 & 5e-05 & 9.47 & 31.635 & 0.089 \\
SN2011fe & M101 & 0.0012 & 0.0008 & 10.004 & 0.006 & 0.97 & 0.002 & 0.008 & 0.00721 & 0.0 & 4e-05 & 10.23 & 29.178 & 0.041 \\
SN2012cg & N4424 & 0.0026 & 0.00146 & 11.996 & 0.008 & 1.013 & 0.007 & 0.071 & 0.00943 & 3e-05 & 6e-05 & 9.7 & 30.844 & 0.128 \\
SN2012fr & N1365 & 0.0051 & 0.00545 & 11.935 & 0.005 & 1.109 & 0.004 & 0.013 & 0.00583 & 0.0 & 2e-05 & 11.25 & 31.378 & 0.056 \\
SN2012ht & N3447 & 0.0047 & 0.00356 & 13.003 & 0.007 & 0.868 & 0.003 & -0.016 & 0.0099 & 1e-05 & 5e-05 & 8.06 & 31.936 & 0.034 \\
SN2013aa & N5643 & 0.0047 & 0.004 & 11.098 & 0.004 & 1.002 & 0.002 & -0.078 & 0.005 & 0.0 & 2e-05 & 9.99 & 30.546 & 0.052 \\
SN2013dy & N7250 & 0.0029 & 0.00389 & 12.747 & 0.004 & 1.091 & 0.007 & 0.209 & 0.005 & 0.0 & 2e-05 & 9.09 & 31.628 & 0.125 \\
SN2015F & N2442 & 0.0053 & 0.00489 & 12.753 & 0.006 & 0.887 & 0.003 & 0.058 & 0.00721 & 1e-05 & 4e-05 & 10.97 & 31.45 & 0.064 \\
\enddata
\end{deluxetable*}
\end{longrotatetable}

\begin{longrotatetable}

\begin{deluxetable*}{llllllllllllllll}
\tablecaption{TRGB calibration SN Ia light-curve fitting with \texttt{SNooPy}, host galaxy stellar mass, and distance modulus.\label{trgbtable}}
\tablewidth{700pt}
\tabletypesize{\scriptsize}
\tablehead{
\colhead{$Name$} & \colhead{$Host$} & \colhead{$z_{cmb}$} & 
\colhead{$z_{hel}$} & \colhead{$B_{max}$} & 
\colhead{$eB_{max}$} & \colhead{$s_{BV}$} & 
\colhead{$es_{BV}$} & \colhead{$(B-V)$} & 
\colhead{$e(B-V)$} & \colhead{$cov(B_{max}, \ s_{BV})$} & \colhead{$cov(B_{max}, \ B-V)$}  & \colhead{$M_*$} & \colhead{$\mu$}&\colhead{$e\mu$} \\ 
&  & &
 & \colhead{(mag)} & 
\colhead{(mag)} & & 
 & \colhead{(mag)} & 
\colhead{(mag)} &  &   & \colhead{$(M_{\odot})$} & \colhead{(mag)}&\colhead{(mag)} \\
} 
\startdata
SN1980N & N1316 & 0.0055 & 0.00587 & 12.458 & 0.009 & 0.904 & 0.006 & 0.07 & 0.01204 & 0.0 & 8e-05 & 11.79 & 31.465 & 0.04 \\
SN1981B & N4536 & 0.0072 & 0.00603 & 11.979 & 0.005 & 0.941 & 0.019 & 0.068 & 0.00781 & 1e-05 & 2e-05 & 10.47 & 30.969 & 0.05 \\
SN1981D & N1316 & 0.0055 & 0.00587 & 12.554 & 0.034 & 0.793 & 0.033 & 0.173 & 0.03847 & 0.00017 & 0.00116 & 11.79 & 31.465 & 0.04 \\
SN1989B & N3627 & 0.0036 & 0.00242 & 12.311 & 0.01 & 0.888 & 0.024 & 0.364 & 0.01166 & -1e-05 & 0.0001 & 11.02 & 30.226 & 0.04 \\
SN1994D & N4526 & 0.0026 & 0.00149 & 11.84 & 0.006 & 0.824 & 0.005 & -0.047 & 0.00849 & 1e-05 & 4e-05 & 11.0 & 31.003 & 0.07 \\
SN1994ae & N3370 & 0.0051 & 0.004 & 13.142 & 0.005 & 1.043 & 0.006 & 0.06 & 0.00707 & 2e-05 & 2e-05 & 9.69 & 32.278 & 0.05 \\
SN1995al & N3021 & 0.0059 & 0.005 & 13.343 & 0.01 & 1.043 & 0.017 & 0.138 & 0.01345 & 5e-05 & 0.0001 & 9.87 & 32.226 & 0.04 \\
SN1998bu & N3368 & 0.0037 & 0.00248 & 12.066 & 0.006 & 0.971 & 0.004 & 0.345 & 0.00721 & 0.0 & 4e-05 & 11.26 & 30.318 & 0.04 \\
SN2001el & N1448 & 0.0037 & 0.0039 & 12.822 & 0.008 & 0.959 & 0.005 & 0.144 & 0.01 & 2e-05 & 6e-05 & 10.69 & 31.325 & 0.06 \\
SN2002fk & N1309 & 0.0066 & 0.00712 & 13.202 & 0.006 & 0.977 & 0.004 & -0.091 & 0.00849 & 1e-05 & 4e-05 & 9.94 & 32.504 & 0.07 \\
SN2006dd & N1316 & 0.0055 & 0.00587 & 12.228 & 0.005 & 0.951 & 0.003 & -0.088 & 0.00707 & 1e-05 & 2e-05 & 11.79 & 31.465 & 0.04 \\
SN2007af & N5584 & 0.0063 & 0.00546 & 13.184 & 0.005 & 0.949 & 0.002 & 0.065 & 0.00707 & 0.0 & 2e-05 & 9.82 & 31.827 & 0.1 \\
SN2007on & N1404 & 0.0062 & 0.00649 & 13.032 & 0.009 & 0.588 & 0.003 & 0.116 & 0.01204 & 1e-05 & 8e-05 & 11.04 & 31.36 & 0.06 \\
SN2007sr & N4038 & 0.0067 & 0.00547 & 12.794 & 0.047 & 0.988 & 0.017 & 0.145 & 0.0586 & -1e-05 & 0.00221 & 10.05 & 31.686 & 0.05 \\
SN2011fe & M101 & 0.0012 & 0.0008 & 10.004 & 0.006 & 0.97 & 0.002 & 0.008 & 0.00721 & 0.0 & 4e-05 & 10.23 & 29.083 & 0.04 \\
SN2011iv & N1404 & 0.0062 & 0.00649 & 12.463 & 0.008 & 0.699 & 0.007 & 0.031 & 0.01063 & 1e-05 & 6e-05 & 11.04 & 31.36 & 0.06 \\
SN2012cg & N4424 & 0.0026 & 0.00146 & 11.996 & 0.008 & 1.013 & 0.007 & 0.071 & 0.00943 & 3e-05 & 6e-05 & 9.7 & 31.005 & 0.06 \\
SN2012fr & N1365 & 0.0051 & 0.00545 & 11.935 & 0.005 & 1.109 & 0.004 & 0.013 & 0.00583 & 0.0 & 2e-05 & 11.25 & 31.365 & 0.06 \\
SN2013aa & N5643 & 0.0047 & 0.004 & 11.098 & 0.004 & 1.002 & 0.002 & -0.078 & 0.005 & 0.0 & 2e-05 & 9.99 & 30.48 & 0.08 \\
SN2017cbv & N5643 & 0.0047 & 0.004 & 11.124 & 0.007 & 1.115 & 0.006 & 0.018 & 0.00922 & 0.0 & 5e-05 & 9.99 & 30.48 & 0.08 \\
\enddata
\end{deluxetable*}
\end{longrotatetable}

\begin{longrotatetable}
\begin{deluxetable*}{lllllllllllllll}
\tablecaption{SBF calibration (\citealt{jensen21}) SN Ia light-curve fitting with \texttt{SNooPy}, host galaxy stellar mass, and distance modulus.\label{irsbftable}}
\tablewidth{700pt}
\tabletypesize{\scriptsize}
\tablehead{
\colhead{$Name$} & \colhead{$z_{cmb}$} & 
\colhead{$z_{hel}$} & \colhead{$B_{max}$} & 
\colhead{$eB_{max}$} & \colhead{$s_{BV}$} & 
\colhead{$es_{BV}$} & \colhead{$(B-V)$} & 
\colhead{$e(B-V)$} & \colhead{$cov(B_{max}, \ s_{BV})$} & \colhead{$cov(B_{max}, \ B-V)$}  & \colhead{$log \ M_*$} & \colhead{$\mu$}&\colhead{$e\mu$} \\ 
&  & 
 & \colhead{(mag)} & 
\colhead{(mag)} & & 
 & \colhead{(mag)} & 
\colhead{(mag)} &  &   & \colhead{$(M_{\odot})$} & \colhead{(mag)}&\colhead{(mag)} \\
} 
\startdata
CSP15aae & N5490 & 0.017 & 0.01618 & 16.762 & 0.008 & 0.505 & 0.004 & 0.454 & 0.01063 & 0.0 & 6e-05 & 11.14 & 34.267 & 0.08 \\
PTF13ebh & N0890 & 0.0125 & 0.01326 & 15.043 & 0.007 & 0.636 & 0.004 & 0.123 & 0.00922 & 1e-05 & 5e-05 & 10.86 & 33.296 & 0.081 \\
SN1970J & N7619 & 0.012 & 0.0132 & 14.278 & 0.084 & 0.704 & 0.071 & -0.144 & 0.12458 & 0.0039 & 0.00706 & 11.34 & 33.341 & 0.081 \\
SN1995D & N2962 & 0.0077 & 0.00656 & 13.265 & 0.008 & 1.171 & 0.009 & -0.053 & 0.01 & -2e-05 & 6e-05 & 10.597 & 32.532 & 0.084 \\
SN1997E & N2258 & 0.0135 & 0.01354 & 15.192 & 0.007 & 0.831 & 0.011 & 0.045 & 0.00922 & 3e-05 & 5e-05 & 11.199 & 33.781 & 0.094 \\
SN1999ej & N0495 & 0.0128 & 0.01372 & 15.435 & 0.005 & 0.78 & 0.022 & 0.017 & 0.0064 & 1e-05 & 2e-05 & 10.99 & 34.049 & 0.081 \\
SN2000cx & N0524 & 0.0069 & 0.00793 & 13.122 & 0.006 & 0.842 & 0.005 & 0.053 & 0.00781 & 1e-05 & 4e-05 & 10.929 & 32.212 & 0.09 \\
SN2002cs & N6702 & 0.0153 & 0.01577 & 15.202 & 0.013 & 1.196 & 0.013 & 0.037 & 0.01703 & 8e-05 & 0.00017 & 11.324 & 33.992 & 0.087 \\
SN2002ha & N6964 & 0.0131 & 0.01405 & 14.743 & 0.005 & 0.833 & 0.005 & -0.065 & 0.00707 & 1e-05 & 2e-05 & 11.021 & 33.685 & 0.096 \\
SN2003hv & N1201 & 0.0051 & 0.0056 & 12.447 & 0.049 & 0.788 & 0.019 & -0.08 & 0.0608 & -1e-05 & 0.0024 & 10.565 & 31.347 & 0.074 \\
SN2006ef & N0809 & 0.017 & 0.01787 & 15.438 & 0.067 & 0.857 & 0.02 & -0.024 & 0.08071 & -4e-05 & 0.00449 & 10.35 & 34.376 & 0.109 \\
SN2007cv & IC2597 & 0.0087 & 0.0076 & 15.06 & 0.012 & 0.711 & 0.01 & 0.017 & 0.01697 & 7e-05 & 0.00014 & 11.478 & 33.673 & 0.082 \\
SN2008L & N1259 & 0.0189 & 0.0194 & 15.273 & 0.019 & 0.771 & 0.016 & -0.063 & 0.02484 & 0.00015 & 0.00036 & 10.888 & 34.365 & 0.105 \\
SN2008R & N1200 & 0.0129 & 0.01349 & 15.27 & 0.008 & 0.633 & 0.006 & 0.12 & 0.01063 & 0.0 & 6e-05 & 11.25 & 33.66 & 0.08 \\
SN2008hs & N0910 & 0.0166 & 0.01735 & 16.031 & 0.02 & 0.611 & 0.006 & 0.075 & 0.02441 & 3e-05 & 0.0004 & 11.504 & 34.459 & 0.093 \\
SN2008hv & N2765 & 0.0136 & 0.01254 & 14.741 & 0.009 & 0.869 & 0.004 & -0.046 & 0.0114 & 1e-05 & 8e-05 & 10.55 & 33.725 & 0.083 \\
SN2008ia & ESO125-G006 & 0.0225 & 0.02198 & 15.881 & 0.011 & 0.836 & 0.006 & -0.035 & 0.01556 & 1e-05 & 0.00012 & 10.77 & 34.896 & 0.195 \\
SN2010Y & N3392 & 0.0113 & 0.0109 & 15.005 & 0.013 & 0.658 & 0.006 & 0.022 & 0.0164 & 2e-05 & 0.00017 & 10.58 & 33.861 & 0.088 \\
SN2014bv & N4386 & 0.0057 & 0.00559 & 14.007 & 0.018 & 0.621 & 0.015 & 0.214 & 0.0222 & -9e-05 & 0.00032 & 10.48 & 32.427 & 0.08 \\
SN2015bp & N5839 & 0.0047 & 0.00407 & 13.775 & 0.01 & 0.703 & 0.006 & 0.082 & 0.01281 & 3e-05 & 0.0001 & 9.979 & 32.369 & 0.078 \\
SN2016ajf & N1278 & 0.0198 & 0.0203 & 16.922 & 0.05 & 0.488 & 0.015 & 0.556 & 0.07507 & 0.00011 & 0.0025 & 11.436 & 34.202 & 0.106 \\
SN2019ein & N5353 & 0.0084 & 0.00776 & 14.178 & 0.018 & 0.863 & 0.01 & -0.014 & 0.0222 & 0.00011 & 0.00032 & 11.484 & 32.711 & 0.076 \\
\enddata
\end{deluxetable*}
\end{longrotatetable}

\begin{longrotatetable}
\begin{deluxetable*}{llllllllllllllll}
\tablecaption{SBF calibration (\citealt{khetan21}) SN Ia light-curve fitting with \texttt{SNooPy}, host galaxy stellar mass, and distance modulus.\label{sbftable}}
\tablewidth{700pt}
\tabletypesize{\scriptsize}
\tablehead{
\colhead{$Name$} & \colhead{$Host$} & \colhead{$z_{cmb}$} & 
\colhead{$z_{hel}$} & \colhead{$B_{max}$} & 
\colhead{$eB_{max}$} & \colhead{$s_{BV}$} & 
\colhead{$es_{BV}$} & \colhead{$(B-V)$} & 
\colhead{$e(B-V)$} & \colhead{$cov(B_{max}, \ s_{BV})$} & \colhead{$cov(B_{max}, \ B-V)$}  & \colhead{$log \ M_*$} & \colhead{$\mu$}&\colhead{$e\mu$} \\ 
&  & &
 & \colhead{(mag)} & 
\colhead{(mag)} & & 
 & \colhead{(mag)} & 
\colhead{(mag)} &  &   & \colhead{$(M_{\odot})$} & \colhead{(mag)}&\colhead{(mag)} \\
} 
\startdata
SN1970J & N7619 & 0.012 & 0.0132 & 14.278 & 0.084 & 0.704 & 0.071 & -0.144 & 0.12458 & 0.0039 & 0.00706 & 11.34 & 33.341 & 0.081 \\
SN1980N & N1316 & 0.0055 & 0.00587 & 12.458 & 0.009 & 0.904 & 0.006 & 0.07 & 0.01204 & 0.0 & 8e-05 & 11.514 & 31.59 & 0.05 \\
SN1981D & N1316 & 0.0055 & 0.00587 & 12.554 & 0.034 & 0.793 & 0.033 & 0.173 & 0.03847 & 0.00017 & 0.00116 & 11.514 & 31.59 & 0.05 \\
SN1983G & N4753 & 0.0032 & 0.00206 & 12.863 & 0.081 & 1.005 & 0.037 & 0.188 & 0.08538 & -0.00015 & 0.00656 & 11.148 & 31.92 & 0.197 \\
SN1992A & N1380 & 0.0059 & 0.00626 & 12.539 & 0.006 & 0.787 & 0.008 & 0.039 & 0.00721 & 2e-05 & 4e-05 & 10.931 & 31.632 & 0.075 \\
SN1992bo & E352-057 & 0.0177 & 0.01851 & 15.787 & 0.009 & 0.744 & 0.008 & -0.01 & 0.0114 & 3e-05 & 8e-05 & 10.395 & 34.27 & 0.15 \\
SN1994D & N4526 & 0.0026 & 0.00149 & 11.84 & 0.006 & 0.824 & 0.005 & -0.047 & 0.00849 & 1e-05 & 4e-05 & 10.996 & 31.32 & 0.12 \\
SN1995D & N2962 & 0.0077 & 0.00656 & 13.265 & 0.008 & 1.171 & 0.009 & -0.053 & 0.01 & -2e-05 & 6e-05 & 10.597 & 32.532 & 0.084 \\
SN1996X & N5061 & 0.0079 & 0.00689 & 13.075 & 0.016 & 0.917 & 0.013 & -0.033 & 0.01887 & 6e-05 & 0.00026 & 11.057 & 32.26 & 0.19 \\
SN1997E & N2258 & 0.0135 & 0.01354 & 15.192 & 0.007 & 0.831 & 0.011 & 0.045 & 0.00922 & 3e-05 & 5e-05 & 11.199 & 33.781 & 0.094 \\
SN1998bp & N6495 & 0.0102 & 0.01043 & 15.384 & 0.01 & 0.549 & 0.009 & 0.265 & 0.01281 & 1e-05 & 0.0001 & 10.462 & 33.1 & 0.15 \\
SN2000cx & N0524 & 0.0069 & 0.00793 & 13.122 & 0.006 & 0.842 & 0.005 & 0.053 & 0.00781 & 1e-05 & 4e-05 & 10.929 & 32.21 & 0.09 \\
SN2003hv & N1201 & 0.0051 & 0.0056 & 12.447 & 0.049 & 0.788 & 0.019 & -0.08 & 0.0608 & -1e-05 & 0.0024 & 10.565 & 31.347 & 0.074 \\
SN2006dd & N1316 & 0.0055 & 0.00587 & 12.228 & 0.005 & 0.951 & 0.003 & -0.088 & 0.00707 & 1e-05 & 2e-05 & 11.514 & 31.59 & 0.05 \\
SN2007on & N1404 & 0.0062 & 0.00649 & 13.032 & 0.009 & 0.588 & 0.003 & 0.116 & 0.01204 & 1e-05 & 8e-05 & 10.932 & 31.526 & 0.072 \\
SN2008Q & N524 & 0.0069 & 0.00793 & 13.471 & 0.013 & 0.791 & 0.023 & -0.04 & 0.01703 & 0.00016 & 0.00017 & 10.929 & 31.922 & 0.212 \\
SN2011iv & N1404 & 0.0062 & 0.00649 & 12.463 & 0.008 & 0.699 & 0.007 & 0.031 & 0.01063 & 1e-05 & 6e-05 & 10.932 & 31.526 & 0.072 \\
SN2012cg & N4424 & 0.0026 & 0.00146 & 11.996 & 0.008 & 1.013 & 0.007 & 0.071 & 0.00943 & 3e-05 & 6e-05 & 9.706 & 31.02 & 0.18 \\
SN2014bv & N4386 & 0.0057 & 0.00559 & 14.007 & 0.018 & 0.621 & 0.015 & 0.214 & 0.0222 & -9e-05 & 0.00032 & 10.48 & 32.427 & 0.08 \\
SN2015bp & N5839 & 0.0047 & 0.00407 & 13.775 & 0.01 & 0.703 & 0.006 & 0.082 & 0.01281 & 3e-05 & 0.0001 & 9.979 & 32.369 & 0.078 \\
SN2016coj & N4125 & 0.0048 & 0.00448 & 13.099 & 0.009 & 0.807 & 0.009 & 0.0 & 0.01273 & 3e-05 & 8e-05 & 11.083 & 31.922 & 0.258 \\
SN2017fgc & N0474 & 0.0067 & 0.00772 & 14.021 & 0.014 & 1.17 & 0.013 & 0.378 & 0.01664 & -1e-05 & 0.0002 & 10.568 & 32.536 & 0.133 \\
SN2018aoz & N3923 & 0.0069 & 0.0058 & 12.746 & 0.009 & 0.875 & 0.005 & 0.012 & 0.01082 & 2e-05 & 8e-05 & 11.204 & 31.795 & 0.101 \\
SN2020ue & N4636 & 0.0043 & 0.00313 & 11.983 & 0.01 & 0.714 & 0.011 & -0.089 & 0.01281 & 6e-05 & 0.0001 & 10.803 & 30.83 & 0.13 \\
\enddata
\end{deluxetable*}
\end{longrotatetable}

% 

%\section{Details on Error}

\bibliography{bibSyedUddin}{}
\bibliographystyle{aasjournal}

\end{document}